\documentclass[aps,twocolumn,superscriptaddress,floatfix]{revtex4-2}

\usepackage{amsmath,amssymb}
\usepackage{graphicx}
\usepackage{bm}
\usepackage[colorlinks=true,allcolors=blue]{hyperref}
\usepackage{xcolor}

\begin{document}

\title{Hamiltonian fingerprints of a collective mode across the BCS--BEC crossover}

\author{Yogeshwar Prasad}
\email{yogeshwar@snu.ac.kr}
\affiliation{Center for Condensed Matter Theory, Department of Physics, Indian Institute of Science, Bangalore 560012, India}
\affiliation{Department of Physics, Hanyang University, Seoul 04763, Korea}
\affiliation{Research Institute of Basic Sciences, Seoul National University, Seoul 08826, Korea}

\date{\today}

\begin{abstract}
Collective excitations can survive profound changes in the microscopic degrees
of freedom that generate them. We track a layer-odd collective response from an
exact weak-coupling hybridisation anchor into the paired regime of an
attractive Hubbard bilayer and find that its Hamiltonian sensitivities evolve
in a manner \emph{consistent with} a crossover from single-particle interlayer
hybridisation towards collective pair superexchange. At weak coupling, an exact
identity for identical tunnel-coupled layers fixes a relative-phase Gaussian
kernel zero at $2t_h$, independent of the intralayer hopping. Sign-free
determinant quantum Monte Carlo reveals an interaction-driven softening of the
many-body spectral scale without analytic continuation, while Gaussian response
theory supplies the continuous pole interpretation and shows a shift from
layer-density to relative-pair-phase character. In the paired regime, the
characteristic response scale acquires a growing fractional sensitivity to the
intralayer hopping and an interaction dependence partway towards the $1/U$ law.
These results show how a collective excitation can be tracked across a
fermion-to-composite-boson crossover while constraining changes in its
microscopic drive.
\end{abstract}

\maketitle


A well-defined collective excitation can persist across a large change of
coupling even when the microscopic process responsible for it is entirely
replaced. The BCS--BEC crossover is the canonical setting: a single superfluid
connects weakly bound, overlapping Cooper pairs to tightly bound composite
bosons~\cite{Eagles1969,Leggett1980,NozieresSchmittRink1985,Randeria2014,Bloch2008},
realised concretely in the attractive Hubbard model~\cite{Toschi2005}, and
collective frequencies have long served as sensitive probes of this crossover
in trapped gases~\cite{Astrakharchik2005}. Far less explored is whether a
specific \emph{finite-frequency} collective mode can be followed continuously
across the crossover, as recently demonstrated for the Holstein model within
dynamical mean-field theory~\cite{ParkChoi2024}, and, above all, whether
coupling-resolved Hamiltonian sensitivities can reveal how its microscopic
origin changes. A classic precedent is the isotope effect in superconductivity,
where changing the ionic mass revealed the lattice contribution to pairing by
selectively shifting the phonon sector of the Hamiltonian. At the few-body
level, double-well dynamics have likewise been observed to cross from direct
tunnelling to the $4J^2/U$ superexchange scale~\cite{Trotzky2008}. Here we
extend this logic to a collective excitation of a many-body system.

\begin{figure*}[t]
  \centering
  \includegraphics[width=\textwidth]{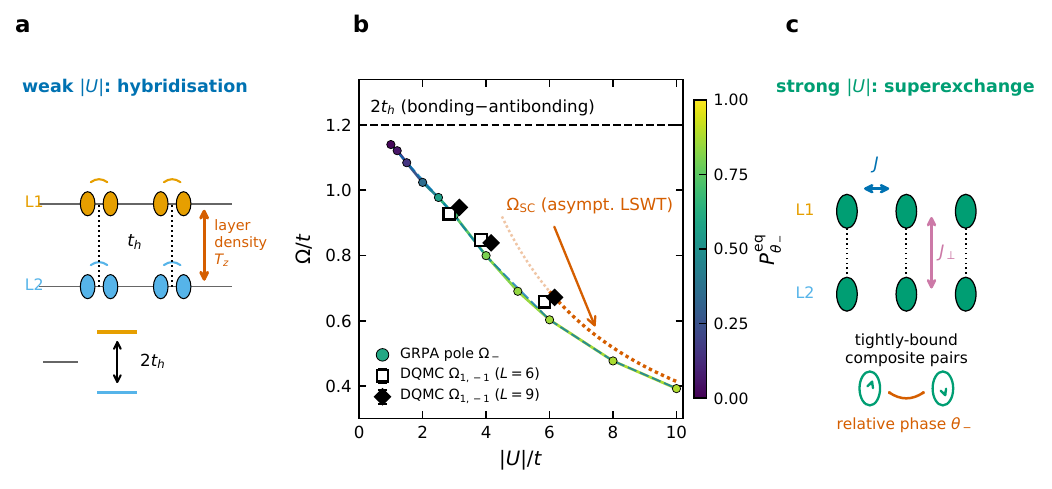}
  \caption{\textbf{Layer-odd response across the crossover.}
  (a)~Weak coupling: single-particle interlayer hybridisation
  (bonding--antibonding) with kernel anchor $2t_h$. (b)~The continuous Gaussian
  (GRPA) pole $\Omega_-(|U|)$ softens from near $2t_h$ towards the
  strong-coupling scale, \emph{coloured by its response composition}
  (density-dominated $\to$ pair-phase-dominated; equal-time metric, Supplement
  Sec.~S10). Continuation-free DQMC moment scales at $U=3,4,6$ and $L=6,9$ are
  overlaid; the dashed blue curve shows the corresponding GRPA moment scale
  (Methods). (c)~Strong coupling: tightly bound pairs exchange through pair
  superexchange, with leading-LSWT scale
  $\Omega_{\rm SC}=\sqrt{zJJ_\perp}F(n)$ (Supplement Sec.~S8). DQMC points are
  at $\beta=12$ (error bars: $1\sigma$ bin-jackknife); the GRPA branch and
  strong-coupling scale are at $T=0$. All results use $n=1.10$ and $t_h=0.6$.}
  \label{fig:hero}
\end{figure*}

We study this in a concrete, sign-free, numerically controlled platform: an
AA-stacked honeycomb bilayer with intralayer hopping $t$ (set to one as the
unit of energy), interlayer hybridisation $t_h$, and on-site attraction
$|U|$~\cite{Micnas1990,Robaszkiewicz1981} [Hamiltonian in Methods,
Eq.~\eqref{eq:H}].
The relevant excitation is the \emph{layer-odd} mode, probed through
fluctuations of the population imbalance $T_z=(N_1-N_2)/2$; the experimentally
measured imbalance is $Q=N_1-N_2=2T_z$. Because only interlayer tunnelling
changes $T_z$, its dynamics are tied directly to the interlayer-current
quadrature, $[H,T_z]=2it_h T_y$, an exact identity underlying the moment
relations used below. Throughout we fix $n=1.10$ and $t_h=0.6$ (remaining
parameters in Methods; conventions in Supplement Sec.~S1).

The mode has two parametrically distinct analytic descriptions
(Sec.~\ref{sec:endpoints}): a single-particle hybridisation feature with a
$t$-independent Gaussian kernel anchor at $2t_h$ at weak coupling, and a
\emph{collective pair-transfer} excitation whose strong-coupling effective
Hamiltonian contains intralayer superexchange. We track the layer-odd response
between these limits using DQMC at $U=3,4,6$ (Sec.~\ref{sec:survive}) and test
whether its microscopic origin changes through coupling-resolved responses of
its characteristic spectral scale. We call these responses \emph{Hamiltonian
fingerprints}, quantified by the hopping and interaction sensitivities
$\Lambda_t$ and $\Lambda_U$. Across the paired regime, the continuation-free
many-body scale softens at both lattice sizes, while its measured fingerprints
move away from the exact $t$-independent weak-coupling reference towards the
Hamiltonian dependences of the pair-superexchange regime, consistent with a
change in microscopic drive.

The evidence combines an exact weak-coupling kernel anchor, a controlled
second-order effective theory at strong coupling, a continuous Gaussian
random-phase-approximation (GRPA) interpolation between them, and
nonperturbative DQMC at $U=3,4,6$ for both $L=6$ and $L=9$, without analytic
continuation. Related interlayer and relative-phase modes occur in multiband
and two-band superconductors~\cite{Leggett1966,%
SharapovGusyninBeck2002,Anishchanka2007,BurnellHuLin2010,Blumberg2007,Cuozzo2024,%
IskinSadeMelo2005}, including superconducting bilayers~\cite{Hackner2023}, as
well as in bilayer and attractive Fermi--Hubbard cold-atom
systems~\cite{Gall2021,Hartke2023,Rydow2025} and coherently coupled Bose
gases~\cite{AbadRecati2013}.


\section{Weak-coupling anchor and pair-superexchange theory}
\label{sec:endpoints}

The two limits differ qualitatively in their dependence on intralayer motion.
At weak coupling, the Gaussian relative-phase anchor is exactly independent of
$t$; at strong coupling, the second-order pair-effective Hamiltonian depends
explicitly on $t$. Figure~\ref{fig:hero} summarises this evolution from the
$2t_h$ hybridisation scale to pair superexchange.

\subsection{Exact weak-coupling anchor at $2t_h$}
\label{sec:anchor}

At weak pairing, the elementary fermionic states are bonding and antibonding
combinations of the two layers, split by the interlayer hopping by $2t_h$.
This hybridisation scale fixes an exact zero of the diagonal relative-phase
Gaussian kernel,
\begin{equation}
  K_{\theta_-\theta_-}(\omega=2t_h,\,\mathbf q{=}0)=0,
  \label{eq:kernelzero}
\end{equation}
for any single-particle dispersion $h_0(\mathbf k;t)$, provided the two layers
are identical and the pairing saddle is layer-symmetric. The result therefore
applies to two identically coupled copies of an arbitrary system and is not
specific to the honeycomb band structure (Supplement Sec.~S3).

The zero belongs to the diagonal relative-phase kernel, not to the full
observable response: mixing with the other layer-odd channels can shift the
collective pole away from $2t_h$. Nevertheless, the anchor itself contains no
intralayer hopping, so its $t$-sensitivity vanishes exactly. At $U=0$, where
the distinction between kernel anchor and observable response disappears, all
$\mathbf q=0$ $T_z$ spectral weight lies at the bonding--antibonding
splitting, giving $\Omega_{1,-1}=2t_h$ and $\Lambda_t=0$ exactly.

At finite interaction, density-channel mixing already moves the Gaussian pole
strongly: it runs from $\Omega_-\simeq1.140$ near the hybridisation scale to
$0.392$ at $|U|=10$, a factor of $2.9$, and is already $23\%$, $33\%$ and
$50\%$ below $2t_h$ at $|U|=3,4,6$. The Gaussian pole is therefore \emph{not}
pinned near $2t_h$ at any finite coupling, and no ``protected value'' should be
read into the anchor. Beyond Gaussian order the interaction torque
$-i[H_U,T_y]\neq0$ removes any exact pinning to $2t_h$. The existence of a
layer-odd resonance at the interlayer scale was established
previously~\cite{FKK1996}. The additional result used here is that the
Gaussian anchor is independent of intralayer motion for arbitrary dispersion.
A measured $t$-dependence of the layer-odd many-body response therefore
excludes a purely single-particle hybridisation description and provides the
contrast exploited in Sec.~\ref{sec:discriminator}.

\subsection{Pair-superexchange effective theory}
\label{sec:strongcoupling}

In the strongly paired regime, electrons form tightly bound pairs, and
single-fermion motion costs the pair-breaking energy $|U|$. The layer-odd
excitation then becomes a relative-phase oscillation of the two pair
condensates, a bilayer Josephson-plasma (Leggett-type) mode. Its dynamics
arise through virtual pair breaking: both intralayer and interlayer pair
motion are generated at second order in the fermion hopping.

Mapping the paired sector onto an $\eta$-pseudospin (hard-core-boson)
magnet~\cite{Yang1989,Zhang1990} makes this explicit. To second order,
\begin{equation}
  H_{\rm eff}^{(2)}=\frac{4t^2}{|U|}\,\mathcal{H}(r,n),\qquad
  \Omega_{\rm eff}=\frac{4t^2}{|U|}\,\Phi(r,n),\qquad r=\frac{t_h^2}{t^2},
  \label{eq:Heff}
\end{equation}
with intralayer and rung exchange couplings $J=4t^2/|U|$ and
$J_\perp=4t_h^2/|U|$. Within this effective theory, all excitation energies
therefore carry the overall $1/|U|$ scaling characteristic of virtual pair
breaking. Their dependence on $t$ is non-universal because changing $t$
modifies both the overall prefactor and the ratio $r$. However,
the effective Hamiltonian contains intralayer exchange explicitly, making the
measured $t$-dependence of the layer-odd response a direct diagnostic of
physics beyond an isolated interlayer scale.

Linear spin-wave theory gives one semiclassical evaluation of $\Phi$,
\begin{equation}
  \Omega_{\rm LSWT}=\sqrt{z\,J\,J_\perp}\,F(n)
  =\frac{4\sqrt{z}\,t\,t_h}{|U|}\,F(n),\qquad z=3,
  \label{eq:OmegaSC}
\end{equation}
where $F(n)$ is a weak density correction (Supplement Sec.~S8). This gives
$\Phi_{\rm LSWT}\!\propto\!\sqrt{r}$ up to the weak $F(n)$ dependence, and hence
$\Lambda_t^{\rm LSWT}\simeq1$.
Because the effective pseudospin is $S=\tfrac12$, quantum corrections are not
parametrically controlled: Eq.~\eqref{eq:OmegaSC} is therefore a semiclassical
reference, not an exact quantum endpoint. For the general effective model,
$\Lambda_t=2-2\,d\ln\Phi/d\ln r$, so its exact quantum value need not equal
one, or even be positive.

The key contrast is therefore physical rather than numerical: the
weak-coupling hybridisation anchor is $t$-independent, whereas the
pair-effective dynamics involve intralayer exchange. A resolved
$t$-dependence excludes both a purely local $2t_h$ scale and an isolated-rung
scale $\Omega_{\rm rung}=J_\perp$ (Sec.~\ref{sec:discriminator}). The same
leading structure appears on the square lattice with $z=4$ (Supplement
Sec.~S12), showing that it reflects the strong-coupling pair dynamics rather
than the honeycomb geometry.

\section{Layer-odd response across the crossover}
\label{sec:survive}
\begin{figure*}[t]
  \centering
  \includegraphics[width=\textwidth]{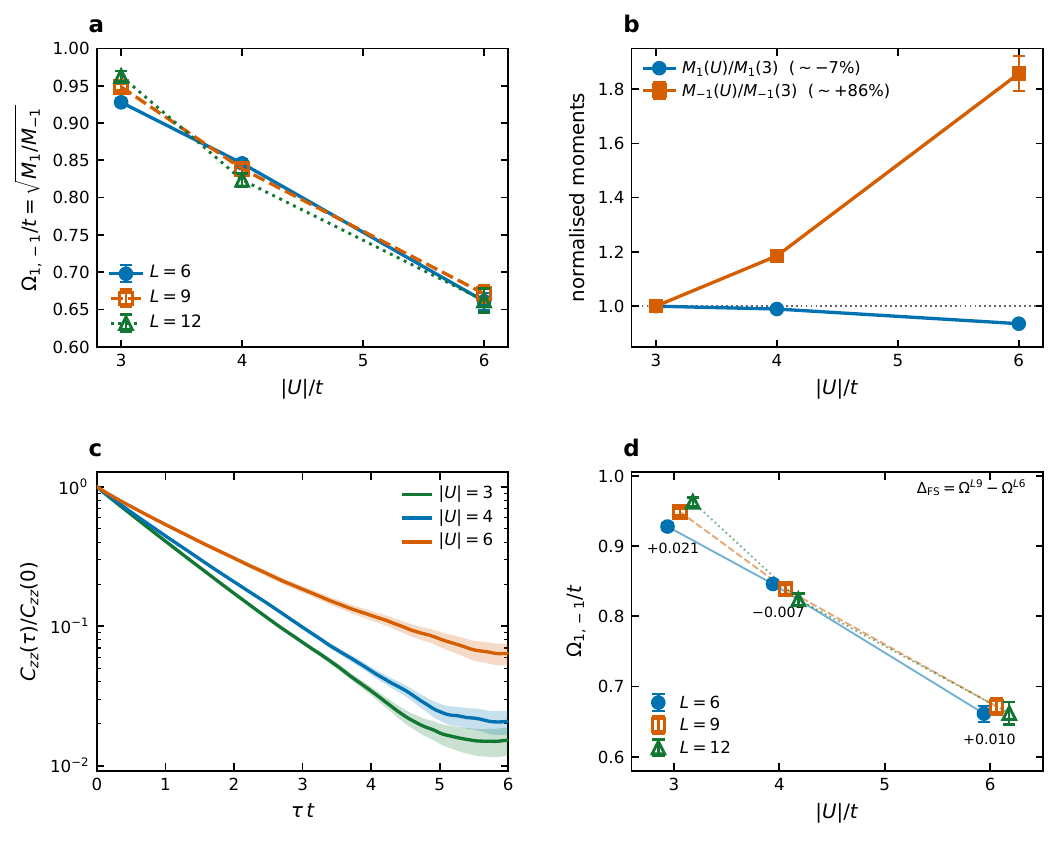}
  \caption{\textbf{Continuation-free softening from exact moments.}
  (a)~The moment scale $\Omega_{1,-1}=\sqrt{M_1/M_{-1}}$ softens with $|U|$ at
  both lattice sizes. (b)~Normalised moments at $L=6$: $M_{-1}$, which
  emphasises low-frequency spectral weight, rises strongly while
  $M_1=t_h\langle T_x\rangle$ remains nearly flat, driving the softening.
  (c)~KMS-folded correlator $C_{zz}(\tau)/C_{zz}(0)$ at $U=3,4,6$: the slower
  decay at $U=6$ directly signals a shift towards lower-frequency weight.
  (d)~Finite-size comparison of $\Omega_{1,-1}$ at $L=6$ and $L=9$: increasing
  the lattice size does not erase the softening (Supplement Sec.~S6). Error
  bars are $1\sigma$ bin-jackknife uncertainties; shaded bands in (c) show the
  corresponding $1\sigma$ statistical uncertainty.}
  \label{fig:trajectory}
\end{figure*}

\subsection{Continuous Gaussian branch}

The kernel zero of Sec.~\ref{sec:anchor} provides the weak-coupling anchor.
Within GRPA~\cite{Anderson1958}, the layer-imbalance response contains a
single non-crossing collective branch across the full coupling range. Its
pole softens monotonically from $\Omega_-\simeq1.14$ near the hybridisation
scale at $|U|=1$ to $\Omega_-\simeq0.39$ at $|U|=10$, while the $T_z$ residue
increases with coupling. The branch remains below the layer-imbalance-active
two-quasiparticle threshold throughout. For $|U|\lesssim2.5$ it overlaps the
absolute pair continuum, which is dark to $T_z$; at stronger coupling it lies
below that continuum. At large $|U|$, the GRPA branch approaches the
pair-superexchange scale of Eq.~\eqref{eq:OmegaSC}. This agreement is an
internal consistency between the Gaussian fermionic description and the
strong-coupling effective-theory calculation, not a many-body confirmation of
the strong-coupling law (Supplement Sec.~S8).

The nonperturbative question is whether a corresponding low-frequency
layer-odd response persists in the paired regime. The DQMC window $U=3,4,6$
lies in the locally paired part of the crossover: at $L=9$,
$D/D_{\rm uncorr}$, with $D_{\rm uncorr}=(n/2)^2$ the uncorrelated value,
increases from $1.26$ to $1.54$ across these couplings,
while the absolute double occupancy moves towards the composite-boson limit
$D=n/2$ (Supplement Sec.~S11). The resulting many-body softening is
established below directly from exact spectral moments.

\subsection{Continuation-free DQMC trajectory}
\label{sec:traj}

DQMC can detect the softening without reconstructing the spectrum, in the
spirit of frequency-moment analyses of imaginary-time QMC
data~\cite{Dornheim2023moments}. With
$M_p=\int_0^\infty\omega^p A_{zz}(\omega)\,d\omega$, two moments of the
layer-imbalance response are exact at all $\beta$ (Supplement Sec.~S4).
Writing $T_x=\tfrac12\sum_{i\sigma}(c^\dagger_{1i\sigma}c_{2i\sigma}+\text{H.c.})$
for the interlayer-coherence operator, with $H_h=-2t_h T_x$, they are
\begin{equation}
  M_1 = t_h\langle T_x\rangle = -\tfrac12\langle H_h\rangle,\qquad
  M_{-1} = \tfrac12\int_0^\beta C_{zz}(\tau)\,d\tau.
\end{equation}
The first moment is fixed by the mean tunnelling energy, whereas $M_{-1}$
emphasises low-frequency spectral weight. Their ratio defines the
characteristic scale $\Omega_{1,-1}=\sqrt{M_1/M_{-1}}$, requiring no analytic
continuation, pole fit or frequency window. At $L=6$, sign-free DQMC gives
$\Omega_{1,-1}=0.928(6),\,0.847(6),\,0.658(11)$ for $U=3,4,6$, respectively,
a monotonic softening with both intervals strongly resolved
[Fig.~\ref{fig:trajectory}(a)]. The same monotonic ordering holds at the two
larger sizes, $0.9484(48),\,0.8386(69),\,0.6714(114)$ at $L=9$ and
$0.9629(68),\,0.8238(85),\,0.6620(159)$ at $L=12$.

The individual moments show where that softening comes from. $M_{-1}$ rises
strongly with coupling while $M_1=t_h\langle T_x\rangle$ remains nearly flat
[Fig.~\ref{fig:trajectory}(b)], so spectral weight is redistributed towards
lower frequency. The same evolution is visible directly in the folded
imaginary-time correlator [Fig.~\ref{fig:trajectory}(c)] and also occurs at
$L=9$ (Supplement Secs.~S4 and~S6). This moment-level redistribution is the
primary nonperturbative result.

\subsection{Finite-size behaviour: $L=6$, $9$ and $12$}

Increasing the lattice size does not erase or strongly reduce the softening.
At $L=9$, the end-to-end change is $\Delta_{3\to6}^{L9}=0.276(12)$, compared
with $0.266(13)$ at $L=6$, and $0.301(17)$ at $L=12$: the end-to-end
softening is stable across $L=6,9,12$ within the present statistical
precision. Three sizes establish the absence of an obvious reversal, not a
thermodynamic extrapolation. While the individual characteristic scales shift by at
most about $2\%$ [Fig.~\ref{fig:trajectory}(d)]. The per-coupling shifts have
no common sign, so no systematic size drift underlies the trajectory. Full
per-seed, warm-up and finite-size checks are given in Supplement Sec.~S6.

\section{Density-to-phase character transfer}
\label{sec:character}
\begin{figure*}[t]
  \centering
  \includegraphics[width=\textwidth]{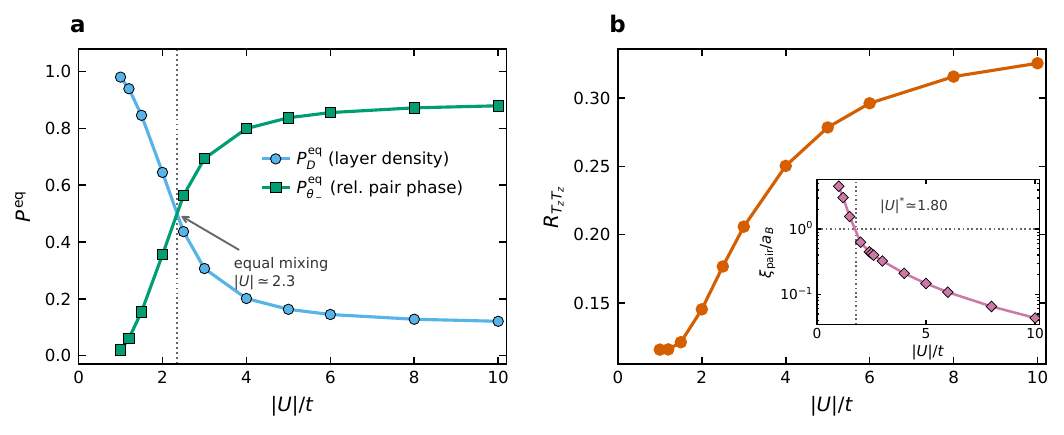}
  \caption{\textbf{GRPA equal-time response composition of the layer-odd branch.}
  (a)~The equal-time-normalised response shifts from layer-density to
  relative-pair-phase character with increasing $|U|$, with equal mixing near
  $|U|\simeq2.3$. The weights are defined using the equal-time covariance
  metric and are therefore metric-dependent (Supplement Sec.~S10). (b)~The
  absolute $T_z$ residue increases with $|U|$, so the branch remains visible in
  the layer-imbalance response. Inset: the mean-field pair size
  $\xi_{\rm pair}$ falls through one Bravais unit near $|U|^\star\simeq1.8$,
  marking the Cooper-pair to compact-pair crossover in approximately the same
  coupling window (Supplement Sec.~S11). The character assignment is
  Gaussian-level; independently, the GRPA-blind DQMC reconstruction establishes
  a dominant low-frequency many-body feature (Supplement Sec.~S7). All results
  use $n=1.10$.}
  \label{fig:character}
\end{figure*}

Having established nonperturbatively that the layer-odd many-body response
softens across the paired regime, we next ask how the continuously tracked
Gaussian branch changes its internal character. Within Gaussian response, the
layer-odd pole becomes progressively more relative-phase-like across the
crossover. Quantitatively, its pole residue,
normalised by the equal-time covariance metric to remove dependence on
internal-field rescalings, shifts from the layer-density channel $T_z$
towards the relative-pair-phase channel $\theta_-$. The corresponding weights
satisfy $P_{\theta_-}^{\rm eq}=1-P_D^{\rm eq}$, with equal mixing near
$|U|\simeq2.3$ [Fig.~\ref{fig:character}(a)]. This composition is
metric-dependent: under static-susceptibility normalisation the contrast
collapses towards equal weights (Supplement Sec.~S10). The absolute $T_z$
residue increases with $|U|$, so the branch remains visible in the
layer-imbalance response; within GRPA, that response also remains close to
single-mode saturation across the coupling range.

This character assignment is a Gaussian-theory result. The many-body data
enter independently: GRPA-blind spectral reconstruction (Methods) places
substantial reconstructed weight progressively lower in frequency with
increasing $|U|$,
consistent with the exact-moment trajectory, with $W_{\rm low}\geq0.92$ at
all three DQMC couplings (Supplement Sec.~S7). Synthetic identifiability
tests show that these $W_{\rm low}$ values imply strong low-frequency
concentration within the tested spectral family, while the precise peak
position and intrinsic linewidth are not identifiable. GRPA therefore
supplies the continuous pole interpolation and its microscopic character,
whereas DQMC independently establishes a concentrated low-frequency many-body
response. DQMC does not determine the GRPA eigenvector character.

\section{Hopping sensitivity of the layer-odd response}
\label{sec:discriminator}
\begin{figure*}[t]
  \centering
  \includegraphics[width=\textwidth]{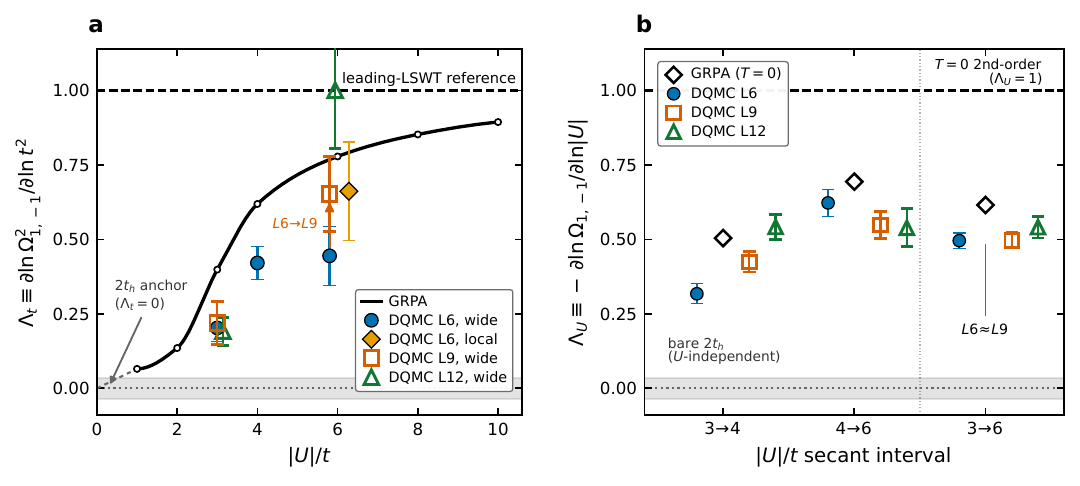}
  \caption{\textbf{Hamiltonian fingerprints across the crossover.}
  (a)~Hopping sensitivity $\Lambda_t\equiv\partial\ln\Omega_{1,-1}^2/\partial\ln t^2$
  versus $|U|/t$: the layer-odd scale increasingly senses intralayer motion,
  with $\Lambda_t$ rising at both lattice sizes ($\approx\!2\sigma$ end-to-end
  at $L=6$, $\approx\!3\sigma$ at $L=9$) away from the $t$-independent $2t_h$
  reference ($\Lambda_t=0$) and towards the leading-LSWT regime
  ($\Lambda_t^{\rm LSWT}\simeq1$, not an exact quantum target). The black curve is GRPA
  evaluated with the same moment observable and finite-difference stencil;
  DQMC error bars are $1\sigma$ bin-jackknife uncertainties.
  (b)~Interaction sensitivity $\Lambda_U\equiv-\partial\ln\Omega_{1,-1}/\partial\ln|U|$
  for the $3\!\to\!4$, $4\!\to\!6$, and $3\!\to\!6$ secants: $\Lambda_U$ moves
  from the $U$-independent hybridisation value $0$ towards the second-order
  $1/|U|$ reference $\Lambda_U=1$, with the $3\!\to\!6$ value agreeing between
  $L=6$ and $L=9$. Full values, significances, stencil definitions and the
  $U^2$-normalised variant $D_t$ are given in Methods and Supplement Sec.~S9.}
  \label{fig:discriminator}
\end{figure*}

The central diagnostic asks whether the layer-odd response acquires
sensitivity to motion within each layer. The two limits differ sharply in
this respect (Sec.~\ref{sec:endpoints}), so we vary $t$ at fixed $|U|$,
$t_h$ and filling $n$, and measure how the characteristic scale responds.
Leading spin-wave theory predicts positive stiffening, but the exact quantum
sign is not fixed a priori. We therefore interpret the measured
$t$-sensitivity as a test of intralayer Hamiltonian dependence, not as a
unique identification of superexchange. We quantify it by
\begin{equation}
  \Lambda_t\equiv\frac{\partial\ln\Omega_{1,-1}^2}{\partial\ln t^2},
  \label{eq:Lambdat}
\end{equation}
the logarithmic derivative of the same moment scale used throughout. Since
$\Omega_{1,-1}^2$ is the mean-square frequency under the
low-frequency-weighted measure $(A/\omega)\,d\omega$, $\Lambda_t$ measures
how intralayer hopping changes that characteristic spectral scale. A purely
$2t_h$-controlled response gives $\Lambda_t=0$, whereas leading spin-wave
theory gives $\Lambda_t\simeq1$ for pair-superexchange dynamics
(Sec.~\ref{sec:endpoints}). Varying $t$ at fixed absolute $|U|$, rather than
fixed $U/t$, ensures that a nonzero $\Lambda_t$ reflects genuine
intralayer-hopping dependence rather than trivial rescaling.

We determine $\Lambda_t$ using two independent finite-difference stencils
(Methods). For the wide stencil it rises from $0.203(46)$ at $U=3$ to
$0.445(99)$ at $U=6$ for $L=6$, and from $0.220(72)$ to $0.653(126)$ for
$L=9$. The end-to-end increase is resolved at approximately $2\sigma$ and
$3\sigma$, respectively [Fig.~\ref{fig:discriminator}(a); full stencil and
size comparisons in Supplement Sec.~S9]. The $L=12$ stencil gives
$0.192(47)$ at $U=3$, $0.369(55)$ at $U=4$ and $1.001(195)$ at $U=6$, the noisiest of the
three sizes but consistent with the smaller sizes and with the same rise across the window.
At $U=4$, where the $L=9$ $t$-scan is unavailable, the two sizes that do exist agree:
$0.421(55)$ at $L=6$ and $0.369(55)$ at $L=12$.
All three sizes therefore show increasing
$\Lambda_t$ across the DQMC window, with no evidence of finite-size
suppression; with three sizes we establish the absence of an obvious reversal,
not thermodynamic convergence.

GRPA, evaluated with the same moment observable and finite-difference
stencils, shows the same increasing trend towards the leading-LSWT regime.
At $L=6$ and $L=9$ the DQMC values are lower quantitatively, although the
$L=9$, $U=6$ point moves closer to the Gaussian result. The $L=12$,
$U=6$ value $1.001(195)$ is no longer below the Gaussian $0.779$, but it is
statistically consistent with the Gaussian result, and remains the noisiest
point, so we draw no conclusion from the sign of that particular offset. Both beyond-Gaussian and
finite-size effects can contribute to the remaining offset (Supplement Sec.~S9).

The layer-odd many-body response develops a stronger fractional sensitivity
to intralayer hopping as pairing increases, consistent with the Hamiltonian
dependence generated by the pair-superexchange effective theory. This
excludes a purely local $2t_h$ or isolated-rung description, for which
$\Lambda_t$ would vanish. The exact quantum mode need not stiffen with $t$
(Sec.~\ref{sec:endpoints}); what the $t$-scan establishes is that the
layer-odd many-body response increasingly senses intralayer dynamics.

A complementary, LSWT-independent test probes the $1/|U|$ dependence
associated with virtual pair breaking. Within the second-order effective
theory, every zero-temperature excitation energy carries the overall $1/|U|$
scale [Eq.~\eqref{eq:Heff}], giving the asymptotic reference
$-\partial\ln\Omega_{\rm exc}/\partial\ln|U|=1$. We define
\begin{equation}
  \Lambda_U\equiv-\frac{\partial\ln\Omega_{1,-1}}{\partial\ln|U|}.
\end{equation}
Across $U=3\to6$, DQMC gives $\Lambda_U=0.488(26)$ for $L=6$, $0.498(26)$
for $L=9$ and $0.541(36)$ for $L=12$
[Fig.~\ref{fig:discriminator}(b)]. The end-to-end values therefore
agree across the three sizes, although the sub-interval secants are not
separately size-converged (Supplement Sec.~S9). The measured exponent lies
partway between the $U$-independent hybridisation reference and the
asymptotic second-order value of one, as expected in the pre-asymptotic
crossover regime.

A $\beta=12$ Gaussian control changes $\Lambda_t$, $\Lambda_U$ and
$\Omega_{1,-1}$ by less than $10^{-4}$ (Methods), so Gaussian thermal
corrections cannot account for the DQMC--GRPA offsets. Beyond-Gaussian
correlations and residual finite-size effects can both contribute. Together,
$\Lambda_t$ and $\Lambda_U$ probe complementary Hamiltonian dependences of
the pair-effective regime: $\Lambda_U$ excludes a $U$-independent $2t_h$
scale but cannot distinguish among $1/|U|$ mechanisms, whereas $\Lambda_t$,
which an isolated rung cannot generate, rules against rung-local dynamics
without uniquely identifying the microscopic mechanism.

Finally, the positive hopping sensitivity
$\partial\Omega_{1,-1}^2/\partial t^2>0$ is a property of the many-body
spectral distribution encoded by the moment ratio, not necessarily of a
quasiparticle-pole position. The imaginary-time data do not require the peak
itself to shift (forward test in Methods). We therefore report a Hamiltonian
sensitivity of the spectral distribution, not a resolved shift of the
underlying pole.

\section{Mechanism change and experimental outlook}
\label{sec:discussion}

A collective excitation can remain trackable across a
fermion-to-composite-boson crossover even as the microscopic process driving
it changes. Across the DQMC window, the measured sensitivities move away from
the weak-coupling hybridisation reference and towards the Hamiltonian
dependences of the pair-effective regime, while the response remains in the
same layer-odd symmetry sector. The nonperturbative result is therefore that
the many-body spectral response acquires the characteristic hopping and
interaction dependences expected from pair dynamics. Continuity of the pole
and its density-to-phase character evolution are supplied by Gaussian
response theory; DQMC does not by itself establish a transmutation of the
eigenmode. Likewise, the DQMC window establishes local pairing rather than
two-dimensional phase coherence, whose determination would require a separate
Berezinskii--Kosterlitz--Thouless
analysis~\cite{Prasad2022,ZhaoParamekanti2006,Iskin2019HoneycombStiffness}.
The sensitivities therefore constrain the microscopic mechanism towards
lattice-mediated pair dynamics without uniquely identifying it.

More broadly, the same situation can arise whenever an exact symmetry label
allows an excitation to remain in one observable sector while its microscopic
composition changes. Here that label is layer parity: the $T_z$ response
shifts towards lower frequency while the Gaussian equal-time composition
evolves from layer density towards relative pair phase
(Fig.~\ref{fig:character}). This distinction between spectral continuity and
microscopic continuity is not specific to the honeycomb bilayer. The
weak-coupling kernel anchor holds for arbitrary layer-identical intralayer
dispersion, while any layer-symmetric further-neighbour hopping only enriches
the superexchange network at strong coupling, so the contrast underlying the
fingerprints is not tied to the nearest-neighbour model studied here
(Supplement Sec.~S13). The broader
pairing context includes fermionic superfluidity emerging from a
band-insulating state and related semiconductor-to-superconductor crossover
physics~\cite{KohmotoTakada1990,NozieresPistolesi1999,PrasadShenoyPRA89}.
Beyond pairing, bilayer honeycomb systems also host layer- and
stacking-controlled topological phenomena~\cite{Ghadimi2024,Mondal2026}.
Related collective modes occur across multiband superconductors, strongly correlated
fluids and cold-atom systems~\cite{Leggett1966,SharapovGusyninBeck2002,%
Anishchanka2007,BurnellHuLin2010,Blumberg2007,Cuozzo2024,IskinSadeMelo2005,%
Hackner2023,Gall2021,Hartke2023,Rydow2025,Feynman1954,Girvin1986}, while
bilayer and attractive-Hubbard platforms already provide the ingredients
needed to manipulate layer-resolved
dynamics~\cite{Gall2021,Hartke2023,Rydow2025,Tarruell2012,Esslinger2010,%
Gall2020,Fontenele2024}.

The diagnostic is experimentally direct because the layer-odd branch remains
visible in the population-imbalance channel. A weak interlayer-bias pulse
followed by the ring-down of $N_1-N_2$, or layer-odd Bragg or radio-frequency
spectroscopy~\cite{Levy2007,Zibold2010,Veeravalli2008,Chin2004}, can
determine the characteristic frequency as the Hamiltonian couplings are
varied. In the paired regime this response is the bilayer analogue of an
interlayer Leggett
mode~\cite{Leggett1966,Blumberg2007,IskinSadeMelo2005,CeaBenfatto2016}.
Detailed temperature, damping and entropy requirements are given in
Supplement Secs.~S13--S14.

\section{Methods}
\label{sec:methods}

\emph{Determinant quantum Monte Carlo.} We use sign-free auxiliary-field
DQMC~\cite{BSS1981,Hirsch1985}, crosschecked our results with ALF~\cite{ALF2017,ALF2022}
(release 2.6, master commit \texttt{3af8e24d}; three-point $\Delta\tau$ check
in Supplement Sec.~S5), for the attractive Hubbard bilayer~\cite{Prasad2024}
\begin{align}
  H =\;& -t\!\!\sum_{\ell,\langle ij\rangle,\sigma}\!\! c^\dagger_{\ell i\sigma}c_{\ell j\sigma}
       -t_h\!\sum_{i\sigma}\!\big(c^\dagger_{1i\sigma}c_{2i\sigma}+\text{H.c.}\big)
       \nonumber\\
     & -|U|\sum_{\ell i}\big(n_{\ell i\uparrow}-\tfrac12\big)\big(n_{\ell i\downarrow}-\tfrac12\big)
       -\mu\sum_{\ell i}(n_{\ell i}-1),
  \label{eq:H}
\end{align}
with layers $\ell=1,2$, honeycomb sublattices $A,B$, intralayer coordination
$z=3$, and $t=1$ as the unit of energy.
Production runs cover $|U|=3$--$6$; runs at $|U|=8,10$ developed long autocorrelation times
and are therefore excluded (Supplement Sec.~S5).
Bin autocorrelation was tested directly by rebinning the per-bin moment series
within each seed and recomputing the jackknife. At $|U|=3,4,6$ no inflation is
resolvable: the error is flat under rebinning at all three sizes, and against a
within-seed shuffle null preserving seed structure and bin counts every
deviation is below $1.8\sigma$. The integrated autocorrelation time of the
combination controlling $\Omega_{1,-1}$ is $\tau_{\rm int}\lesssim0.9$, so a
residual inflation of order $10$--$15\%$ on individual chains is not excluded;
a conservative envelope covering it changes no conclusion. The identical
test recovers error inflation factors of $\times1.45$ at $|U|=8$ and $\times2.0$
at $|U|=10$ (at $+3\sigma$ and $+9\sigma$ against the same null), which is what
excludes those couplings.
Periodic $L=6,9,12$ Bravais meshes contain $2L^2=72$, $162$ and $288$ sites per layer and
$4L^2=144$, $324$ and $576$ sites in the bilayer, respectively; all three sizes are multiples
of three, so $K$ and $K'$ are represented exactly.
The $L=6$ and $L=9$ runs are converged production ensembles ($180$--$376$ pooled
post-warm-up bins per point); the $L=12$ runs are complete and carry two
independent seeds with $45$--$47$ post-warm-up bins each,
at densities $n=1.100$, $1.104$ and $1.098$. They are drawn with open symbols
throughout to mark the smaller seed count, and no conclusion rests on them alone.
We use a density-channel Hubbard--Stratonovich decoupling in a single flavour,
for which the simulations are sign-problem-free, with $\langle{\rm sign}\rangle=1$
throughout~\cite{Scalettar1989}.
Runs use $\beta=12$, Trotter step $\Delta\tau=0.05$, stabilisation interval $N_{\rm wrap}=12$,
and time-displaced measurements ($L_\tau=1$); a custom estimator provides $\langle T_y^2\rangle$
for the $M_2$ cross-check (Supplement Sec.~S4).
The chemical potential is calibrated separately at each coupling to $n=1.10$,
giving $\mu^\star=0.265,0.221,0.160$ for $U=3,4,6$, respectively.
For the $t$-scan, $\mu$ is recalibrated at every $t$ to hold the filling fixed.

\emph{Continuation-free estimators.} $M_1=t_h\langle T_x\rangle$, obtained
from the equal-time Green function, and
$M_{-1}=\tfrac12\int_0^\beta C_{zz}(\tau)\,d\tau$, obtained from the
time-displaced layer-imbalance correlator, are exact at all $\beta$. We report
$\Omega_{1,-1}=\sqrt{M_1/M_{-1}}$ as the primary characteristic scale;
uncertainties are obtained by bin jackknife after discarding warm-up bins. The thermal equal-time value $C_{zz}(0)$ is not used as a
moment because it is not the spectral zeroth moment $M_0$ at finite
temperature. An effective-mass gap extracted from the folded arccosh estimator
over $\tau\in[0.6,2.0]$ is used only as corroboration; its nonlinear-pooling
and bin-count biases are quantified in Supplement Sec.~S6.

\emph{Reconstruction and identifiability.} GRPA-blind spectral reconstructions
use covariance-whitened non-negative least squares (NNLS)/Tikhonov and,
independently, MaxEnt with a flat default model, both with the
finite-temperature kernel; no GRPA-informed spectral prior is used at any
coupling. Identifiability is tested by propagating predeclared synthetic
spectra through each coupling's measured covariance and reconstructing them
with the same procedure. Within the tested family, the reconstructions recover
the injected $W_{\rm low}$ values without bias and distinguish strong
low-frequency concentration from high-frequency continuum backgrounds
(Supplement Sec.~S7). $W_{\rm low}$ is an internal reconstruction diagnostic,
not a literal fraction of the exact spectral weight.

\emph{Hopping-sensitivity protocol.} $\Lambda_t$ is obtained from
finite-difference logarithmic slopes using two independent stencils: a
\emph{wide} secant, $t=0.9\!\leftrightarrow\!1.1$, and a \emph{local} secant,
$t=0.95\!\leftrightarrow\!1.05$. Their agreement tests sensitivity to the
finite-difference stencil. In Fig.~\ref{fig:discriminator}(a), the $U=6$
points are displaced horizontally for legibility: the two wide-stencil points
share the same physical abscissa and are linked by an arrow indicating the
$L=6\to L=9$ comparison, while the local-stencil point is offset; the two
$U=3$ wide-stencil points likewise share an abscissa. The sensitivities are
jackknifed from the measured moments. A $U^2$-normalised variant $D_t$, equal
to one for the leading-LSWT law, together with full stencil and convergence
tables, is given in Supplement Sec.~S9.

A covariance-weighted forward test determines how far the $t$-dependence can
be interpreted at the pole level. Fitting the measured $C(\tau)$ at $t=0.9$
and $1.1$ with a peak-plus-continuum model, allowing the width, weight and
high-frequency tail to vary, leaves a common peak position statistically
acceptable ($\Delta\chi^2\simeq1.2$ for one degree of freedom). An independent
MaxEnt reconstruction shows a small upward shift of the dominant low-energy
feature, whereas the NNLS reconstruction is bimodal at the lower-hopping
endpoint and does not resolve a unique peak position (Supplement Sec.~S9).

\emph{Finite-$\beta$ Gaussian control.} Evaluating the same moment estimators
in GRPA at $\beta=12$ changes $\Lambda_t$, $\Lambda_U$ and $\Omega_{1,-1}$ by
less than $10^{-4}$. Gaussian thermal corrections are activation-suppressed,
with $\beta\Delta\gtrsim8$ across the $t$-scan endpoints (Supplement Sec.~S9).

\emph{Analytic theory.} The exact $2t_h$ kernel zero and the second-order
effective Hamiltonian, including its LSWT evaluation, are derived in
Supplement Secs.~S3 and~S8. The Gaussian/GRPA response and equal-time
response-composition metric are detailed in Secs.~S2 and~S10.


\emph{Acknowledgements.} Y.P.\ thanks Vijay B.\ Shenoy, Syed R.\ Hassan and
Manjari Gupta for many fruitful discussions. Y.P.\ also thanks the Institute of
Mathematical Sciences, Chennai, for hospitality during visits when part of the
mean-field analysis was carried out.

\emph{Funding.} Y.P.\ would like to acknowledge the CSIR, India for financial
support. This work was supported by the National Research Foundation of
Korea (NRF) through the Basic Science Research Programs (Grant Nos.\
NRF-2022R1A2C1011646, NRF-2022M3H3A1085772, RS-2024-00416036, RS-2025-03392969,
and RS-2026-25490114). This work was also supported by the Creation of the
Quantum Information Science R\&D Ecosystem through the NRF, funded by the
Ministry of Science and ICT of the Korean government (Grant No.\
RS-2023-NR068116), and by the Quantum Simulator Development Project for Materials
Innovation through the NRF, funded by the Ministry of Science and ICT, Republic
of Korea (Grant No.\ RS-2023-NR119931).

\emph{Data availability.} Source data underlying all main-text and
Supplementary figures, including the DQMC moments, spectral-reconstruction
outputs and GRPA results, together with the corresponding figure-generation
scripts, accompany this submission and will be deposited in a public
repository upon publication.

\emph{Code availability.} Custom code used for the Gaussian/GRPA calculations,
including the Bogoliubov--de~Gennes saddle, strong-coupling spin-wave analysis,
determinant-QMC measurements and analysis, and figure generation is available
from the corresponding author upon reasonable request.

\emph{Competing interests.} The author declares no competing interests.


\clearpage
\onecolumngrid
\setcounter{section}{0}
\setcounter{equation}{0}
\setcounter{figure}{0}
\setcounter{table}{0}
\renewcommand{\thesection}{S\arabic{section}}
\renewcommand{\theequation}{S\arabic{equation}}
\renewcommand{\thefigure}{S\arabic{figure}}
\renewcommand{\thetable}{S\arabic{table}}
\begin{center}
{\large\bfseries Supplementary Material for\\[2pt]
``Hamiltonian fingerprints of a collective mode across the BCS--BEC crossover''}
\end{center}

\title{Supplement: Hamiltonian fingerprints of a collective mode across the BCS--BEC crossover}
\author{Yogeshwar Prasad}
\email{yogeshwar@snu.ac.kr}
\affiliation{Center for Condensed Matter Theory, Department of Physics, Indian Institute of Science, Bangalore 560012, India}
\affiliation{Department of Physics, Hanyang University, Seoul 04763, Korea}
\affiliation{Research Institute of Basic Sciences, Seoul National University, Seoul 08826, Korea}
\date{\today}
\maketitle
\tableofcontents

\section{Model, conventions, and normalisation}
\label{sm:model}

All quantities in the main text and in this Supplement use the conventions
stated here.

\paragraph{Hamiltonian.} Spin-$\tfrac12$ fermions occupy two AA-stacked honeycomb
layers ($\ell=1,2$; sublattices $A,B$; intralayer coordination $z=3$),
\begin{align}
  H =\;& -t\!\!\sum_{\ell,\langle ij\rangle,\sigma}\!\! c^\dagger_{\ell i\sigma}c_{\ell j\sigma}
       -t_h\!\sum_{i\sigma}\!\big(c^\dagger_{1i\sigma}c_{2i\sigma}+\mathrm{H.c.}\big)
       \nonumber\\
     & -|U|\sum_{\ell i}\big(n_{\ell i\uparrow}-\tfrac12\big)\big(n_{\ell i\downarrow}-\tfrac12\big)
       -\mu\sum_{\ell i}(n_{\ell i}-1),
  \label{eq:Hsm}
\end{align}
with intralayer nearest-neighbour hopping $t$, coherent interlayer hybridisation
$t_h$, on-site attraction $-|U|<0$, and chemical potential $\mu$. This is the
Hamiltonian defined in the Methods of the main text. The particle--hole-symmetric form of the interaction and of the
$\mu$-term is chosen so that $\mu=0$ is exactly half-filling ($n=1$) at every $|U|$;
$\mu$ is the particle--hole-shifted chemical potential ($\mu\equiv\mu_{\rm phys}+|U|/2$).
Unless stated otherwise we work at $t=1$, $t_h=0.6$, filling $n=1.10$
($d\equiv n-1=0.10$), and inverse temperature $\beta=12$; the DQMC runs use periodic
boundary conditions on $L\times L$ Bravais meshes with $L=6,9$ (Sec.~\ref{sm:dqmc}).

\paragraph{Site and cell counting.} The bilayer unit cell contains four orbitals
(two sublattices $\times$ two layers). An $L\times L$ lattice therefore has
$N_{\rm cell}=L^2$ bilayer cells, $2L^2$ honeycomb sites per layer, and
$N_{\rm sites}=4L^2$ lattice sites (orbitals) in the bilayer ($72$ per layer and $144$ total at
$L=6$; $162$ per layer and $324$ total at $L=9$). ``Per orbital'' divides by $N_{\rm sites}$, ``per cell'' by $L^2$;
each intensive quantity states which normalisation it uses.

\paragraph{Basis ordering.} Decoupling the attraction in the pairing channel gives a
Bogoliubov--de Gennes (BdG) Hamiltonian that is $8\times8$ in the combined
Nambu $\otimes$ layer $\otimes$ sublattice space, with the Nambu spinor ordered as
\begin{equation}
  \Psi_{\mathbf k}=\big(c_{1A\uparrow},c_{1B\uparrow},c_{2A\uparrow},c_{2B\uparrow},
  c^\dagger_{1A\downarrow},c^\dagger_{1B\downarrow},c^\dagger_{2A\downarrow},c^\dagger_{2B\downarrow}\big)^{\!\top}_{\mathbf k}.
  \label{eq:spinor}
\end{equation}
The uniform $s$-wave gap $\Delta$ and $\mu$ are determined self-consistently on an
$N_k\times N_k$ mesh ($N_k=72$; the response-composition calculation uses the same mesh).
The saddle preserves layer parity at equal fillings, which underlies the selection
rule of the damping analysis (Sec.~\ref{sm:damp}).

\paragraph{Layer pseudospin.} The layer degree of freedom carries an SU(2) algebra
generated by
\begin{equation}
  T_z=\tfrac12(N_1-N_2),\quad
  T_x=\tfrac12\!\sum_{i\sigma}\!\big(c^\dagger_{1i\sigma}c_{2i\sigma}+\mathrm{H.c.}\big),\quad
  T_y=\tfrac1{2i}\!\sum_{i\sigma}\!\big(c^\dagger_{1i\sigma}c_{2i\sigma}-\mathrm{H.c.}\big),
  \label{eq:Tdef}
\end{equation}
with $[T_x,T_y]=iT_z$, $[T_y,T_z]=iT_x$, $[T_z,T_x]=iT_y$. The interlayer
hybridisation is a transverse field, $H_h=-2t_h T_x$, and $T_z$ (the half
layer-imbalance) is the operator probed throughout. Since $t$, $-|U|$ and $\mu$ are
all layer-diagonal, only $H_h$ fails to commute with $T_z$, giving the operator
identity
\begin{equation}
  [H,T_z]=[H_h,T_z]=2i\,t_h\,T_y,
  \label{eq:HTz}
\end{equation}
which underlies the exact moment relations of Sec.~\ref{sm:moments}. Layer symmetry
at equal fillings gives $\langle T_z\rangle=0$, so the $T_z$ correlator is connected.

\paragraph{Unhalved imbalance and response normalisation.} The physically measured
relative particle number is $Q\equiv N_1-N_2=2T_z$, so every response carries the
definitional factor
\begin{equation}
  \chi_{QQ}=4\,\chi_{T_zT_z},\qquad M_1[Q]=4\,M_1[T_z]=-2\langle H_h\rangle .
  \label{eq:Qconv}
\end{equation}
Data streams reported in the unhalved variable $Q$ are divided by the appropriate
power of two before entering any figure. Separately, the GRPA code carries an
internal channel factor $1/C_N$ ($C_N=4$) inside $\chi_{\rm code}$; the only
remaining conversion to the physical $T_z$ susceptibility is a single spin/Nambu
factor of two, fixed by matching the exact $U=0$ Lehmann
representation of $\chi_{T_zT_z}$ (ratio $2.0000$ to machine precision):
\begin{equation}
  \boxed{\;\chi^{\rm phys}_{T_zT_z}=2\,\chi_{\rm code}\;}.
  \label{eq:norm}
\end{equation}
The definitional factor $4$ of Eq.~\eqref{eq:Qconv} and the code factor $2$ of
Eq.~\eqref{eq:norm} are distinct and are never conflated: for the unhalved imbalance
$\chi^{\rm phys}_{QQ}=4\chi^{\rm phys}_{T_zT_z}=8\chi_{\rm code}$.

\paragraph{Spectral functions and moments.} For a Hermitian operator $O$ we use the
single-sided (positive-frequency) spectral function
$A_O(\omega)=-\tfrac1\pi\,\mathrm{Im}\,\chi^R_{OO}(\omega)$, $\omega>0$, and moments
\begin{equation}
  M_p[O]=\int_0^\infty\!\omega^p\,A_O(\omega)\,d\omega .
  \label{eq:momdef}
\end{equation}
These conventions are used identically in Secs.~\ref{sm:moments}, \ref{sm:dqmc}, and
\ref{sm:char}.

\section{Mean field and Gaussian (GRPA) response}
\label{sm:grpa}
\begin{figure}[t]
  \centering
  \includegraphics[width=0.86\linewidth]{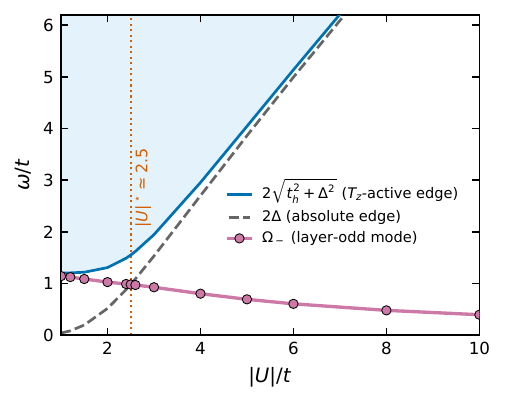}
  \caption{Layer-odd GRPA branch $\Omega_-(|U|)$ (markers) against the two
  two-quasiparticle continuum edges: the absolute pair edge $2\Delta$ (dashed) and the
  $T_z$-active inter-parity edge $2\sqrt{t_h^2+\Delta^2}$ (solid; the $T_z$-active
  continuum is shaded). The mode stays below the $T_z$-active edge at every coupling,
  so it is kinematically dark to linear $T_z$ decay; it crosses the absolute edge
  $2\Delta$ at $|U|^\star\simeq2.5$ (dotted), below which it is embedded in the
  absolute pair continuum but still dark to the $T_z$ channel, and above which it
  lies below the absolute pair continuum. GRPA at $n=1.10$.}
  \label{fig:sm:continua}
\end{figure}

Gaussian random-phase-approximation (GRPA) fluctuations are constructed about
the self-consistent Bogoliubov--de Gennes saddle of Sec.~\ref{sm:model}. We use
$t=1$, $t_h=0.6$, filling $n=1.10$ ($d=n-1=0.10$), and the $8\times8$
Nambu\,$\otimes$\,layer\,$\otimes$\,sublattice basis of Eq.~\eqref{eq:spinor}.
The principal DQMC comparison uses $|U|=3,4,6$, while the full $|U|\in[1,10]$
sweep determines the GRPA branch in Fig.~\ref{fig:sm:continua}.

\paragraph{Self-consistent saddle.} Decoupling the attraction in the pairing
channel and imposing a uniform, layer-symmetric $s$-wave order parameter
$\Delta_1=\Delta_2\equiv\Delta$ closes the BdG problem. Diagonalising the
$8\times8$ Bloch Hamiltonian $\mathcal H_{\rm BdG}(\mathbf k)$ into quasiparticle
branches $E_{\nu}(\mathbf k)$, the gap and number conditions
\begin{align}
  \Delta &= \frac{|U|}{N_k^2}\sum_{\mathbf k}
            \big\langle c_{\ell\alpha,-\mathbf k\downarrow}c_{\ell\alpha,\mathbf k\uparrow}\big\rangle_{\rm saddle},
  \label{sm:grpa:gap}\\
  n &= 1+\frac{1}{N_k^2}\sum_{\mathbf k}
        \big\langle n_{\ell\alpha,\mathbf k}-1\big\rangle_{\rm saddle},
  \label{sm:grpa:num}
\end{align}
with $n_{\ell\alpha,\mathbf k}=\sum_\sigma
c^\dagger_{\ell\alpha,\mathbf k\sigma}c_{\ell\alpha,\mathbf k\sigma}$ (both
momentum-resolved averages are independent of the layer--sublattice pair
$\ell\alpha$ for the uniform, layer-symmetric saddle), are iterated to
convergence on the production $N_k\times N_k$ mesh ($N_k=72$; the
response-composition calculation of Sec.~\ref{sm:char} uses the same mesh), fixing
$\Delta$ and the particle--hole-shifted $\mu$ at each $|U|$. The saddle preserves
layer parity at equal fillings, so $\langle T_z\rangle=0$ and the layer channel
enters only through fluctuations.

\paragraph{Odd-sector Gaussian kernel.} Expanding the action to quadratic order in
the fluctuations of the pairing field and of the density/layer bilinears organises
the response into decoupled parity sectors. The layer-imbalance operator $T_z$
lives entirely in the layer-\emph{odd} sector, which is spanned by three collective
coordinates: the amplitude and phase quadratures of the layer-odd pair field,
\begin{equation}
  \delta\Delta_-=\Delta\,(\sigma_-+i\,\theta_-),
  \label{sm:grpa:quad}
\end{equation}
together with $T_z$ itself. Here $\sigma_-$ and $\theta_-$ are the relative
(out-of-phase) amplitude and phase between the two layers, and $\theta_-$ is the
relative condensate phase that carries the $2t_h$ physics discussed below. In the
basis $(\sigma_-,\theta_-,T_z)$ the dynamical GRPA kernel is a $3\times3$ matrix
\begin{equation}
  K(\omega,\mathbf q{=}0)=K^{\rm HS}+\Pi(\omega,\mathbf q{=}0),
  \label{sm:grpa:kernel}
\end{equation}
where $K^{\rm HS}$ is the instantaneous Hubbard--Stratonovich part (diagonal in
the amplitude/phase quadratures, set by $1/|U|$) and $\Pi$ is the BdG polarisation
matrix assembled from the saddle Green's functions. Each entry is a bubble
$\Pi_{ab}=\langle\!\langle \mathcal V_a;\mathcal V_b\rangle\!\rangle$ of the
saddle-generated vertices, which in the bonding/antibonding parity frame all carry
the layer-flipping factor $\sigma_x$: the odd phase attaches
$\sigma_x\otimes\Gamma_\phi$, the odd amplitude $\sigma_x\otimes\Gamma_\Delta$, and
$T_z\to\sigma_x\otimes\tau_z$. The diagonal relative-phase element
$K_{\theta_-\theta_-}$ is the object of the exact theorem of
Sec.~\ref{sm:kernelzero}, which shows that it vanishes at $\omega=2t_h$,
$\mathbf q=0$ for any single-particle dispersion when $U_1=U_2$; we use that result
here without reproving it.

\paragraph{Layer-imbalance response and its pole.} The physical layer-imbalance
susceptibility follows by dressing the $T_z$ bubble with the full odd-sector
kernel,
\begin{equation}
  \chi_{T_zT_z}(\omega)=\Pi_{T_zT_z}(\omega)
     +\sum_{ab}\Pi_{T_z a}(\omega)\,[K^{-1}(\omega)]_{ab}\,\Pi_{b T_z}(\omega),
  \label{sm:grpa:chi}
\end{equation}
so that its collective poles are the frequencies at which the odd kernel becomes
singular,
\begin{equation}
  \det K(\Omega,\mathbf q{=}0)=0,\qquad K(\Omega)\,v=0 ,
  \label{sm:grpa:det}
\end{equation}
with $v$ the zero-eigenvalue vector of the mode. Near an isolated pole, the
susceptibility has the simple-pole form $\chi_{T_zT_z}(\omega)\simeq
Z_{T_z}/(\Omega^2-\omega^2)$, whose weight is proportional to the $T_z$--$T_z$
entry of the rank-one residue dyad constructed in Sec.~\ref{sm:char}; that section also
defines its normalisation with respect to the equal-time metric. Off-diagonal
elements of the odd-sector kernel couple $T_z$ to $\sigma_-$ and $\theta_-$, so
the observable pole $\Omega_-$ need not coincide with the diagonal
relative-phase zero at $2t_h$. At particle--hole symmetry the relevant
amplitude--phase mixing vanishes and the two coincide
(Sec.~\ref{sm:kernelzero}). The absolute normalisation of $\chi_{T_zT_z}$
follows Eq.~\eqref{eq:norm}, and its oscillator strength satisfies the exact
first-moment sum rule of Sec.~\ref{sm:moments}.

\paragraph{The layer-odd branch.} Solving Eq.~\eqref{sm:grpa:det} across coupling
gives a single, non-crossing layer-odd branch $\Omega_-(|U|)$. It softens
monotonically from the hybridisation anchor $\Omega_-\simeq1.14$ near $2t_h$ at
$|U|=1$ towards $\Omega_-\simeq0.39$ at $|U|=10$, approaching the LSWT evaluation of
the strong-coupling $\eta$-magnet effective theory of Sec.~\ref{sm:sw}. The branch stays below the
layer-imbalance-active two-quasiparticle edge throughout. That $T_z$-active
threshold is set by the minimum inter-parity pair energy,
\begin{equation}
  \omega^{T_z}_{\rm 2QP}=2\sqrt{t_h^2+\Delta^2},
  \label{sm:grpa:thresh}
\end{equation}
which rises from $1.20$ at $|U|=1$ to $9.39$ at $|U|=10$; the layer-imbalance
bubble has no spectral weight below it, so the mode is kinematically dark to
linear $T_z$ decay at Gaussian order (Sec.~\ref{sm:damp}). The lower
absolute pair edge $2\Delta$ crosses the branch at $|U|^\star\simeq2.5$, where
$\Omega_-=2\Delta$. For $|U|\lesssim|U|^\star$ the mode overlaps the absolute
pair continuum but remains dark to the $T_z$-active channel; for stronger
coupling it lies below the absolute pair continuum.

\paragraph{Numerical verification.} An independent GRPA implementation
reproduces $\Omega_-(|U|)$ to better than $0.1\%$ across the full coupling
range. The branch and response-composition calculations, both evaluated on
the $N_k=72$ mesh, agree to the quoted precision.

\section{Exact Gaussian theorem: the $\theta_-$ kernel zero at $2t_h$}
\label{sm:kernelzero}

We prove that the diagonal relative-phase Gaussian kernel has an exact zero at
$\omega=2t_h$, $\mathbf q=0$, for an \emph{arbitrary} single-particle dispersion
under the identical-copy, layer-symmetric conditions stated below.

\paragraph{Setup.} Two identical layers give the two-copy normal Bloch Hamiltonian
\begin{equation}
  H_N(\mathbf k)=\sigma_0\otimes h_0(\mathbf k)-t_h\,\sigma_x\otimes\openone-\mu,
\end{equation}
with Pauli matrices $\sigma$ acting in layer space. Layer-symmetric $s$-wave pairing
$\Delta$ closes the BdG problem. The relative and common phase fluctuations are
$\theta_\pm=(\theta_1\pm\theta_2)/\sqrt2$, with saddle-generated vertices
$V_+=\sigma_0\otimes\Gamma_\phi$, $V_-=\sigma_z\otimes\Gamma_\phi$, where
$\Gamma_\phi=\tfrac{i}{2}[\tau_z,H_0]$ ($\tau$ = Nambu).

\paragraph{Band-resolved bubble.} Diagonalising $h_0$ into bands $\varepsilon_n$,
the uniform on-site pairing is band-diagonal with the same gap in every band,
$\Delta_n=\Delta$; the theorem requires only that $\Delta_n$ be identical in the
two $t_h$-split copies of each band, which the layer symmetry guarantees. Define
$\xi_{a,n}=\varepsilon_n+t_h-\mu$, $\xi_{b,n}=\varepsilon_n-t_h-\mu$ (so
$\xi_a-\xi_b=2t_h$) and $E_{\kappa n}=\sqrt{\xi_{\kappa n}^2+\Delta_n^2}$. For
bosonic Matsubara frequency $i\Omega_m$ the odd-sector bubble is
\begin{equation}
  \Pi_-(i\Omega_m,\mathbf k)=\sum_n\frac{\mathcal N_n\,(E_a+E_b)}
  {E_aE_b\big[(i\Omega_m)^2-(E_a+E_b)^2\big]},\quad
  \mathcal N_n=E_aE_b+\xi_a\xi_b+\Delta_n^2 .
  \label{eq:bubble}
\end{equation}

\paragraph{The collapse identity.} Using $\xi_a-\xi_b=2t_h$,
\begin{equation}
  (E_a+E_b)^2-(2t_h)^2=2\big(E_aE_b+\xi_a\xi_b+\Delta_n^2\big)=2\mathcal N_n .
\end{equation}
After analytic continuation $i\Omega_m\to\omega+i0^+$, evaluating
Eq.~\eqref{eq:bubble} at $\omega=2t_h$ cancels the factor $\mathcal N_n$:
\begin{equation}
  \Pi_-(2t_h,\mathbf k)=-\tfrac12\sum_n\Big(\tfrac1{E_a}+\tfrac1{E_b}\Big)
  =\Pi_+(0,\mathbf k),
  \label{eq:pointwise}
\end{equation}
i.e.\ the odd bubble at $2t_h$ equals the even (Goldstone) bubble at $\omega=0$,
\emph{pointwise} in $\mathbf k$ and for arbitrary $h_0$.

\paragraph{Operator/Ward route.} Equivalently, let
$Q_+=\sigma_0\otimes\tau_z$ denote the global U(1) charge generator in the
layer\,$\otimes$\,Nambu notation of this section (distinct from the layer
imbalance $Q=N_1-N_2$ of Sec.~\ref{sm:model}), and
$L=\sigma_x\otimes\tau_z+i\sigma_y\otimes\openone$; then
\begin{equation}
  [\mathcal H,Q_+]=2iV_+,\qquad [\mathcal H,L]+2t_h\,L=2iV_- ,
  \label{eq:ward}
\end{equation}
so the relative-phase response inherits the Goldstone (even) structure shifted by
$2t_h$.

\paragraph{Kernel zero.} For identical interactions $U_1=U_2$ the instantaneous
Hubbard--Stratonovich contributions are identical in the even and odd phase
sectors, $K^{\rm HS}_-=K^{\rm HS}_+$. With $K=K^{\rm HS}+\Pi$,
\begin{equation}
  K_{\theta_-\theta_-}(2t_h,\mathbf q{=}0)=K_{\theta_+\theta_+}(0,\mathbf q{=}0)=0
\end{equation}
by the gap equation (the Goldstone condition). Hence the diagonal $\theta_-$ kernel
vanishes at $2t_h$.

\paragraph{Three statements (kept distinct).}
(i)~The \emph{diagonal} $\theta_-$ kernel has the exact zero above.
(ii)~The \emph{observable} pole in the $T_z$ response is a zero of the \emph{full} Gaussian
determinant, where $\theta_-$ mixes with the density/layer channels; this mixing
displaces the pole from $2t_h$ already within Gaussian theory and gives it
$|U|$-dependence.
(iii)~Beyond Gaussian order there is \emph{no} all-orders protection: the interaction
torque $-i[H_U,T_y]\neq0$, so no symmetry pins the physical mode to $2t_h$. Thus the
exact result is a Gaussian kernel identity: it neither pins the full observable pole
to $2t_h$ nor provides all-orders protection beyond Gaussian theory.

\section{Exact finite-temperature moments}
\label{sm:moments}

The many-body trajectory of the main text is built from spectral moments of the
imaginary-time layer-imbalance correlator that are \emph{exact} at every temperature,
requiring no analytic continuation, pole fitting, or frequency window. We use the
bosonic (KMS-symmetric) correlator
\begin{equation}
  C_{zz}(\tau)=\int_0^\infty\!d\omega\,A_{zz}(\omega)\,
  \frac{\cosh[\omega(\beta/2-\tau)]}{\sinh(\beta\omega/2)},\qquad 0\le\tau\le\beta,
  \label{eq:Czz}
\end{equation}
with $A_{zz}\ge0$ and $\langle T_z\rangle=0$ (Sec.~\ref{sm:model}). For the finite
systems and parameters studied, the $T_z$ response has no elastic $\delta(\omega)$
contribution; we denote its lowest non-zero spectral scale by $\Delta_{\rm odd}>0$,
which makes all moments used below finite.

\paragraph{First moment (exact f-sum).} For a Hermitian $O$ the single-sided first
moment equals the thermal double-commutator average at any temperature,
\begin{equation}
  M_1[O]=\int_0^\infty\!\omega\,A_O(\omega)\,d\omega=\tfrac12\big\langle[O,[H,O]]\big\rangle_\beta ,
  \label{eq:M1gen}
\end{equation}
a commutator (odd-moment) identity that carries \emph{no} thermal factor. Only $H_h$
fails to commute with $T_z$; using Eq.~\eqref{eq:HTz} and the algebra
Eq.~\eqref{eq:Tdef},
\begin{align}
  [H,T_z]&=-2t_h[T_x,T_z]=2i\,t_h\,T_y,\\
  [T_z,[H,T_z]]&=2i\,t_h[T_z,T_y]=2t_h\,T_x,
\end{align}
so that
\begin{equation}
  \boxed{\;M_1[T_z]=\tfrac12\big\langle[T_z,[H,T_z]]\big\rangle
  =t_h\langle T_x\rangle=-\tfrac12\langle H_h\rangle\;}
  \label{eq:M1Tz}
\end{equation}
using $\langle T_x\rangle=-\langle H_h\rangle/2t_h$. Equation~\eqref{eq:M1Tz} is an
operator identity valid for any $|U|$ and independent of any mean-field decoupling;
its value is set by the interlayer coherence $\langle T_x\rangle$, itself
$|U|$-dependent. It provides two independent DQMC estimators of $M_1$: the equal-time
$t_h\langle T_x\rangle$ and the short-time slope $-\partial_\tau C_{zz}(0^+)$, which
must agree.

\paragraph{Independent confirmation of the f-sum leg.} Equation~\eqref{eq:M1Tz}
can be reached by a second, independent route, which we record because the static
protocol of Sec.~\ref{sm:expt} rests on it. Writing the layer-odd density as
$n_-=2T_z$, the same double-commutator identity gives
$S_1\equiv\langle[n_-,[H,n_-]]\rangle=8M_1[T_z]$. The companion mean-field
analysis of the same bilayer establishes independently that $S_1=-4\langle
H_h\rangle$; combined with $H_h=-2t_h T_x$ this yields $-4\langle H_h\rangle=
8t_h\langle T_x\rangle$ and hence $M_1=t_h\langle T_x\rangle$, identical to
Eq.~\eqref{eq:M1Tz}. The two derivations share no intermediate step. Both legs of
the static route $\Omega_{1,-1}^2=2t_h\langle T_x\rangle/\chi_{zz}(0)$ are
therefore exact: the Kramers--Kronig leg $M_{-1}=\chi_{zz}(0)/2$ is verified
numerically to $10^{-10}$, and the f-sum leg is the operator identity above.

\paragraph{Inverse moment (exact, thermally clean).} Integrating
Eq.~\eqref{eq:Czz} over $\tau\in[0,\beta]$, the kernel integrates to
$(2/\omega)\sinh(\beta\omega/2)$, which cancels the $\sinh(\beta\omega/2)$
denominator, leaving
\begin{equation}
  \boxed{\;M_{-1}[T_z]=\int_0^\infty\!\frac{A_{zz}(\omega)}{\omega}\,d\omega
  =\tfrac12\int_0^\beta\!C_{zz}(\tau)\,d\tau\;}\qquad\text{(exact).}
  \label{eq:Mm1}
\end{equation}
Both $M_1$ and $M_{-1}$ are thus continuation-free, and the geometric spectral scale
\begin{equation}
  \Omega_{1,-1}\equiv\sqrt{\frac{M_1[T_z]}{M_{-1}[T_z]}}
  \label{eq:Om1m1}
\end{equation}
inherits their exactness. This is the primary many-body observable.

\paragraph{Even/odd derivative hierarchy.} The thermal structure of the moments
follows from the even/odd derivative hierarchy of the KMS kernel:
\begin{equation}
  -C_{zz}^{(2r+1)}(0^+)=M_{2r+1}\ \ (\text{odd; no thermal factor}),\qquad
  C_{zz}^{(2r)}(0^+)=\int_0^\infty\!\omega^{2r}A_{zz}\coth(\tfrac{\beta\omega}{2})\,d\omega\ \ (\text{even; }\coth\text{-dressed}).
\end{equation}
In particular the equal-time value is coth-weighted and is \emph{not} $M_0$,
\begin{equation}
  C_{zz}(0)=\int_0^\infty\!A_{zz}(\omega)\coth(\tfrac{\beta\omega}{2})\,d\omega
  \equiv\tilde M_0\ \ge\ M_0,
  \label{eq:tildeM0}
\end{equation}
with equality only as $\beta\to\infty$. For a spectrum gapped below
$\Delta_{\rm odd}$ the thermal excess is exponentially bounded,
\begin{equation}
  0\le\frac{\tilde M_0-M_0}{M_0}\le\coth\!\Big(\frac{\beta\Delta_{\rm odd}}{2}\Big)-1
  =\frac{2}{e^{\beta\Delta_{\rm odd}}-1},
  \label{eq:cothbound}
\end{equation}
i.e.\ a few $\times10^{-3}$ for $\beta\Delta_{\rm odd}\gtrsim6$; the same bound
governs $(\tilde M_2-M_2)/M_2$.

\paragraph{Second moment via $\langle T_y^2\rangle$.} Sandwiching
Eq.~\eqref{eq:HTz} between eigenstates gives the exact spectral identity
$A_{yy}(\omega)=(\omega^2/4t_h^2)A_{zz}(\omega)$, hence the equal-time
current--current variance is a (coth-dressed) second moment,
\begin{equation}
  4t_h^2\langle T_y^2\rangle=\int_0^\infty\!\omega^2 A_{zz}(\omega)\coth(\tfrac{\beta\omega}{2})\,d\omega
  \equiv\tilde M_2\ \xrightarrow{\beta\to\infty}\ M_2,
  \label{eq:M2}
\end{equation}
measured cleanly from the equal-time $\langle T_y^2\rangle$ (custom estimator,
Sec.~\ref{sm:dqmc}) with the correlator relation
$C_{yy}(\tau)=+\tfrac{1}{4t_h^2}\partial_\tau^2 C_{zz}(\tau)$ ($0<\tau<\beta$). $M_2$
is high-frequency-biased and serves as a reconstruction/tail constraint rather than a
concentration measure.

\paragraph{Concentration diagnostics.} With normalised density $p(\omega)=A_{zz}/M_0$,
the single-mode saturation ratio is a mean-ratio,
\begin{equation}
  S=\frac{M_0^2}{M_1\,M_{-1}}=\frac{1}{\langle\omega\rangle_p\,\langle1/\omega\rangle_p}
  =\frac{\omega_{\rm H}}{\omega_{\rm A}}\ \le\ 1,
  \label{eq:Sdef}
\end{equation}
where $\omega_{\rm H}\equiv\langle1/\omega\rangle_p^{-1}$ and
$\omega_{\rm A}\equiv\langle\omega\rangle_p$ are the harmonic and arithmetic mean
frequencies of $p(\omega)$; $S$ saturates ($S=1$) iff the response is
monochromatic (Cauchy--Schwarz / AM--HM inequality). The
exact positive relative-spread functional is
\begin{equation}
  \frac1S-1=\frac{M_1M_{-1}-M_0^2}{M_0^2}
  =\frac{1}{2M_0^2}\!\int\!\!\int\!d\omega\,d\omega'\,A_{zz}(\omega)A_{zz}(\omega')\,
  \frac{(\omega-\omega')^2}{\omega\omega'}\ \ge0,
\end{equation}
whose $1/(\omega\omega')$ denominator softens high-energy weight, making $S$ a better
concentration measure than the variance. When the true spectral zeroth moment $M_0$
is available, $S$ also gives a rigorous bound on spectral concentration: defining
$\Omega\equiv\Omega_{1,-1}$,
\begin{equation}
  W_{\rm in}(\kappa)=\frac{1}{M_0}\!\int_{\Omega/\kappa}^{\kappa\Omega}\!A_{zz}\,d\omega
  \ \ge\ 1-\frac{2\,(S^{-1/2}-1)}{\kappa+\kappa^{-1}-2}.
  \label{eq:Win}
\end{equation}
This bound avoids identifying $S\simeq1$ with a literal single pole. For the
finite-temperature DQMC data below, however, the true $M_0$ is not directly
available, so neither $S$ nor this bound is evaluated numerically.

\paragraph{$L=6$ values.} The exact-moment trajectory at $L=6$, $\beta=12$,
$n=1.10$, $t_h=0.6$ is (Table~\ref{tab:moments}) a clean monotone softening dominated
by the growth of $M_{-1}$ ($+86\%$ from $U=3$ to $U=6$), with $M_1$ contributing only
weakly ($-7\%$); the two intervals are resolved at $\sim10\sigma$
($\Delta_{3\to4}=0.081(8)$) and $\sim15\sigma$ ($\Delta_{4\to6}=0.189(13)$).
We deliberately quote no numerical saturation ratio $S$ from the DQMC data. With the
zeroth moment available only as the thermally dressed equal-time value $C_{zz}(0)$
rather than the true spectral $\int A_{zz}\,d\omega$, the finite-temperature surrogate
$\tilde S=C_{zz}(0)^2/(M_1M_{-1})$ is not bounded above by unity
[Eq.~\eqref{eq:tildeM0}] and would misstate the concentration; the softening evidence
therefore rests on the continuation-free scale $\Omega_{1,-1}$ alone. Equation
\eqref{eq:Sdef} remains an exact spectral inequality at any temperature; we simply do
not evaluate it from DQMC, because the true $M_0=\int A_{zz}\,d\omega$ is not directly
available from the thermal equal-time $C_{zz}(0)$.

\begin{table}[t]
\centering
\caption{Finite-temperature moments of the $T_z$ response (DQMC, $L=6$,
$\beta=12$, $n=1.10$, $t_h=0.6$). $M_1$ and $M_{-1}$ are quoted per bilayer cell,
$N_{\rm cell}=L^2$ (Sec.~\ref{sm:model}), with $T_z=(N_1-N_2)/2$. The
characteristic scale $\Omega_{1,-1}=\sqrt{M_1/M_{-1}}$ is
normalisation-independent; uncertainties are $1\sigma$ bin-jackknife. The column
$C_{zz}(0)\equiv\tilde M_0$ is the coth-dressed equal-time correlator, not the
spectral zeroth moment $M_0=\int A_{zz}(\omega)\,d\omega$; it satisfies
$\tilde M_0\ge M_0$ and is included only as a corroborative quantity. $M_1$,
$M_{-1}$ and $\Omega_{1,-1}$ require no analytic continuation.}
\label{tab:moments}
\begin{tabular}{ccccc}
\hline\hline
$|U|/t$ & $M_1=t_h\langle T_x\rangle$ & $M_{-1}$ & $C_{zz}(0){=}\tilde M_0$ & $\Omega_{1,-1}=\sqrt{M_1/M_{-1}}$\\
\hline
$3$ & $0.1626$ & $0.189$ & $0.166$ & $0.928(6)$\\
$4$ & $0.1609$ & $0.224$ & $0.177$ & $0.847(6)$\\
$6$ & $0.1521$ & $0.351$ & $0.206$ & $0.658(11)$\\
\hline\hline
\end{tabular}
\end{table}

\section{DQMC implementation and validation}
\label{sm:dqmc}

\paragraph{Algorithm and decoupling.}
All many-body results use sign-free auxiliary-field determinant quantum Monte Carlo
(DQMC), with the implementation cross-checked against ALF~\cite{ALF2017,ALF2022}
for the AA-stacked bilayer honeycomb of Eq.~\eqref{eq:Hsm}.
The on-site attraction is decoupled in the \emph{density} channel using a discrete
Hubbard--Stratonovich field coupled to $n_{\ell i}-1$. For reproducibility, the
corresponding ALF parameters are $N_{\rm SUN}=2$, $N_{\rm FL}=1$, and
\texttt{Mz}$=$false. In this channel the two spin species share the same auxiliary
field and the fermion determinant is a perfect square, so the Monte Carlo weight is
non-negative at all couplings and densities studied. The average sign is
$\langle\mathrm{sign}\rangle=1$ (measured phase $\sim10^{-15}$) throughout,
including away from half-filling. The fermion sign problem is therefore absent over
the entire parameter range studied.

\paragraph{Discretisation and stabilisation.} Imaginary time uses the symmetric
(checkerboard) Trotter decomposition at $\Delta\tau=0.05$ ($\beta=12$, i.e.\
$L_\tau^{\rm slices}=\beta/\Delta\tau=240$). All production runs use this single
time step; the associated $O(\Delta\tau^2)$ Trotter-plus-quadrature systematic is
quantified by a dedicated three-point discretisation check at
$\Delta\tau\in\{0.1,0.05,0.025\}$ ($t=1$, $L=6$, $U=3$ and $6$, production
$\mu^\star$): halving the time step shifts $\Omega_{1,-1}$
by $-0.005(8)$ at $U=3$ and $+0.021(15)$ at $U=6$, both consistent with zero
($0.6\sigma$ and $1.4\sigma$); the $U6$ variation is consistent with the
chain-to-chain scatter of the six independent $\Delta\tau=0.025$ chains
(per-chain $\Omega_{1,-1}$ spanning $0.64$--$0.72$), with no equilibration
failure detected in any individual chain. The coarser step $\Delta\tau=0.1$
differs from production by $+0.004(10)$ ($U3$) and $+0.030(17)$ ($U6$), so no
resolved $\Delta\tau$ trend emerges at either coupling at the present
precision; the trajectory difference $\Omega(U6)-\Omega(U3)$ moves by
$+0.026(17)$ ($1.5\sigma$), dominated by the same $U6$ chain scatter. At $U3$,
where the three time steps are consistent with a smooth $\Delta\tau^2$
dependence, a quadratic fit bounds the residual bias of the $\Delta\tau=0.05$
production value at $+0.004$, below the statistical error. The end-to-end
exponent re-derived entirely at $\Delta\tau=0.025$ is $\Lambda_U=0.443(23)$,
within $1.5\sigma$ of the production $0.488(26)$. The $t$-scan diagnostic $\Lambda_t$ involves a difference
between two hoppings whose Trotter errors need not cancel identically; a dedicated
$\Delta\tau=0.025$ $t$-differential check at $U=6$, repeating the full scan protocol
at both endpoints ($t=0.9,1.1$, production $\mu^\star=0.131/0.200$), gives
$\Lambda_t^{(0.025)}(U6)=0.543(92)$: consistent with the production $0.445(99)$
($0.7\sigma$), and nonzero at $\sim\!6\sigma$, showing that the finite-$\Delta\tau$
discretisation does not account for the observed hopping sensitivity.
The endpoint values themselves make the control transparent:
\begin{center}
\begin{tabular}{cccc}
\hline\hline
$t$ & $\Omega_{1,-1}$ ($\Delta\tau=0.05$) & $\Omega_{1,-1}$ ($\Delta\tau=0.025$) & shift\\
\hline
$0.9$ & $0.6503(111)$ & $0.6285(90)$ & $-0.022(14)$\\
$1.1$ & $0.7110(72)$  & $0.7009(82)$  & $-0.010(11)$\\
\hline\hline
\end{tabular}
\end{center}
both endpoint discretisation shifts are small and have the same sign, so their
common component cancels in the logarithmic slope and only their differential
shift contributes; both are several times smaller than the $\simeq0.061$ endpoint
splitting that carries the measured sensitivity. The finer time step does not shift the filling either: the
achieved densities at $\Delta\tau=0.025$ are $n=1.1006(3)$, $1.1024(7)$,
$1.1059(10)$, and $1.1033(5)$ for the $U3$, $U6$, and the two $U6$ $t$-endpoints
respectively, within about $0.6\%$ of the target filling. Using the
corresponding GRPA density sensitivity, $|\partial\ln\Omega_-/\partial\ln n|\simeq0.13$,
these filling offsets would produce shifts of order $10^{-3}$, well below the
statistical uncertainty. Hence a discretisation-induced contribution of magnitude
comparable to $\Lambda_t$ would require compensating endpoint shifts far larger
than observed. Green-function
stabilisation (UDV
re-orthogonalisation) is performed every $N_{\rm wrap}=12$ slices, and time-displaced
correlators are measured ($\mathrm{Ltau}=1$), which is required for
$C_{zz}(\tau)$ and hence $M_{-1}$ [Eq.~\eqref{eq:Mm1}]. A custom estimator
evaluates the equal-time $\langle T_y^2\rangle$ for the $M_2$ cross-check
[Eq.~\eqref{eq:M2}]. We use $L=6$ and $L=9$ periodic $L\times L$ Bravais
meshes, corresponding to $2L^2=72$ and $162$ sites per honeycomb layer and
$4L^2=144$ and $324$ sites in the bilayer, respectively. Both $L$ are multiples of three, so the
honeycomb Dirac valleys $K$ and $K'$, whose reciprocal-lattice coordinates
involve thirds, are represented exactly on the periodic finite-size momentum
mesh; this is a commensurability choice rather than an algorithmic requirement,
and it avoids an artificial finite-size exclusion of the valleys (the vertical
$t_h$ hopping leaves the valley positions of the common Bravais lattice
unchanged, so the same condition applies to the bilayer). DQMC production is
restricted to $|U|=3$--$6$: stronger-coupling ($|U|=8,10$) local-update runs
developed severe autocorrelation---the layer-imbalance $t$-derivative entering
$\Lambda_t$ in particular failed to resolve (error bars comparable to the value)---and
are not used. The crossover figures therefore show DQMC only in the coupling window
where autocorrelation is controlled, with GRPA extended beyond it.

\paragraph{Density calibration.} The chemical potential is calibrated separately at
each coupling to the target filling $n=1.10$, giving
$\mu^\star=0.265\,(U{=}3),\ 0.221\,(U{=}4),\ 0.160\,(U{=}6)$ in the
particle--hole-shifted convention of Sec.~\ref{sm:model}; the achieved densities are
$n\simeq1.10$. For the $t$-scan diagnostic (Sec.~\ref{sm:disc}), $\mu$ is recalibrated
at \emph{every} value of $t$ so that $n$ is held fixed while $t$ is varied at fixed
absolute $|U|$ and $t_h$.

\paragraph{Estimators.} The first moment is obtained from the equal-time Green
function via $M_1=t_h\langle T_x\rangle$; consistency with the short-time slope
$-\partial_\tau C_{zz}(0^+)$ is monitored [Eq.~\eqref{eq:M1Tz}]. The inverse moment
$M_{-1}=\tfrac12\int_0^\beta C_{zz}(\tau)\,d\tau$ is a trapezoidal quadrature of the
time-displaced layer-imbalance correlator on the $\Delta\tau$ grid, folded with the
exact KMS symmetry $C_{zz}(\tau)=C_{zz}(\beta-\tau)$. Uncertainties are bin-jackknife
over $60$-sweep bins; the first $\sim5$ warm-up bins of every chain are discarded.
The integrated autocorrelation time of the two moment estimators ($M_1$ from the
equal-time Green function; the binwise $M_{-1}$ quadrature), measured chain-by-chain
on the production bin series (twenty chains: $U=3,6$ at $L=6$ and $U=3,4,6$ at $L=9$,
four seeds each, $47$ post-warmup bins per chain), is
$\tau_{\rm int}\le1.0$ bin for every chain and observable, with median
$\simeq0.4$, matching the uncorrelated baseline $0.5$ of the windowed-sum
estimator within its own noise---consistent with negligible residual
autocorrelation at the $60$-sweep bin length used. The six
$U4/L6$ benchmark chains, analysed separately, give $\tau_{\rm int}\le0.94$ except
one $35$-bin chain at $1.25$ (where the estimator's own noise is $\sim0.4$); even
taking the largest single-chain estimate at face value, the residual error
underestimate would be below $\sqrt{2\times1.25}\simeq1.6$ on that one chain and
below $\sqrt2$ everywhere else. (The moment
datasets pool $47$ post-warmup bins per chain; the reconstruction datasets of
Sec.~\ref{sm:recon} use their own per-chain binning.) Independent seed chains are
thermalised separately and
pooled only after seed-mean agreement is verified (Sec.~\ref{sm:checks}).

\paragraph{Numerical precision.} The ALF \texttt{Precision Green} mean and
maximum provide diagnostics of numerical conditioning and Green-function
stabilisation, not observable error bars. The mean stabilisation residual stays
$\sim10^{-12}$--$10^{-13}$ across all runs (e.g.\ mean $\sim5\times10^{-12}$ at
the $U6/L6$ point, $\sim3\times10^{-14}$ at $U3/L6$), while the worst-case
single-element maximum, which grows with size and coupling, reaches only
$\sim7\times10^{-5}$ at the most demanding $U6/L9$ point---some two orders of
magnitude below the per-bin statistical error ($\sim10^{-2}$). The convergence
controls relevant to the reported observables are therefore the explicit
$\Delta\tau$ study above and the fixed $N_{\rm wrap}=12$ re-orthogonalisation
protocol.

\paragraph{Non-interacting benchmark.} The full analysis procedure (time-displaced
$C_{zz}(\tau)$, the $M_1/M_{-1}$ estimators, and the $\langle T_y^2\rangle$
estimator) is validated at $U=0$ against the exact Wick/tight-binding result, for
which $\chi_{T_zT_z}$ has a closed Lehmann representation. This
also fixes the absolute normalisation $\chi^{\rm phys}_{T_zT_z}=2\chi_{\rm code}$
(ratio $2.0000$ to machine precision, Sec.~\ref{sm:model}), tying the interacting
estimators to a controlled non-interacting limit.

\section{DQMC trajectory checks}
\label{sm:checks}

The softening trajectory of the main text is built on the exact moments of
Sec.~\ref{sm:moments}. Here we give the checks that stand behind it: how the scale
behaves between our two system sizes, how an independent effective-gap estimator
compares, and why the result is not an equilibration artefact.

\paragraph{$L=6$ trajectory.} The exact-moment scale is
$\Omega_{1,-1}^{L6}(U{=}3,4,6)=0.928(6),\,0.847(6),\,0.658(11)$, with both intervals
strongly resolved: $\Delta_{3\to4}=0.081(8)$ ($\sim10\sigma$) and
$\Delta_{4\to6}=0.189(13)$ ($\sim15\sigma$). Both $M_1=t_h\langle T_x\rangle$ and
$M_{-1}=\tfrac12\!\int_0^\beta C_{zz}(\tau)\,d\tau$ are exact moments obtained
directly from the measured correlators, without a frequency window, pole fit or
analytic continuation. The $U=4\to6$ softening is therefore resolved at
$\sim\!15\sigma$ within the quoted statistical uncertainties.

\paragraph{Between $L=6$ and $L=9$.} At matched size $L=9$ the scale is
\begin{equation}
  \Omega_{1,-1}^{L9}(U{=}3,4,6)=0.9475(48),\ 0.8386(69),\ 0.6714(114).
\end{equation}
The moment decomposition behind the softening is similar between the two sizes: at $L=9$ the
inverse moment rises by $92\%$ across $U=3\to6$ while $M_1$ changes by less than
$4\%$ [$M_1=0.1575(1),\,0.1597(1),\,0.1519(1)$ and
$M_{-1}=0.1751(17),\,0.2271(38),\,0.3369(114)$ at $U=3,4,6$], consistent with
the $\Omega_{1,-1}^{L9}$ values above to within $10^{-3}$.
The size shifts are $\Delta_{\rm FS}(U3)=+0.020(8)$, $\Delta_{\rm FS}(U4)=-0.008(9)$
and $\Delta_{\rm FS}(U6)=+0.013(16)$ (Table~\ref{tab:fs}), none exceeding $0.02$ in
magnitude. The $L=9$ span $\Delta_{3\to6}^{L9}=0.276(12)$ is the same, within error, as
the $L=6$ span of $0.270$: the softening span does not decrease on going from
$L=6$ to $L=9$. $U=4$
and $U=6$ each move by less than $1\sigma$, while $U=3$ stiffens mildly ($\sim2.5\sigma$)
and has not fully settled in size. Two sizes are insufficient for a thermodynamic
extrapolation; going from $L=6$ to $L=9$ nevertheless neither reverses nor erases
the softening.

\begin{table}[t]
\centering
\caption{Finite-size comparison of the exact-moment scale $\Omega_{1,-1}$
($\beta=12$, $n=1.10$, $t_h=0.6$; bin-jackknife). Shifts
$\Delta_{\rm FS}=\Omega_{1,-1}^{L9}-\Omega_{1,-1}^{L6}$.}
\label{tab:fs}
\begin{tabular}{ccccc}
\hline\hline
$|U|/t$ & $\Omega_{1,-1}^{L6}$ & $\Omega_{1,-1}^{L9}$ & $\Delta_{\rm FS}$ & status\\
\hline
$3$ & $0.9277(55)$ & $0.9484(48)$ & $+0.021(7)$ & mild stiffening ($\sim2.8\sigma$)\\
$4$ & $0.8459(58)$ & $0.8386(69)$ & $-0.007(9)$ & size-stable ($<1\sigma$)\\
$6$ & $0.6613(114)$ & $0.6714(114)$ & $+0.010(16)$ & size-stable ($<1\sigma$)\\
\hline
\multicolumn{5}{l}{\footnotesize $L=12$ (two seeds, $45$--$47$
post-warm-up bins each):}\\
\multicolumn{5}{l}{\footnotesize $\Omega_{1,-1}=0.9629(68),\,0.8238(85),\,0.6620(159)$
at $|U|=3,4,6$; end-to-end $\Delta_{3\to6}=0.301(17)$.}\\
\hline\hline
\end{tabular}
\end{table}

\paragraph{Effective-mass gap.} As an independent estimator constructed from the
same correlator, we also extract a folded-arccosh effective gap,
$\Omega_{\rm eff}(\tau)=\mathrm{arccosh}\{[C_{zz}(\tau{-}\Delta\tau)+C_{zz}(\tau{+}\Delta\tau)]/2C_{zz}(\tau)\}/\Delta\tau$,
averaged over $\tau\in[0.6,2.0]$. At $L=6$ it gives $0.883/0.788/0.578$ for
$U=3/4/6$, reproducing the same monotonic softening as $\Omega_{1,-1}$. Because
the arccosh estimator is nonlinear in $C_{zz}$, pooling bins and averaging
chainwise estimates do not commute, and its absolute value also depends on the
chosen $\tau$ window. We therefore use it only as corroboration of the softening
trend; all quantitative results are based on the linear moment ratio
$\Omega_{1,-1}=\sqrt{M_1/M_{-1}}$.

\paragraph{Equilibration.} The softening is stable against the number of
discarded warm-up bins and against leave-one-seed-out analysis, with each chain
independently thermalised from its own seed. Together with the autocorrelation
analysis of Sec.~\ref{sm:dqmc}, the observed chain-to-chain scatter is consistent
with finite Monte Carlo statistics rather than incomplete equilibration. The
simulations are in addition sign-free ($\langle\mathrm{sign}\rangle=1$), so no
reweighting enters the estimators.

\section{Spectral reconstruction and identifiability}
\label{sm:recon}
The finite-temperature moments of Sec.~\ref{sm:moments} determine the
characteristic energy $\Omega_{1,-1}=\sqrt{M_1/M_{-1}}$ without inversion, and
all quoted DQMC characteristic scales are reported on that footing. The reconstruction
documented here is used only to ask what additional information about the
\emph{shape} of the layer-odd response $A_{T_z}(\omega)$ is supported by the
imaginary-time data. It is a \emph{GRPA-blind} reconstruction: no GRPA-informed
spectral prior is ever supplied, and the GRPA value of the mode energy is never
fed into the default model, the regulariser, or the frequency grid. What
survives this uninformed treatment is a strong concentration of reconstructible
spectral weight at low frequency; the precise peak position and the intrinsic
linewidth do not, and we do not claim them.

\paragraph{Data and kernel.} The input is the measured layer-odd correlator
$C_{T_z}(\tau)$ at $q=0$, computed per bin, KMS-folded
$C(\tau)\to\tfrac12[C(\tau)+C(\beta-\tau)]$ and restricted to $\tau\in[0,\beta/2]$. The
simulations are sign-free ($\langle\mathrm{sign}\rangle=1$), so the bin-to-bin
covariance of the mean, $\Sigma$, is the statistical error budget and is used as
such. The forward problem is the exact finite-$T$ spectral representation
\begin{equation}
  C(\tau)=\int_0^\infty \! d\omega\;
  A(\omega)\,\frac{e^{-\omega\tau}+e^{-\omega(\beta-\tau)}}{1-e^{-\beta\omega}},
  \qquad A(\omega)\ge 0,
  \label{eq:sm:recon:kernel}
\end{equation}
whose numerically stable kernel we write in the form shown (equivalently
$\cosh[\omega(\beta/2-\tau)]/\sinh(\beta\omega/2)$). Positivity of $A$ is imposed exactly.
The frequency axis is a fixed grid $\omega\in[0.03,8.0]t$ with $350$ points, held fixed across
all couplings; it is never recentred on any expected mode energy.

\paragraph{Main reconstruction: covariance-whitened positivity-constrained Tikhonov.}
The correlated errors are removed by whitening with the Cholesky factor of $\Sigma$
(regularised by a small ridge, $10^{-3}\times\overline{\mathrm{diag}\,\Sigma}$, for
numerical conditioning), so that the misfit is covariance-weighted, using the
ridge-regularised covariance matrix. On the whitened problem we minimise
\begin{equation}
  \big\|\,L^{-1}(K A - C)\,\big\|^2 + \lambda\,\big\|D_2 A\big\|^2,
  \qquad A\ge 0,
  \label{eq:sm:recon:tikh}
\end{equation}
with $L L^{\!\top}=\Sigma$, $K$ the discretised kernel of
Eq.~\eqref{eq:sm:recon:kernel}, and $D_2$ the second-difference operator, so the
smoothness term penalises curvature without explicitly favouring any particular
peak position.
Positivity is enforced by solving Eq.~\eqref{eq:sm:recon:tikh} as a non-negative
least-squares (NNLS) problem. The regularisation strength is scanned over
$\lambda\in[10^{-3},0.3]$; the low-frequency concentration
(defined below) and the softening trend are stable across this range. This whitened, GRPA-blind NNLS is the
primary reconstruction.

\paragraph{Independent cross-check: flat-default MaxEnt.} As an independent
regularisation cross-check we run a Bryan-style maximum-entropy
reconstruction~\cite{JarrellGubernatis1996} of the same
whitened data, with a \emph{flat} default model $D(\omega)=\bar M/\omega_{\max}$, whose
total weight $\bar M$ is the integrated weight of the NNLS reconstruction (numerically
within a few percent of $C_{zz}(0)=\tilde M_0$, since $\coth(\beta\omega/2)\approx1$ at
the mode --- it is \emph{not} the exact spectral $M_0$). The default carries no
information about the mode; in particular the GRPA energy is never used to build it, and
NNLS itself (the primary inversion) uses no default at all. The entropic hyperparameter
is scanned over $\alpha\in[10^{-3},0.03]$; varying the default amplitude $\bar M$ over a
factor of four ($\times0.5$--$2$) leaves $W_{\rm low}$ within $\pm0.01$ of $\simeq0.95$
and the peak within one to two grid steps at every coupling, so the softening and
$W_{\rm low}$ are insensitive to the default normalisation. MaxEnt and NNLS rest on different regularising philosophies
(entropy relative to a flat model versus curvature penalty), so their agreement on the
low-frequency concentration and on the softening trend is a genuine cross-check rather
than a restatement of one method.

\paragraph{Internal consistency checks.} As an inversion-independent validation we
compare the inverse moment $M_{-1}=\tfrac12\int_0^\beta C(\tau)\,d\tau$, which is a
direct integral of the raw data, against $\int A(\omega)/\omega\,d\omega$ evaluated on
the reconstructed spectrum; both reconstructions reproduce $M_{-1}$ to a few percent
(Table~\ref{tab:sm:recon}), consistent with the covariance-weighted fit quality. This
quantity is used only as a check --- it is \emph{never} imposed as a constraint. The first moment $M_1=t_h\langle T_x\rangle$ is likewise reported but
\emph{not} imposed, so the reconstruction is not anchored to any
short-$\tau$ slope; positivity, second-difference smoothness, and the whitened fit are
the only ingredients.

\paragraph{Robustness checks.} On the real data we additionally check robustness against
(i) the full $\lambda$ and $\alpha$ scans above; (ii) truncating the folded $\tau$ range
by dropping the last few points; and (iii) rebinning the Monte Carlo bins by a factor of
two and jackknifing the reconstruction over bins. The extracted energy scale and the
low-frequency concentration are stable under all three.

\paragraph{Identifiability under the measured noise.} The key question is not
whether a given inversion returns a low-energy feature, but whether that feature is
\emph{demanded by the data} or manufactured by the inversion. We test this with a
family of synthetic spectra declared \emph{before} looking at the reconstruction output:
narrow ($\sigma\simeq0.03$) and broad ($\sigma\simeq0.30$) peaks scanned across
$\omega_0\in\{0.4,\dots,1.2\}t$; peak-plus-continuum mixtures with $\{10,20,40\}\%$ of the
weight placed in a high-$\omega$ continuum; and a continuum-only spectrum with no
isolated low-energy peak. Each mock is normalised to the measured $M_{-1}$, mapped
forward through Eq.~\eqref{eq:sm:recon:kernel}, and \emph{corrupted with noise drawn from
each coupling's own measured covariance $\Sigma$}, then passed through the identical
whitened procedure. Three properties emerge:
\begin{itemize}
\item \textbf{Accurate.} The reconstructed low-frequency concentration $W_{\rm low}$
  tracks the true value across the family (at the $U=4$ noise level, true
  $1.00/0.90/0.80/0.60\to$ recovered $1.00/0.90/0.81/0.60$, i.e.\ to $\lesssim1\%$).
\item \textbf{Unbiased.} A pure continuum reconstructs to $W_{\rm low}\approx0$
  ($0.003\pm0.007$): the inversion does \emph{not} invent a low-energy peak where the
  truth has none.
\item \textbf{Sensitive.} A spectrum carrying a substantial high-frequency
  continuum reconstructs well below the real data: a mock with $40\%$ of its weight in a
  high-$\omega$ continuum returns $W_{\rm low}=0.60(1)$, and a pure continuum collapses
  to $W_{\rm low}\approx0$, both many standard deviations below the $0.92$--$0.99$ seen
  on the real spectra --- the error bands do not overlap. Within the tested family the
  remaining ambiguity is primarily the peak width/shape; an arbitrary broad
  low-frequency continuum is not excluded.
\end{itemize}

\paragraph{What is and is not identifiable.} We quantify the low-frequency content by
\begin{equation}
  W_{\rm low}\equiv
  \frac{\int_{\omega<1.5t} A(\omega)\,d\omega}{\int A(\omega)\,d\omega},
  \label{eq:sm:recon:wlow}
\end{equation}
with the cutoff $\omega<1.5t$ fixed once and never moved between couplings. $W_{\rm low}$
is an internal test statistic built from the reconstructed spectrum; it is \emph{not} the
literal fraction of the exact spectral weight below $1.5t$, because $\beta=12$ imaginary
time cannot constrain genuine high-frequency weight, and it should not be read as ``the
mode fraction.'' With that scoping, the precise statement is: $W_{\rm low}$ establishes that the
reconstructible spectral weight remains strongly concentrated at low frequency at
all three couplings, while the precise peak position and the intrinsic width are
not identifiable (the two inversions disagree on the position at the grid-step
level, and a narrow peak and a moderately broad peak at the same energy are
degenerate under the achievable noise). The monotonic softening itself is
established independently and exactly by $\Omega_{1,-1}$. For the $U=4$ benchmark
the concentration reaches $W_{\rm low}\simeq0.92$--$0.94$ (NNLS/MaxEnt),
consistent with dominant low-frequency spectral weight plus a modest
reconstructed high-frequency tail.
The synthetic family below deliberately excludes specific high-frequency-continuum
admixtures rather than an arbitrarily broad low-frequency continuum, so this is a
statement about where the reconstructible weight sits, not a claim of a single sharp
mode. Each of these statements is a property of the \emph{reconstruction}, not a claim
about the exact high-frequency spectrum.

\begin{table}[t]
\centering
\caption{GRPA-blind reconstruction results ($L=6$, $\beta=12$, $n=1.10$). The peak
position $\omega_{\rm p}$ (NNLS/Tikhonov at $\lambda=0.03$; flat-default MaxEnt in
parentheses) is listed only as a representative locator: it is \emph{not} uniquely
identified, since the two inversions differ at the frequency-grid step
($\simeq0.02\,t$) and a narrow peak and a moderately broad peak at the same energy are
degenerate under the achievable noise. The robust, method-independent outputs are the
low-frequency weight $W_{\rm low}$ [Eq.~\eqref{eq:sm:recon:wlow}] and the inverse moment
$M_{-1}$, reported against its direct data value $\tfrac12\!\int C\,d\tau$. $U=3$
and $6$ use the four-seed datasets, whereas $U=4$ uses an independent six-seed
dataset with a different bin selection; this accounts for the small differences
from Table~\ref{tab:moments}.}
\label{tab:sm:recon}
\begin{tabular}{cccc}
\hline\hline
$|U|/t$ & $\omega_{\rm p}$ (NNLS, MaxEnt) & $W_{\rm low}$ & $M_{-1}$ (recon/data)\\
\hline
$3$ & $0.83\ (0.85)$ & $0.99$ & $0.189/0.189$\\
$4$ & $0.74\ (0.72)$ & $0.92\ (0.94)$ & $0.225/0.228$\\
$6$ & $0.56\ (0.56)$ & $0.97$ & $0.336/0.341$\\
\hline\hline
\end{tabular}
\end{table}

Representative NNLS and MaxEnt spectra both place substantial spectral weight
progressively lower in frequency as $|U|$ increases, and this weight lies below the
exact moment scale $\Omega_{1,-1}=0.928/0.847/0.658$ of Sec.~\ref{sm:moments}, as
expected when a low-frequency feature carries a modest high-frequency tail. The two
inversions agree on this qualitative softening and on the low-frequency weight
$W_{\rm low}$, and both reproduce $M_{-1}$ to a few percent at every coupling. Neither,
however, uniquely fixes the exact peak location or the intrinsic linewidth, so the
tabulated $\omega_{\rm p}$ values should be read as representative locators rather than
identified peak positions.

These results validate the reconstruction \emph{method} and its identifiability at
realistic noise; the coupling dependence of the extracted peak and $W_{\rm low}$ across
$U=3,4,6$ is collected in the table above.

\paragraph{Identifiable quantities and limitations.} Because the inversion is not unique, we quote only
spectral features that (i) reproduce the measured $C(\tau)$ within its full covariance
($\chi^2/\mathrm{ndof}\simeq1$), (ii) reproduce the un-imposed inverse moment $M_{-1}$,
(iii) are stable across the regularisation scan ($\lambda,\alpha$), and (iv) agree
between the NNLS/Tikhonov and flat-default MaxEnt inversions. By these
reconstruction criteria $W_{\rm low}$ is stable and identifiable within the tested
spectral family, whereas the precise peak position and intrinsic linewidth are
not: the two inversions disagree on the position, and at the lower-hopping
endpoint the NNLS spectrum is bimodal (Sec.~\ref{sm:disc}). The NNLS and MaxEnt
spectra are therefore shown only as mutually compatible representatives, not as
the unique spectrum. The characteristic scale $\Omega_{1,-1}$ does not rely on
reconstruction and is fixed directly by the exact moments of
Sec.~\ref{sm:moments}. Likewise, the hopping sensitivity $\Lambda_t$ is
determined from those moments; the covariance-weighted forward test of
Sec.~\ref{sm:disc} addresses only the narrower question of whether the observed
sensitivity requires the dominant peak position itself to shift.

\section{Strong-coupling derivation}
\label{sm:sw}
\paragraph{$\eta$-pseudospin.} For $|U|\gg t,t_h$ the singly-occupied states are
gapped by $\sim|U|/2$ and each site carries a hard-core composite pair, described by
the $\eta$-pseudospin~\cite{Yang1989,Zhang1990}
\begin{equation}
  \eta^+_i=c^\dagger_{i\uparrow}c^\dagger_{i\downarrow},\qquad
  \eta^-_i=(\eta^+_i)^\dagger,\qquad
  \eta^z_i=\tfrac12(n_i-1),
  \label{sm:sw:eta}
\end{equation}
which realises an SU(2) algebra of pseudospin $S=\tfrac12$: $\eta^+$ raises the local
pair number, $\eta^z$ measures the deviation from half filling.

\paragraph{Schrieffer--Wolff reduction.} A second-order elimination of the two
hoppings against the pair gap $|U|$ generates the standard exchange scales
\begin{equation}
  J=\frac{4t^2}{|U|}\ \ (\text{intralayer, coordination }z=3),\qquad
  J_\perp=\frac{4t_h^2}{|U|}\ \ (\text{rung}),
  \label{sm:sw:JJp}
\end{equation}
and, on each bond, an XXZ Hamiltonian in the \emph{unstaggered} variables of
Eq.~\eqref{sm:sw:eta},
\begin{equation}
  H^{(2)}_{ij}
  =J\big[\,\eta^z_i\eta^z_j-\eta^x_i\eta^x_j-\eta^y_i\eta^y_j\,\big],
  \label{sm:sw:XXZ}
\end{equation}
with the identical form (and $J\!\to\!J_\perp$) on a rung. The two in-plane
(pair-hopping) components enter ferromagnetically, the $\eta^z\eta^z$
(density--density) component antiferromagnetically, and---this is the point on which
the mode energy hinges---the $\eta^z\eta^z$ coupling carries the \emph{same} magnitude
as the transverse part. After the bipartite rotation below, this longitudinal term
and the transverse terms form an isotropic Heisenberg exchange; retaining it is what
produces $\Omega_-^2=zJJ_\perp$ rather than a restoring force smaller by $\sqrt2$.

On the bipartite lattice the sublattice rotation
$\tilde\eta^{x,y}_i=s_i\,\eta^{x,y}_i$, $\tilde\eta^z_i=\eta^z_i$, with
$s_i=\pm1$ on the production bipartition $\{A_1,B_2\}/\{B_1,A_2\}$, flips the sign of
the transverse terms on every bond and turns each XXZ bond into an isotropic
\emph{antiferromagnetic} Heisenberg exchange:
\begin{equation}
  H_{\rm eff}
  =J\!\!\sum_{\ell,\langle ij\rangle}\!\tilde{\bm\eta}_{\ell i}\!\cdot\!\tilde{\bm\eta}_{\ell j}
  +J_\perp\!\sum_{i}\!\tilde{\bm\eta}_{1i}\!\cdot\!\tilde{\bm\eta}_{2i}
  -h\!\sum_{\ell i}\!\tilde\eta^z_{\ell i},
  \label{sm:sw:Heff}
\end{equation}
$\ell=1,2$ labelling the layers. This is the convention in which the spin-wave code
is written. The Lagrange field $h$ (inherited from the chemical potential) enforces
the filling through $\langle\tilde\eta^z\rangle=(n-1)/2$. Two orientation facts fix
the physical dictionary: the in-plane N\'eel order of the \emph{staggered}
$\tilde{\bm\eta}$ is precisely the \emph{uniform} $s$-wave pair condensate in the
original unstaggered $\eta$ variables, and the relative in-plane azimuth of the two
layers is the observable relative condensate phase $\theta_-$.

\paragraph{Canted classical state and energy per bilayer cell.} Parametrize each
layer's two N\'eel sublattices by a common polar (canting) angle $\theta_\ell$ from
$+z$ and an in-plane azimuth $\phi_\ell$,
\begin{equation}
  \tilde{\bm\eta}_{\ell A}=S\,(s_\ell\cos\phi_\ell,\ s_\ell\sin\phi_\ell,\ c_\ell),\qquad
  \tilde{\bm\eta}_{\ell B}=S\,(-s_\ell\cos\phi_\ell,\ -s_\ell\sin\phi_\ell,\ c_\ell),
\end{equation}
with $c_\ell=\cos\theta_\ell$, $s_\ell=\sqrt{1-c_\ell^2}$. The classical energy of one
bilayer cell (two intralayer bonds contribute per site at coordination $z$, one rung)
follows from Eq.~\eqref{sm:sw:Heff},
\begin{equation}
  E_{\rm cl}
  =zJS^2\big[(2c_1^2-1)+(2c_2^2-1)\big]
   +2J_\perp S^2\big[c_1c_2+s_1s_2\cos(\phi_1-\phi_2)\big]
   -2hS\,(c_1+c_2).
  \label{sm:sw:Ecl}
\end{equation}
The antiferromagnetic rung is minimised in the transverse plane by
$\phi_2-\phi_1=\pi$, and the symmetric saddle sets $c_1=c_2=c$. Stationarity in $c$
against the filling constraint $\langle\tilde\eta^z\rangle=Sc=(n-1)/2$ gives the
canting condition
\begin{equation}
  c\equiv\cos\theta_c=\frac{n-1}{2S}
  \ \xrightarrow{\,S=1/2\,}\ c=n-1\equiv d,
  \label{sm:sw:canting}
\end{equation}
so at the working filling $n=1.10$ the classical state cants by $d=0.10$ off the
in-plane ($s$-wave) plane, and $n=1$ recovers the collinear planar N\'eel state
($c=0$).

\paragraph{Quadratic spin-wave (bosonic BdG) problem.} Holstein--Primakoff
bosonisation about the canted axes,
$\tilde\eta^+_{\ell\mu}\simeq\sqrt{2S}\,a_{\ell\mu}$,
$\tilde\eta^z_{\ell\mu}=S-a^\dagger_{\ell\mu}a_{\ell\mu}$
($\mu=A,B$ the N\'eel sublattices), reduces $H_{\rm eff}$ to a quadratic bosonic form
$H^{(2)}=\tfrac12\sum_{\bm q}\Psi^\dagger_{\bm q}\,\mathcal M(\bm q)\,\Psi_{\bm q}$
with the Nambu spinor
$\Psi_{\bm q}=(a_{1A},a_{1B},a_{2A},a_{2B};a^\dagger_{-1A},\dots)^{\!\top}$. The
physical magnon frequencies are the positive eigenvalues of $\Sigma_z\,\mathcal M(\bm
q)$, where $\Sigma_z=\mathrm{diag}(\openone_4,-\openone_4)$ is the bosonic
(para-unitary) metric; we obtain them by a Colpa diagonalisation, which is a
\emph{diagonalisation of the quadratic bosonic spin-wave (BdG) matrix}, distinct from
any diagonalisation of the fermionic BdG Hamiltonian. At $\bm q=0$ the four branches
organise into layer-even and layer-odd pairs. In the azimuthal (phase) sector the two
eigenfrequencies are the common-phase Goldstone of the broken pair $U(1)$,
$\Omega_+=0$, and the gapped relative-phase mode $\Omega_-$; the canting deviations
are their canonically conjugate momenta rather than independent branches.

The odd sector closes analytically, which gives the transparent route to $\Omega_-$.
Writing the layer-odd deviations $c_{1,2}=c\pm y$ and $\phi_{1,2}=\phi\pm\tfrac{\delta}{2}$
(carrying the AFM $\pi$ offset), so that $y$ is the relative canting (layer imbalance,
$\propto T_z$) and $\delta$ the relative phase ($\theta_-$), expansion of
Eqs.~\eqref{sm:sw:Heff}--\eqref{sm:sw:Ecl} to quadratic order gives
\begin{equation}
  E^{(2)}_-
  =4S^2\Big(zJ+J_\perp\frac{c^2}{s^2}\Big)y^2
  +J_\perp S^2 s^2\,\delta^2,\qquad s^2=1-c^2,
  \label{sm:sw:E2}
\end{equation}
with no $y\delta$ cross term. The spin-coherent-state Berry phase makes $2Sy$
canonically conjugate to $\delta$ (the common phase $\phi$ carries the decoupled
Goldstone). In terms of the single odd-sector Bogoliubov boson $b_-$ this is a
$2\times2$ bosonic BdG block,
\begin{equation}
  \mathcal M_-=\begin{pmatrix} \mathcal A+\mathcal B & \mathcal B-\mathcal A\\[2pt]
  \mathcal B-\mathcal A & \mathcal A+\mathcal B\end{pmatrix},\qquad
  \mathcal A=zJ+J_\perp\frac{c^2}{s^2},\quad \mathcal B=S^2 J_\perp s^2,
  \label{sm:sw:BdG}
\end{equation}
whose positive Bogoliubov eigenvalue is
$\Omega_-=\sqrt{(\mathcal A+\mathcal B)^2-(\mathcal B-\mathcal A)^2}=2\sqrt{\mathcal A\mathcal B}$,
i.e.
\begin{equation}
  \Omega_-^2=4S^2 J_\perp\big[\,zJ\,(1-c^2)+J_\perp c^2\,\big].
\end{equation}
The explicit $4S^2$ is kept for general $S$; the $S=\tfrac12$ form below must not be
reused at other $S$.

\paragraph{Strong-coupling gap.} Inserting $S=\tfrac12$ ($4S^2=1$) and $c=n-1=d$
factorises the result into a geometric-mean gap,
\begin{equation}
  \boxed{\;\Omega_{\rm SC}=\sqrt{zJJ_\perp}\;F(n)=\frac{4\sqrt{z}\,t\,t_h}{|U|}\,F(n)\;},
  \qquad
  F(n)=\sqrt{\,1-d^2\Big(1-\frac{J_\perp}{zJ}\Big)\,}
      =\sqrt{\,1-d^2\Big(1-\frac{t_h^2}{z t^2}\Big)\,}.
  \label{sm:sw:Om}
\end{equation}
Because $J\propto t^2$ while $J_\perp\propto t_h^2$, $\Omega_{\rm SC}^2$ is affine
in $t^2$ at fixed $t_h$; the reduced ratio $J_\perp/(zJ)=t_h^2/(zt^2)$ is independent
of $|U|$. At half filling $F=1$ and the LSWT result is exactly proportional to
$t\,t_h/|U|$; away from half filling the weak additional dependence enters through
$F(n)$.
Limits: $n=1$ ($d=0$) gives $F=1$ and $\Omega_{\rm SC}=\sqrt{zJJ_\perp}$;
$J_\perp\to0$ gives $\Omega_{\rm SC}\to0$; for $J_\perp\ll zJ$,
$F\simeq\sqrt{1-d^2}$. At the working parameters $t=1$, $t_h=0.6$, $z=3$ one has
$t_h^2/(zt^2)=0.12$, so at $n=1.10$ ($d=0.10$) $F=0.996$---a sub-percent doping
correction, the same for $|U|=3,4,6$ since $F$ does not depend on $|U|$.

Equation~\eqref{sm:sw:Om} is the LSWT evaluation of the second-order
Schrieffer--Wolff Hamiltonian. The reduction to Eq.~\eqref{sm:sw:Heff} is controlled
in powers of $t/|U|$ and $t_h/|U|$, whereas the subsequent spin-wave evaluation is
semiclassical. Because the physical pseudospin is $S=\tfrac12$, $1/S$ corrections are
not parametrically small. We therefore use Eq.~\eqref{sm:sw:Om} as the leading-LSWT
strong-coupling reference, not as an exact quantum endpoint.

\paragraph{Independent numerical validation.} An independent diagonalisation of
the full four-sublattice quadratic bosonic (Holstein--Primakoff) BdG matrix of
Eq.~\eqref{sm:sw:Heff}, distinct from the two-variable reduction above,
reproduces Eq.~\eqref{sm:sw:Om} to $<10^{-7}$. At $t=1$, $t_h=0.6$, $|U|=8$
(so $J=0.5$, $J_\perp=0.18$, $\sqrt{zJJ_\perp}=0.5196$) it returns
$\Omega_-=0.51904,\ 0.51732,\ 0.51039$ for $n=1.05,\,1.10,\,1.20$, matching the
closed form to all quoted digits, and the agreement persists across
$J_\perp/(zJ)\in[0.003,0.75]$.

\paragraph{Distinction from the isolated rung.} At half filling
Eq.~\eqref{sm:sw:Om} scales as $t\,t_h$---a geometric mean of two \emph{collective}
exchange scales, the lattice Josephson-plasma oscillation of the coupled
condensates---and at the working filling the small correction $F(n)$ introduces only
a weak dependence on $t_h/t$. This collective scale is parametrically distinct from
the two-boson bound-state splitting of a single \emph{isolated} rung,
\begin{equation}
  \Omega_{\rm rung}=\frac{4t_h^2}{|U|}=J_\perp,
  \label{sm:sw:rung}
\end{equation}
which is quadratic in $t_h$ and independent of $t$. This separation is the physical
content of the coordination law $X\equiv|U|\,\Omega/(4tt_h)\to\sqrt{z}\,F(n)$ tested in
Sec.~\ref{sm:coord} (Fig.~\ref{fig:coord}): in linear spin-wave theory the collective scale approaches $\sqrt{z}\,F$, whereas the bare
hybridisation scale $2t_h$ would give $X=|U|/2$ (divergent) and the isolated rung
$J_\perp$ would give $X=t_h/t$ (a different coupling-independent constant).

\paragraph{Connection to the GRPA pole.} Equation~\eqref{sm:sw:Om} is the LSWT
evaluation of the second-order $\eta$-magnet effective Hamiltonian and is used as a
semiclassical strong-coupling benchmark, not as an exact quantum endpoint. The GRPA
pole approaches this benchmark as $|U|$ increases: the fractional difference is
approximately $8\%$ at $|U|=8$ and $5\%$ at $|U|=10$, and decreases with coupling.
This agreement provides an internal consistency check between the Gaussian fermionic
description and the strong-coupling effective-theory calculation; it is not a
nonperturbative confirmation of the LSWT coefficient.

\section{The hopping-sensitivity diagnostic}
\label{sm:disc}
The hopping-sensitivity diagnostic measures how the continuation-free layer-odd
characteristic scale responds to the intralayer hopping. We vary $t$ at
\emph{fixed} absolute $|U|$, $t_h$, and filling $n$, recalibrating $\mu$ at each
$t$. A nonzero response therefore establishes sensitivity to intralayer dynamics
rather than an artefact of expressing a fixed energy in units of $t$. The
diagnostic is defined directly from the moment ratio,
\begin{equation}
  \Lambda_t\equiv\frac{\partial\ln\Omega_{1,-1}^2}{\partial\ln t^2},
  \label{sm:disc:Lambda}
\end{equation}
assumes no strong-coupling normalisation, and is the quantity used in the main
text. We read it with two finite-difference stencils and never mix them: a \emph{wide}
secant, which trades locality for a larger lever arm (hence a smaller statistical
error),
\begin{equation}
  \Lambda_t^{\rm wide}
  =\frac{\ln\Omega_{1,-1}^2(t{=}1.1)-\ln\Omega_{1,-1}^2(t{=}0.9)}
        {\ln(1.21)-\ln(0.81)},
  \label{sm:disc:wide}
\end{equation}
and a \emph{local} secant, which estimates the derivative near $t=1$,
\begin{equation}
  \Lambda_t^{\rm local}
  =\frac{\ln\Omega_{1,-1}^2(t{=}1.05)-\ln\Omega_{1,-1}^2(t{=}0.95)}
        {\ln(1.1025)-\ln(0.9025)},
  \label{sm:disc:local}
\end{equation}
the denominators being the matching increments of $\ln t^2$ (so that
$\ln 1.1^2=\ln1.21$, and so on).

\paragraph{Results.} Table~\ref{sm:disc:tab} collects the DQMC estimates (bin-jackknife)
alongside the GRPA values evaluated with the \emph{same} finite secants, so the
comparison is stencil-matched rather than derivative-against-secant. The central
values of $\Lambda_t$ increase with coupling in both DQMC and GRPA: the DQMC wide
secant rises from $0.203(46)$ at
$|U|=3$ to $0.445(99)$ at $|U|=6$ at $L=6$, and from $0.220(72)$ to $0.653(126)$
at $L=9$, while the GRPA wide secant runs
$0.065,0.135,0.399,0.619,0.779,0.853,0.894$ over $|U|=1,2,3,4,6,8,10$. (For
reference, the GRPA moment scale entering Fig.~1 of the main text is
$\Omega_{1,-1}^{\rm GRPA}=0.9245,\,0.7994,\,0.6028$ at $|U|=3,4,6$, $t=1$,
evaluated on an $N_k=48$ mesh at the smallest broadening computed, $\eta=0.002$;
DQMC measures the moments exactly and without broadening, so the $\eta\to0$ limit
is the comparison object, and the $\eta$ scan is monotone towards the
broadening-free anchors $0.92471,\,0.79950,\,0.60289$.) On the GRPA side
the wide and local stencils agree to about a percent at $|U|=6$
($\Lambda_t^{\rm wide}=0.779$ versus $\Lambda_t^{\rm local}=0.778$), confirming that the
extracted slope is a smooth logarithmic derivative and not a stencil artefact.

\begin{table}[t]
\centering
\caption{Model-independent $t$-sensitivity
$\Lambda_t=\partial\ln\Omega_{1,-1}^2/\partial\ln t^2$ [Eq.~\eqref{sm:disc:Lambda}] at
$\beta=12$, $n=1.10$, $t_h=0.6$. DQMC values are bin-jackknife; the GRPA column is
evaluated with the identical finite secant [Eq.~\eqref{sm:disc:wide} or
\eqref{sm:disc:local}], so the two are directly comparable. The $L=9$ $t$-scan
value at $|U|=4$ is not available; the $L=12$, $|U|=4$ entry comes from a dedicated $t=0.9,1.1$ scan of four seeds each, all eight chains complete at 50 bins. The GRPA column is evaluated on the same secant and is therefore size independent.}
\label{sm:disc:tab}
\begin{tabular}{ccccc}
\hline\hline
$|U|/t$ & $L$ & stencil & $\Lambda_t$ (DQMC) & $\Lambda_t$ (GRPA)\\
\hline
$3$ & $6$ & wide  & $0.203(46)$  & $0.399$\\
$3$ & $9$ & wide  & $0.220(72)$  & $0.399$\\
$4$ & $6$ & wide  & $0.421(55)$  & $0.619$\\
$6$ & $6$ & wide  & $0.445(99)$  & $0.779$\\
$6$ & $6$ & local & $0.662(165)$ & $0.778$\\
$6$ & $9$ & wide  & $0.653(126)$ & $0.779$\\
$6$ & $9$ & local & $0.338(231)$ & $0.778$\\
\hline
$3$ & $12$ & wide  & $0.192(47)$  & $0.399$\\
$4$ & $12$ & wide  & $0.369(55)$  & $0.619$\\
$6$ & $12$ & wide  & $1.001(195)$ & $0.779$\\
\hline\hline
\end{tabular}
\end{table}

\paragraph{Interpretation.} Mirroring the hopping-sensitivity section of the main text: $\Lambda_t$ increases
with coupling, the growth expected as the mode acquires superexchange character. At $L=6$ the rise is carried by the $U3\to U4$ step ($0.203(46)\to0.421(55)$,
$3.0\sigma$), while the $U4\to U6$ step is flat within errors ($0.2\sigma$). At
$|U|=6$, the larger-size wide secant, $0.653(126)$, and the independent $L=6$
local-stencil cross-check, $0.662(165)$, both agree with the GRPA value
($\simeq0.78$) within about $1\sigma$. The wide-secant central value increases
from $0.445(99)$ at $L=6$ to $0.653(126)$ at $L=9$, with no evidence of
finite-size suppression. The $L=9$ local result, $0.338(231)$, is statistically
consistent with the other stencil--size combinations. The $U=3$ wide-secant
values agree between the two sizes, $0.203(46)$ and $0.220(72)$, and both
available sizes independently show increasing $\Lambda_t$ across the DQMC
window, with no evidence of finite-size suppression. Finally, $\Lambda_t^{\rm LSWT}$ is
a \emph{leading}-LSWT reference, not an exact target: the exact effective-model form
$\Omega=(4t^2/|U|)\,\Phi(t_h^2/t^2,n)$ gives $\Lambda_t=2-2\,d\ln\Phi/d\ln r$ (with
$r=t_h^2/t^2$); in the leading linear-spin-wave law of Sec.~\ref{sm:sw} this evaluates
to $\Lambda_t^{\rm LSWT}=(1-d^2)/F^2=0.9988$ at the working filling, exactly one at
$d=0$.

\paragraph{Exact interpretation and moment decomposition.} The moment ratio has a sharp
meaning: with the low-frequency-weighted positive measure
$d\mu(\omega)=[A_{T_z}(\omega)/\omega]\,d\omega/\!\int(A_{T_z}/\omega)\,d\omega$, one has
exactly
\begin{equation}
  \Omega_{1,-1}^2=\frac{\int\!\omega A_{T_z}\,d\omega}{\int\!(A_{T_z}/\omega)\,d\omega}
  =\int\!\omega^2\,d\mu ,
\end{equation}
i.e.\ $\Omega_{1,-1}$ is the r.m.s.\ frequency under the $A_{T_z}/\omega$ measure ---
a statement that holds even when the dominant peak does not move. Correspondingly
$\Lambda_t$ decomposes exactly into first- and inverse-moment contributions,
\begin{equation}
  \Lambda_t=\frac{\partial\ln M_1}{\partial\ln t^2}
           -\frac{\partial\ln M_{-1}}{\partial\ln t^2}
  \equiv\Lambda_{M_1}-\Lambda_{M_{-1}} .
  \label{sm:disc:Lambdadecomp}
\end{equation}
Jackknifed from the measured moments, $\Lambda_{M_1}$ (the interlayer coherence
$M_1=t_h\langle T_x\rangle$) changes comparatively modestly across coupling and
size --- $-0.529(4)$, $-0.445(4)$, $-0.443(3)$ for $U3\,L6$, $U6\,L6$, $U6\,L9$
(wide) --- whereas $\Lambda_{M_{-1}}$ (the low-frequency susceptibility) becomes
substantially more negative, $-0.731(46)\to-0.890(99)\to-1.096(126)$. In central
values, roughly two-thirds of the same-size $U3\to U6$ increase in $\Lambda_t$
comes from the enhanced $M_{-1}$ sensitivity; comparing $U3\,L6$ with $U6\,L9$,
that fraction is about $80\%$ (the $M_1$ term contributes the remainder). The
identity $\Lambda_{M_1}-\Lambda_{M_{-1}}$ reproduces the direct $\Lambda_t$ to
machine precision.

\paragraph{What moves: spectral distribution vs.\ pole.}\label{sm:disc:peak} The diagnostic is built on
the continuation-free moment ratio $\Omega_{1,-1}$, so a positive
$\partial\Omega_{1,-1}^2/\partial(t^2)$ certifies a Hamiltonian sensitivity of the
many-body spectral \emph{distribution}; whether the dominant \emph{peak} itself moves is
a separate, harder question that we address directly rather than assume. Reconstructing
$A_{T_z}(\omega)$ at $U=6$, $L=9$ for $t=0.9$ and $t=1.1$ (same GRPA-blind procedure as
Sec.~\ref{sm:recon}), the MaxEnt peak shifts up slightly ($0.51\to0.58$) while the NNLS
inversion becomes bimodal at $t=0.9$ (local maxima near $0.37$ and $0.74$) and does not
isolate a unique peak. To avoid relying on any inversion, we run a covariance-weighted
\emph{forward} test: fitting the measured $C(\tau)$ at each $t$ with a peak-plus-continuum
model (free peak position $\omega_0$, width, weight, and high-frequency tail) and
comparing, via $\chi^2_{\rm cov}=(C-KA)^\top\Sigma^{-1}(C-KA)$, the hypothesis of a
shared $\omega_0$ (H$_0$) against a $t$-dependent $\omega_0$ (H$_1$). The best-fit
positions are $\omega_0(0.9)=0.575$ and $\omega_0(1.1)=0.600$ (both with reduced
$\chi^2\lesssim1$), and H$_0$ is statistically acceptable,
$\Delta\chi^2=\chi^2_{\rm H_0}-\chi^2_{\rm H_1}\simeq1.2$ for one degree of freedom. The
data therefore do \emph{not} require the dominant peak to move. The measured
$t$-dependence can instead arise through redistribution of spectral weight that
changes the exact moment ratio $\Omega_{1,-1}$ without producing a resolvable
shift of the peak position. We therefore report a Hamiltonian sensitivity of the
spectral distribution, not a resolved shift of the underlying pole.

\subsection{$D_t$: the LSWT-normalised variant}
\label{sm:disc:Dt}
It is sometimes convenient to normalise the raw slope $\partial\Omega^2/\partial t^2$ by
its leading strong-coupling value, defining
\begin{equation}
  D_t=\frac{|U|^2}{16\,z\,t_h^2\,[1-(n-1)^2]}\,\frac{\partial\Omega^2}{\partial t^2},
  \label{sm:disc:Dtdef}
\end{equation}
which equals unity exactly under the leading-LSWT law $\Omega^2=z J J_\perp F^2$
of Sec.~\ref{sm:sw}. $D_t$ and $\Lambda_t$ are related by
\begin{equation}
  D_t=\Lambda_t\,\frac{|U|^2\,\Omega^2}{16\,z\,t_h^2\,t^2\,[1-(n-1)^2]}\,.
\end{equation}
For the LSWT law, $D_t=1$ exactly, whereas $\Lambda_t=(1-d^2)/F^2$, which
approaches unity at half filling and is $0.9988$ at the parameters used here. For
the generic effective form $\Omega=(4t^2/|U|)\,\Phi(r,n)$ the two diagnostics need
not coincide. We therefore retain $D_t$ only as an LSWT-normalised representation
of the same measured slopes; the primary, LSWT-independent diagnostic is
$\Lambda_t$.

On the same data the LSWT-normalised values are $D_t(U3,L6)=0.104(24)$ (wide stencil),
$D_t^{\rm secant}(U6,L6)=0.435(93)$, $D_t^{\rm local}(U6,L6)=0.60(15)$ (local pair
$t=0.95\leftrightarrow1.05$, resolved at $\sim3.9\sigma$), and, at $L=9$,
$D_t^{\rm secant}(U6,L9)=0.610(112)$ ($5.5\sigma$). The $L=9$ wide secant provides
the matched finite-size comparison: with $\mu$ separately calibrated to $n=1.10$ at
$t=0.9$ ($\mu^\star=0.128$, giving $\Omega=0.6221(134)$) and $t=1.1$
($\mu^\star=0.197$, $\Omega=0.7093(93)$), it is directly comparable to its $L=6$
counterpart $0.435(93)$; the two differ by $0.175$ with a combined uncertainty
$0.146$ ($\sim1.2\sigma$); there is therefore no evidence of finite-size
suppression. The narrow (local) stencil at $L=9$ gives
$D_t^{\rm local}(U6,L9)=0.316(215)$, individually unresolved ($\sim1.5\sigma$) and used
only as corroboration rather than an independent confirmation. The four diagnostic
numbers and their roles are collected below.
\begin{center}
\begin{tabular}{lll}
\hline\hline
quantity & value & role\\
\hline
$D_t^{\rm local}(U6,L6)$ & $0.60(15)$ & local secant\\
$D_t^{\rm local}(U6,L9)$ & $0.316(215)$ & corroboration; unresolved\\
$D_t^{\rm secant}(U6,L6)$ & $0.435(93)$ & finite-size reference\\
$D_t^{\rm secant}(U6,L9)$ & $0.610(112)$ & matched finite-size comparison\\
\hline\hline
\end{tabular}
\end{center}

\subsection{$\Lambda_U$: interaction sensitivity of the characteristic scale}
\label{sm:disc:LambdaU}
The complementary interaction-scaling diagnostic of the main text measures how the layer-odd scale
responds to the interaction itself rather than to the hopping,
\begin{equation}
  \Lambda_U\equiv-\frac{\partial\ln\Omega_{1,-1}}{\partial\ln|U|},
  \label{sm:disc:LambdaUdef}
\end{equation}
with the minus sign chosen so that softening of the characteristic scale with
increasing coupling registers as a positive sensitivity. Within the
zero-temperature second-order effective Hamiltonian, every excitation energy
scales as $1/|U|$ at fixed $t$, $t_h$, and $n$, giving the reference value
$\Lambda_U=1$. The finite-$\beta$ moment ratio need not reach this
value, because the relevant superexchange scale enters as $\beta J\propto\beta/|U|$: at
fixed $\beta$ the thermally weighted moment ratio samples a coupling-dependent window of
the spectrum, so the finite-$\beta$ exponent can differ from its zero-temperature
counterpart. Whether it actually does, at Gaussian order, is answered directly below:
$|\delta\Lambda_U^{\rm GRPA}(\beta{=}12)|<10^{-4}$.

We read $\Lambda_U$ as a log-secant between adjacent (and end-to-end) couplings, computed
from the same moment-ratio dataset as Table~\ref{tab:moments}
($\Omega_{1,-1}=0.928,\,0.847,\,0.658$ at $|U|=3,4,6$ for $L=6$, and
$0.9475,\,0.8386,\,0.6714$ for $L=9$). The GRPA reference in the last column is the
$T=0$ Gaussian value evaluated over the matching coupling interval.
\begin{center}
\begin{tabular}{ccccc}
\hline\hline
interval & $\Lambda_U$ ($L6$) & $\Lambda_U$ ($L9$) & $\Lambda_U$ ($L12$)\textsuperscript{$\dagger$} & GRPA ($T=0$)\\
\hline
$3\to4$ & $0.321(32)$ & $0.428(34)$ & $0.542(43)$ & $0.505$\\
$4\to6$ & $0.607(46)$ & $0.548(47)$ & $0.539(64)$ & $0.694$\\
$3\to6$ & $0.488(26)$ & $0.498(26)$ & $0.541(36)$ & $0.616$\\
\hline\hline
\end{tabular}
\end{center}
The $L=12$ chains carry two seeds with $45$--$47$ post-warm-up bins each.
The adjacent secants share the $U=4$ datum and are therefore statistically
correlated. The end-to-end values agree across the three sizes,
$\Lambda_U^{3\to6}=0.488(26)$ at $L=6$, $0.498(26)$ at $L=9$ and $0.541(36)$
at $L=12$. The subinterval
secants are not separately size-converged. Moreover, the agreement of the
end-to-end values partly reflects comparable fractional finite-size shifts of the
two endpoints entering the logarithmic secant with opposite signs. We therefore
make no claim that the within-window running of $\Lambda_U$ is thermodynamically
converged. The DQMC secants sit systematically below their GRPA counterparts.

\paragraph{Finite-$\beta$ Gaussian control: the offsets are not thermal.}
To test whether temperature explains the DQMC--GRPA offsets, we evaluated the
\emph{identical} moment estimators within the Gaussian theory at the simulation
temperature: the finite-$T$ bubble of Sec.~\ref{sm:damp} (Fermi factors at
$\beta=12$, RPA-dressed static susceptibility for $M_{-1}$, thermal BdG coherence
for $M_1$), with the $\beta\to\infty$ limit of the same construction reproducing the
$T=0$ GRPA values [$\Lambda_t=0.399/0.619/0.779$, $\Lambda_U(3{\to}6)=0.616$] to
$\le10^{-5}$ relative. The effect is negligible: at $\beta=12$ the Gaussian
$\Lambda_t$, $\Lambda_U$, and $\Omega_{1,-1}$ shift from their $T=0$ values by at
most $2\times10^{-4}$ ($\Lambda_t$ at $U=3$, where the $t$-scan endpoint has
$\beta\Delta\approx8$) and by less than $10^{-5}$ everywhere else, because all Gaussian thermal corrections are
activation-suppressed ($\beta\Delta=9.2,\,16.2,\,30.0$ at $U=3,4,6$; the $M_{-1}$
thermal enhancement is $\le10^{-6}$ and \emph{shrinks} with $|U|$). The $T=0$ and
$\beta=12$ GRPA results are therefore numerically indistinguishable at the
precision relevant here, and the measured offsets---$4.3\sigma$, $3.6\sigma$,
$3.4\sigma$ in $\Lambda_t$ and $4.6\sigma$ in $\Lambda_U(3{\to}6)$---cannot be
attributed to Gaussian thermal corrections; beyond-Gaussian correlations,
residual finite-size effects and non-Gaussian finite-temperature effects can all
contribute [the discretisation contribution is bounded by the $\Delta\tau$ check
of Sec.~\ref{sm:dqmc}].

\section{GRPA equal-time-normalised response composition}
\label{sm:char}
The main text reports that the layer-odd branch stays a single non-crossing pole
throughout the crossover, yet transfers its equal-time-normalised response composition
from the layer-density response $T_z$ to the relative-pair-phase response $\theta_-$ as
$|U|$ grows. This
section makes that statement quantitative and, at the same time,
\emph{field-normalisation-invariant}. We show (i)~that the weights read off a
GRPA kernel eigenvector are meaningless, because they depend on the
arbitrary normalisation of the internal fluctuation fields; (ii)~that an
equal-time-normalised response composition built from the pole-residue matrix in the two physical response channels
$(\theta_-,T_z)$, normalised by their equal-time covariance, is invariant under those
rescalings; (iii)~that this residue is rank one, so a single number---a
diagonal-residue ratio---captures the character; and (iv)~that the equal-time
covariance is the informative metric, whereas the static-susceptibility
normalisation self-cancels towards equal channel weights when the pole dominates
the static response, making it uninformative as a character metric in that
regime.

\paragraph{Setup.} At each coupling the layer-odd GRPA kernel $K(\omega,\mathbf q{=}0)$
has a zero-eigenvalue vector $v$ at the layer-odd pole $\Omega_-$, $K(\Omega_-)\,v=0$. Here
$\sigma_-$ and $\theta_-$ are the Hermitian amplitude and phase quadratures of the
layer-odd pair field, $\delta\Delta_-=\Delta(\sigma_-+i\,\theta_-)$ (both odd under
layer exchange), and $T_z=(N_1-N_2)/2$ is the layer-imbalance density. In the basis
$(\sigma_-,\theta_-,T_z)$ the pole-residue matrix is the rank-one dyad
\begin{equation}
  R=-\frac{v\,v^{\dagger}}{|U|^{2}\,\lambda'},\qquad
  \lambda'\equiv v^{\dagger}K'(\Omega_-)\,v<0 ,
  \label{sm:char:residue}
\end{equation}
so that $R$ is Hermitian and positive semidefinite. The diagonal entry
$R_{T_zT_z}$ is the positive-frequency layer-imbalance pole residue plotted in
Fig.~3 of the main text and used in the feasibility estimate of
Sec.~\ref{sm:expt}; like the moments of Sec.~\ref{sm:moments} it is quoted
\emph{per bilayer cell} (divided by $N_{\rm cell}=L^2$), using the same
normalisation as the $T_z$ spectral function entering the exact first-moment sum
rule of Eq.~\eqref{eq:M1Tz}. The character weights defined below are
ratios and are independent of this overall normalisation. We work in the two-dimensional
physical subspace of the response channels $\{\theta_-,T_z\}$ and use two metrics: the
symmetrised \emph{equal-time} covariance, which at $T=0$ reads
\begin{equation}
  G_{0,ab}=\tfrac12\big\langle\{\delta O_a,\delta O_b\}\big\rangle_c
  =\mathrm{Re}\!\int_0^\infty\!A_{ab}(\omega)\,d\omega ,
  \label{sm:char:G0def}
\end{equation}
and the static susceptibility $\chi_{ab}(0)$. Equation~\eqref{sm:char:G0def} is the
real part of the zeroth spectral moment of the bare Gaussian/BdG bubble: it is the
equal-time covariance \emph{within the present GRPA approximation} (a
physically motivated metric, not an exact interacting covariance), and the
numerical calculation forms it directly by taking the real part. The complementary antisymmetric part
$\tfrac12\langle[\delta O_a,\delta O_b]\rangle$ (the conjugate-quadrature commutator)
is generally nonzero but is \emph{not} part of the metric.

\paragraph{The equal-time metric is diagonal in $\{\theta_-,T_z\}$.} The off-diagonal
covariance vanishes by time reversal. The paired mean-field saddle is
time-reversal invariant; under $\Theta$ the relative pairing phase is odd,
$\Theta\,\theta_-\,\Theta^{-1}=-\theta_-$, whereas the layer-imbalance density is even,
$\Theta\,T_z\,\Theta^{-1}=+T_z$ (equivalently, $\theta_-$ and $T_z$ carry opposite
signature under the combined time-reversal/layer-parity operation left unbroken by the
saddle). Hence $\delta\theta_-\,\delta T_z$ is $\Theta$-odd and its symmetrised
equilibrium average vanishes,
\begin{equation}
  G_{0,\theta_- T_z}=\tfrac12\big\langle\{\delta\theta_-,\delta T_z\}\big\rangle_c=0 .
  \label{sm:char:G0diag}
\end{equation}
The numerical calculation builds the full $2\times2$ $G_0$ and whitens with it, finding
$|G_{0,\theta_- T_z}|/\sqrt{G_{0,\theta_-\theta_-}G_{0,T_zT_z}}\lesssim2\times10^{-18}$,
confirming Eq.~\eqref{sm:char:G0diag} numerically. Thus
$G_0=\mathrm{diag}(G_{0,\theta_-\theta_-},G_{0,T_zT_z})$, the whitening $G_0^{-1/2}$ is
diagonal, and (as shown in Sec.~\ref{sm:char:rank}) the whitened residue eigenvector
weight reduces \emph{exactly} to the diagonal-residue ratio. This exact reduction is
the reason a single number suffices; it holds \emph{provided} the off-diagonal
equal-time covariance $G_{0,\theta_- T_z}$ vanishes, which Eq.~\eqref{sm:char:G0diag}
guarantees on symmetry grounds.

\subsection{Failure of the raw auxiliary-field weights}\label{sm:char:raw}
The naive ``mode composition'' obtained by squaring the kernel eigenvector components,
$P_a^{\rm raw}=|v_a|^{2}/\!\sum_b|v_b|^{2}$, is not physical. A GRPA kernel is defined
only up to the congruence generated by rescaling the internal fluctuation fields,
$\bar O_a=\lambda_a^{-1}O_a$ (i.e.\ $O=L\bar O$ with
$L=\mathrm{diag}(\lambda_a)$), under which $K\to\bar K=L^{\dagger}KL$ and the
zero-eigenvalue vector transforms as $\bar v=L^{-1}v$ ($\bar v_a=\lambda_a^{-1}v_a$), while every
physical susceptibility and the pole $\Omega_-$ are unchanged; $P_a^{\rm raw}$ can
therefore be tuned at will. We verify this by applying the rescaling to the two
internal fields $(\theta_-,T_z)$ over the nine combinations
$(\lambda_\theta,\lambda_\rho)\in\{0.1,1,10\}^{2}$ and recomputing the
zero-eigenvalue vector of the congruent kernel. The raw density weight $P_D^{\rm raw}$ sweeps
essentially the entire unit interval---from $\sim\!10^{-5}$ at
$(\lambda_\theta,\lambda_\rho)=(0.1,10)$ to $\sim\!0.9996$ at $(10,0.1)$
(Table~\ref{tab:sm:char:stress}, representative field-dependence values). A
``character'' that can be driven from $\approx0$ to $\approx1$ by an internal field
convention carries no physical information; the raw kernel-eigenvector weights are
therefore discarded from every main-text figure and statement.

\begin{table}[t]
\centering
\caption{Field-rescaling stress test (representative, illustrating the qualitative
field-dependence). The raw density weight $P_D^{\rm raw}$ depends entirely on the
internal field normalisation $(\lambda_\theta,\lambda_\rho)$, whereas the
field-normalisation-invariant weight $P_D^{\rm eq}$ of
Eq.~\eqref{sm:char:invariant} is flat to all listed digits and equals the
series of Eq.~\eqref{sm:char:transfer}.}
\label{tab:sm:char:stress}
\begin{tabular}{cc|cc|cc}
\hline\hline
 & & \multicolumn{2}{c|}{$|U|/t=2$} & \multicolumn{2}{c}{$|U|/t=8$}\\
$\lambda_\theta$ & $\lambda_\rho$ & $P_D^{\rm raw}$ & $P_D^{\rm eq}$ & $P_D^{\rm raw}$ & $P_D^{\rm eq}$\\
\hline
$0.1$ & $0.1$ & $0.181$              & $0.645$ & $0.109$              & $0.128$\\
$0.1$ & $1$   & $0.0022$             & $0.645$ & $0.0012$             & $0.128$\\
$0.1$ & $10$  & $2.2\times10^{-5}$   & $0.645$ & $1.2\times10^{-5}$   & $0.128$\\
$1$   & $0.1$ & $0.957$              & $0.645$ & $0.925$              & $0.128$\\
$1$   & $1$   & $0.181$              & $0.645$ & $0.109$              & $0.128$\\
$1$   & $10$  & $0.0022$             & $0.645$ & $0.0012$             & $0.128$\\
$10$  & $0.1$ & $0.9996$             & $0.645$ & $0.9992$             & $0.128$\\
$10$  & $1$   & $0.957$              & $0.645$ & $0.925$              & $0.128$\\
$10$  & $10$  & $0.181$              & $0.645$ & $0.109$              & $0.128$\\
\hline\hline
\end{tabular}
\end{table}

\subsection{Invariance of the equal-time-normalised response composition}\label{sm:char:inv}
The equal-time-normalised response composition is built from the pole \emph{residue} of
a physical susceptibility, normalised by that channel's own fluctuation scale. For each response
channel $a\in\{\theta_-,T_z\}$ define the metric-normalised residue and the character
weight
\begin{equation}
  r_a=\frac{R_{aa}}{G_{0,aa}},\qquad
  P_a^{\rm eq}=\frac{r_a}{r_D+r_{\theta_-}} ,
  \label{sm:char:invariant}
\end{equation}
with $G_{0,aa}=\langle O_a^{2}\rangle_c$ the equal-time covariance and $r_D\equiv
r_{T_z}$. Under the internal rescaling above, the residue and the metric transform
covariantly, $\bar R=L^{-1}RL^{-\dagger}$ and $\bar G_0=L^{-1}G_0L^{-\dagger}$, so
for diagonal $L$ both diagonal entries scale by the same factor $\lambda_a^{-2}$
and $r_a$ (and hence $P_a^{\rm eq}$) is exactly invariant. This is confirmed numerically to all reported digits: over the same nine
rescalings, $P_D^{\rm eq}$ is flat to $<10^{-9}$ while $P_D^{\rm raw}$ ranges over
the full interval (Table~\ref{tab:sm:char:stress}). The resulting invariant series
(all at $n=1.10$) is the equal-time-normalised composition quoted in the main text ---
a property of this chosen equal-time metric rather than an observer-independent weight
fraction,
\begin{equation}
  P_D^{\rm eq}:\ 0.981\,(1)\to0.645\,(2)\to0.306\,(3)\to0.201\,(4)\to0.144\,(6)
  \to0.128\,(8)\to0.120\,(10),
  \label{sm:char:transfer}
\end{equation}
(coupling $|U|/t$ in parentheses) with the complementary
$P_{\theta_-}^{\rm eq}=1-P_D^{\rm eq}$ rising from $0.019$ to $0.880$ and equal
mixing near $|U|\simeq2.3$: within this equal-time metric, the GRPA branch evolves
from density-dominated to relative-phase-dominated response character across the
crossover.

\subsection{Rank-one physical residue and the whitened eigenvector}\label{sm:char:rank}
Because $\theta_-$ is a phase quadrature and $T_z$ a density, their pole amplitudes are
in quadrature: the off-diagonal residue is imaginary, and in the whitened basis
$\tilde R=G_0^{-1/2}RG_0^{-1/2}$ the $\{\theta_-,T_z\}$ block has the form
\begin{equation}
  \tilde R=\begin{pmatrix} r_{\theta_-} & -i\,b\\[2pt] +i\,b & r_D\end{pmatrix},
  \qquad b^{2}=r_D\,r_{\theta_-} .
  \label{sm:char:quad}
\end{equation}
The identity $b^{2}=r_D r_{\theta_-}$ follows from Eq.~\eqref{sm:char:residue}: any
$2\times2$ principal block of a rank-one dyad has vanishing determinant,
$R_{T_zT_z}R_{\theta_-\theta_-}-|R_{\theta_- T_z}|^{2}=0$, and because the equal-time
metric is diagonal [Eq.~\eqref{sm:char:G0diag}] the positive whitening $G_0^{-1/2}$
preserves this rank without mixing the channels. The normalised range eigenvector of
$\tilde R$---its eigenvector with the nonzero eigenvalue $r_D+r_{\theta_-}$---is then
\begin{equation}
  u=\frac{1}{\sqrt{r_D+r_{\theta_-}}}\big(\sqrt{r_{\theta_-}},\,-i\sqrt{r_D}\big)^{\!\top},
  \qquad
  |u_D|^{2}=\frac{r_D}{r_D+r_{\theta_-}}=P_D^{\rm eq} ,
  \label{sm:char:uweight}
\end{equation}
so the whitened residue-eigenvector character is \emph{analytically identical} to the
diagonal-residue ratio of Eq.~\eqref{sm:char:invariant}: the two independent-looking
definitions of the equal-time-normalised response composition coincide. This coincidence requires the diagonal
whitening---i.e.\ $G_{0,\theta_- T_z}=0$---established in Eq.~\eqref{sm:char:G0diag};
were the off-diagonal covariance nonzero, $G_0^{-1/2}$ would mix the channels and the
eigenvector weight would differ from the diagonal ratio. Numerically, evaluating
Eq.~\eqref{sm:char:quad} across the full coupling grid gives
\begin{equation}
  \max_{|U|}\;\frac{\big|\,b^{2}-r_D r_{\theta_-}\,\big|}{r_D r_{\theta_-}}
  \;=\;2.7\times10^{-16},
  \label{sm:char:rank1}
\end{equation}
i.e.\ the residue is rank one within numerical precision. We note that
Eq.~\eqref{sm:char:rank1} is the floating-point residual of the algebraic identity
built into Eq.~\eqref{sm:char:residue} and is \emph{not} an independent test of
single-pole structure; we therefore do not claim the residue to be ``exactly rank
one.'' The independent evidence that the pole is a single simple pole is
that the analytic residue Eq.~\eqref{sm:char:residue} agrees with a finite-difference
residue of the GRPA susceptibility, on the diagonal, to $\lesssim10^{-8}$ relative at
every coupling away from the weak-coupling edge.

\subsection{Equal-time vs.\ static metric: why static normalisation is uninformative}\label{sm:char:metric}
One might instead normalise the residue by the static susceptibility $\chi_{aa}(0)$.
This choice becomes uninformative once the pole dominates the static response.
When the layer-odd pole saturates the response, the
single-pole (Kramers--Kronig) relation gives
\begin{equation}
  \chi_{aa}(0)\simeq-\frac{2R_{aa}}{\Omega_-}
  \quad\Longrightarrow\quad
  \frac{R_{aa}}{-\chi_{aa}(0)}\simeq\frac{\Omega_-}{2},
  \label{sm:char:saturation}
\end{equation}
which is \emph{independent of the channel} $a$. The static-normalised ratio therefore
cancels the channel dependence: both channels converge to $\Omega_-/2$ and the static
character collapses to equal channel weights, $P_a^{\rm static}\to\tfrac12$. This is
exactly what the data show (Table~\ref{tab:sm:char:metric}): the $T_z$ channel
obeys $R_{T_zT_z}/[-\chi_{T_zT_z}(0)]=\Omega_-/2$ to several digits at all couplings
except the weak-coupling edge, the $\theta_-$ channel joins it once the pole dominates
($|U|\gtrsim4$), and consequently $P_D^{\rm static}$ saturates to $\tfrac12$ and
carries \emph{no} density/phase information---in sharp contrast to the equal-time
weight of Eq.~\eqref{sm:char:transfer}, which runs $0.981\to0.120$ over the same
range. The equal-time covariance retains the contrast because it normalises by
the unweighted instantaneous fluctuation scale, whereas the static susceptibility
carries an inverse-frequency weighting that becomes dominated by the same
low-frequency pole in both channels. Accordingly, the static metric is
reported here only as a single-pole-saturation \emph{consistency check} (the approach
$P_a^{\rm static}\to\tfrac12$ confirms that one pole dominates the low-frequency
response); it is neither plotted as a character measure nor averaged with the
equal-time weight.

\begin{table}[t]
\centering
\caption{Static-metric self-cancellation. As the pole saturates the static response,
$R_{aa}/[-\chi_{aa}(0)]\to\Omega_-/2$ becomes channel-independent, driving the static
character $P_D^{\rm static}\to\tfrac12$ (equal channel weights). The equal-time weight
$P_D^{\rm eq}$ (last column) retains the density$\to$phase contrast that the static
metric discards. All at $n=1.10$. Columns 2--5 are evaluated on an $N_k=96$ mesh
(the self-cancellation they demonstrate is mesh-insensitive); the last column is
the $N_k=72$ series plotted in Fig.~3 of the main text, so third-digit
comparisons across that divide carry a $\sim\!1\%$ mesh difference.}
\label{tab:sm:char:metric}
\begin{tabular}{c|ccc|cc}
\hline\hline
$|U|/t$ & $\Omega_-/2$ & $\dfrac{R_{T_zT_z}}{-\chi_{T_zT_z}(0)}$ & $\dfrac{R_{\theta_-\theta_-}}{-\chi_{\theta_-\theta_-}(0)}$ & $P_D^{\rm static}$ & $P_D^{\rm eq}$\\[4pt]
\hline
$1$  & $0.579$ & $0.579$ & $0.003$ & $0.995$ & $0.981$\\
$2$  & $0.535$ & $0.525$ & $0.298$ & $0.638$ & $0.645$\\
$3$  & $0.473$ & $0.473$ & $0.437$ & $0.519$ & $0.306$\\
$4$  & $0.403$ & $0.403$ & $0.397$ & $0.504$ & $0.201$\\
$6$  & $0.303$ & $0.303$ & $0.302$ & $0.500$ & $0.144$\\
$8$  & $0.239$ & $0.239$ & $0.239$ & $0.500$ & $0.128$\\
$10$ & $0.197$ & $0.197$ & $0.197$ & $0.500$ & $0.120$\\
\hline\hline
\end{tabular}
\end{table}

\section{Crossover / competing-order checks}
\label{sm:crossover}
This section documents two supporting checks referenced in the main text: the
BCS--BEC crossover diagnostics that place the pair-size crossover near
$|U|\approx2$, and the competing-order energetics that test whether the uniform
paired saddle is pre-empted by the competing CDW or SF+CDW coexistence states
considered here. Throughout we set the nearest-neighbour
bond as the unit of length, $a_{\rm nn}=1$; the triangular Bravais vector then has
length $a_B=\sqrt3$. Unless stated otherwise the honeycomb hoppings are $t=1$,
$t_h=0.6$.

\subsection{BCS--BEC crossover diagnostics}
\label{sm:cross:diag}
The diagnostics are computed from the self-consistent $8\times8$ BdG state at each
$|U|$ on an $N_k\times N_k$ momentum mesh ($N_k=60$) at the production filling
$n=1.10$. Writing
$V_{a,n}(\mathbf k)$ for the BdG eigenvectors---Nambu components $a=1\ldots4$
($a=5\ldots8$) the particle (hole) amplitudes on the four orbitals $s$---and
$\sum_{n:E_n<0}$ for the sum over negative-energy columns, the momentum-resolved
anomalous pair amplitude on orbital $s$ is the standard $uv^\ast$ combination
\begin{equation}
F_s(\mathbf k)=\!\!\sum_{n:E_n<0}\!\!V_{s,n}(\mathbf k)\,V^\ast_{s+4,n}(\mathbf k),
\qquad F(\mathbf k)=\sum_s F_s(\mathbf k).
\label{sm:cross:F}
\end{equation}

\emph{Double occupancy.} Wick factorisation in the singlet BdG state gives
\begin{equation}
D=\langle n_\uparrow n_\downarrow\rangle
=\frac14\sum_s\!\Big(\langle n_{s\uparrow}\rangle\langle n_{s\downarrow}\rangle+|F_s|^2\Big),
\qquad F_s=\langle c_{s\uparrow}c_{s\downarrow}\rangle
\equiv\frac1{N_k^2}\sum_{\mathbf k}F_s(\mathbf k),
\label{sm:cross:D}
\end{equation}
with the on-site pair amplitude $F_s$ the momentum average of
Eq.~\eqref{sm:cross:F}; $D$ interpolates between the uncorrelated reference
$(n/2)^2$ (anomalous term negligible) and the composite-boson ceiling $n/2$. At
$n=1.10$ these bounds are $D_{\rm uncorr}=0.3025$ and $D_{\rm BEC}=0.5500$; the computed $D$ climbs steadily
across them (Table~\ref{sm:cross:tab:diag}), passing the midpoint near $|U|\approx4$.

\emph{Double occupancy from DQMC.} This mean-field estimate is corroborated by the exact
double occupancy measured directly in the DQMC density--density correlator,
$D=\tfrac12(\langle n^2\rangle-\langle n\rangle)$ with $\langle n^2\rangle$ read from the
on-site $\texttt{Den\_eq}$ (validated bin-by-bin against the measured density). At $\beta=12$,
$L=9$ we find $D(U{=}3,4,6)=0.388(1),\,0.420(1),\,0.470(1)$ (the $L=6$ values agree to
$\lesssim0.3\%$), measured at the achieved per-run fillings $n=1.109,\,1.104,\,1.103$
rather than the nominal $n=1.10$. Normalising by the uncorrelated reference
$D_{\rm uncorr}=(n/2)^2=0.308,\,0.305,\,0.305$ evaluated at those same achieved
densities gives $D/D_{\rm uncorr}=1.26,\,1.38,\,1.54$. The mean-field BdG estimate at the
same filling gives $1.21,\,1.38,\,1.57$ (Table~\ref{sm:cross:tab:diag}), tracking the measured
trend. The monotone rise confirms that the
nonperturbative $U=3$--$6$ points lie progressively deeper in the locally paired regime,
beyond the pair-size crossover near $|U|\simeq2$. We emphasise the scope of this
diagnostic: double occupancy certifies local pairing, not superfluid phase
coherence. Determining phase coherence in two dimensions is a separate
Berezinskii--Kosterlitz--Thouless problem, requiring the superfluid stiffness or
long-distance pair correlations over a range of temperatures and system sizes
with the corresponding finite-size analysis --- a different program from the
one this work uses DQMC for (continuation-free spectral moments and Hamiltonian
sensitivities at one controlled temperature). Such thermodynamic
phase-coherence analyses exist for related attractive-Hubbard
systems~\cite{Prasad2022,ZhaoParamekanti2006,Iskin2019HoneycombStiffness} and
are outside the scope of the present work. Accordingly, the DQMC data establish
local pairing and the layer-odd spectral response at $\beta=12$, while the
relative-condensate-phase interpretation is supplied by GRPA and the
strong-coupling effective theory.

\emph{Mean-field pair size.} The root-mean-square size of the pair wavefunction is
$\xi_{\rm pair}^2=\sum_{\mathbf r}|\mathbf r|^2|F_{\mathbf r}|^2/\sum_{\mathbf r}|F_{\mathbf r}|^2$,
which by Parseval we evaluate from the momentum-space amplitude as
\begin{equation}
\Big(\frac{\xi_{\rm pair}}{a_B}\Big)^{\!2}
=\frac{\sum_{\mathbf k}\big(|\partial_1F|^2+|\partial_2F|^2+\mathrm{Re}\,\partial_1F^\ast\partial_2F\big)}
       {\sum_{\mathbf k}|F(\mathbf k)|^2},
\label{sm:cross:xi}
\end{equation}
with $\partial_i\equiv\partial/\partial k_i$ symmetric finite differences with
respect to the dimensionless reduced coordinates along the two
reciprocal-primitive directions, whose conjugate real-space variables are the
integer cell indices $(n_1,n_2)$. The cross term is not optional: the honeycomb
Bravais lattice is oblique ($\mathbf a_1\!\cdot\!\mathbf a_2=a_B^2/2$, i.e.\ a
$60^\circ$ basis), so the Euclidean pair extent is
$|\mathbf r|^2=a_B^2(n_1^2+n_2^2+n_1n_2)$; dropping the $n_1n_2$ term (a naive
reduced-coordinate rms) overestimates $\xi_{\rm pair}$ by up to $2/\sqrt3\approx1.15$
in the deep-BEC limit. We quote $\xi_{\rm pair}$ in units of the Bravais vector
$a_B=\sqrt3\,a_{\rm nn}$; $\xi_{\rm pair}\sim1$ then means a pair spread across
roughly one unit cell, the natural marker of the crossover.
Independently, the weak-coupling estimate $\xi_{\rm pair}^{\rm BCS}=v_F/\pi\Delta$
uses the mesh average $v_F$ of $|\nabla_{\mathbf k}\varepsilon(\mathbf k)|$ over the
normal-state Fermi surfaces of the bonding/antibonding bands
$\varepsilon_{s\sigma}(\mathbf k)=s\,t_h+\sigma|f(\mathbf k)|$. Both estimates
collapse through unity between $|U|\approx1.5$ and $2.1$---$\xi_{\rm pair}=1$ at
$|U|^\star\simeq1.80$ and $\xi_{\rm pair}^{\rm BCS}=1$ at $|U|\simeq2.10$---and
become numerically comparable in the present calculation. They need not coincide:
$\xi_{\rm pair}^{\rm BCS}=v_F/\pi\Delta$ is a weak-coupling BCS estimate with no
controlled reason to remain accurate in the paired regime. These crossings---the
pair-size crossing $|U|^\star_\xi\simeq1.8$ and the gap crossing
$|U|^\star_{2\Delta}\simeq2.5$ where $2\Delta$ passes $\Omega_-$
(Sec.~\ref{sm:grpa})---bracket the window containing the density-to-phase
crossover of the GRPA branch (equal mixing near $|U|\simeq2.3$,
Sec.~\ref{sm:char}): as
$|U|$ grows the pairs shrink from a Fermi-surface Cooper-pair regime
($\xi_{\rm pair}\gg1$, $D\to D_{\rm uncorr}$) to tightly bound composite bosons
($\xi_{\rm pair}\ll1$, $D\to n/2$).

\begin{table}[h]
\centering
\caption{BCS--BEC crossover diagnostics at $n=1.10$, $N_k=60$.
$\Delta$ is the self-consistent
gap; $\xi_{\rm pair}$ [Eq.~\eqref{sm:cross:xi}, metric-correct] and
$\xi_{\rm pair}^{\rm BCS}$ are
in units of $a_B=\sqrt3\,a_{\rm nn}$; $D$ [Eq.~\eqref{sm:cross:D}] runs between the
uncorrelated reference $0.3025$ and the BEC ceiling $0.5500$. $\Delta$ reproduces
an independent solver implementation to machine precision.}
\label{sm:cross:tab:diag}
\begin{tabular}{ccccc}
\hline\hline
$|U|/t$ & $\Delta$ & $\xi_{\rm pair}$ & $\xi_{\rm pair}^{\rm BCS}$ & $D$\\
\hline
$1.0$  & $0.0185$ & $4.545$ & $15.33$ & $0.3028$\\
$1.5$  & $0.0961$ & $1.563$ & $2.971$ & $0.3066$\\
$2.0$  & $0.2570$ & $0.631$ & $1.116$ & $0.3190$\\
$2.5$  & $0.4904$ & $0.423$ & $0.586$ & $0.3410$\\
$3.0$  & $0.7648$ & $0.325$ & $0.375$ & $0.3675$\\
$4.0$  & $1.3492$ & $0.210$ & $0.211$ & $0.4163$\\
$6.0$  & $2.4977$ & $0.108$ & $0.104$ & $0.4758$\\
$8.0$  & $3.5945$ & $0.065$ & $0.082$ & $0.5044$\\
$10.0$ & $4.6581$ & $0.044$ & $0.068$ & $0.5195$\\
\hline\hline
\end{tabular}
\end{table}

\subsection{Competing-order energetics}
\label{sm:cross:energetics}
Near half-filling the attractive Hubbard model has competing $s$-wave superfluid
(SF) and charge-density-wave (CDW) instabilities, and one must check that the
uniform paired saddle used throughout is not pre-empted by charge order. We use
two complementary criteria: a local RPA-stability eigenvalue and a direct
self-consistent grand-potential comparison.

\emph{CDW stability eigenvalue.} The proximity to charge order is the smallest
eigenvalue
\begin{equation}
\lambda_{\rm CDW}=\lambda_{\min}\!\big[\,\openone-|U|\,\chi^0_{\rm CDW}(0)\,\big]
\label{sm:cross:lambda}
\end{equation}
of the dimensionless static charge-channel GRPA stability matrix, evaluated in the
staggered-density channel with the sublattice/layer pattern
$\mathrm{diag}(+,-,-,+)$---the $\eta$-partner of the uniform pair---where
$\chi^0_{\rm CDW}(0)$ is the corresponding bare static BdG bubble. A positive
$\lambda_{\rm CDW}$ signals a locally stable paired saddle; $\lambda_{\rm CDW}\to0$
is the Thouless point where the CDW channel goes soft and the saddle is unstable.
Across the whole crossover sweep---here reported at $n=1.05$, which sits closer to
half-filling than the production $n=1.10$ and is therefore the more stringent test
for charge order---$\lambda_{\rm CDW}$ stays strictly
positive, falling smoothly from $0.550$ at $|U|=1$ to $0.054$ at $|U|=2$,
$4.6\times10^{-4}$ at $|U|=8$, and $3.2\times10^{-4}$ at $|U|=10$, while the
self-consistent staggered charge density remains zero
($|\rho_{\rm CDW}|\lesssim10^{-15}$) and the pair-phase Goldstone stays gapless.
The paired saddle is therefore locally stable throughout; the residual smallness
of $\lambda_{\rm CDW}$ at strong coupling reflects the near-degeneracy with the
$\eta$-partner, not an instability.

\emph{Direct grand-potential comparison.} Local stability does not by itself fix
the global ground state, so we solve the $8\times8$ BdG problem self-consistently
for three ans\"atze---uniform $s$-wave pairing $\Delta$ (SF); a staggered onsite
CDW order parameter $W$, the $\eta$-pairing partner of $\Delta$ implemented as a
staggered potential $W\,\mathrm{diag}(+,-,-,+)$ on the bipartition
$\{A_1,B_2\}/\{B_1,A_2\}$ (this is a staggered \emph{onsite} density order, not an
$s$-wave bond CDW)---and coexistence, comparing the $T=0$ grand potential per
orbital
\begin{equation}
\Omega=\frac1{2N_{\rm sites}}\!\!\sum_{\mathbf k,\nu:\,E_{\mathbf k\nu}<0}\!\!
E_{\mathbf k\nu}+\frac{\Delta^2+W^2}{|U|}+\text{const}
\label{sm:cross:omega}
\end{equation}
at fixed chemical potential. The factor $\tfrac12$ removes the Nambu
particle-hole double counting; ``const'' collects the ansatz-independent additive
terms from the particle-hole shift of the interaction ($-|U|/4$ per orbital) and
the $-\mu\sum_{\ell i}(n_{\ell i}-1)$ reference ($+\mu$ per orbital), which are
identical for the three ans\"atze at fixed $\mu$ and cancel exactly in every
difference $\Omega_{\rm CDW}-\Omega_{\rm SF}$ below.

\emph{Validation against $\eta$-SU(2).} At half-filling ($n=1$, i.e.\ $\mu=0$ in
the particle-hole-shifted convention) the $\eta$-SU(2) symmetry forces
$\Omega_{\rm SF}=\Omega_{\rm CDW}$ exactly. The solver reproduces this degeneracy
to machine precision---$|\Omega_{\rm SF}-\Omega_{\rm CDW}|\lesssim8\times10^{-12}$,
with $\Delta=W=3.613$ at $|U|=8$ and $\Delta=W=4.682$ at $|U|=10$---confirming the
free-energy bookkeeping and the exact cancellation of the additive constants.

\emph{Result.} The CDW is incompressible: its self-consistent solution stays
pinned at $n=1$ as $\mu$ increases (a charge gap), whereas the SF dopes
continuously. At the chemical potentials that give $n=1.05$ and $1.10$ in the SF
phase, the SF grand potential is the lowest of the three
(Table~\ref{sm:cross:tab:energetics}), the margin exceeding the solver/mesh
tolerance ($\sim10^{-9}$) by six to seven orders of magnitude, unchanged under
mesh refinement ($N_k=36\to48$ to all quoted digits). The coexistence ansatz
relaxes essentially onto the pure SF ($W\to0.033$ at $|U|=8$, $n=1.10$) and lands
marginally \emph{above} it ($\Omega_{\rm coex}-\Omega_{\rm SF}\approx+2\times10^{-6}$),
so no energetically favoured SF+CDW coexistence state is found within this
mean-field ansatz. The direct comparison is performed at the strongest couplings
examined, $|U|=8$ and $10$, where the SF/CDW $\eta$-partner near-degeneracy is
most severe and the test is therefore most stringent; at both tested dopings
$n=1.05$ and $1.10$ the self-consistent energetics selects the uniform paired
state. Together with the positive CDW stability eigenvalue throughout the sweep,
these calculations find no evidence that the nearby CDW instability pre-empts
the paired saddle in the regime considered. The margin increases between the two tested dopings, consistent with the
incompressible CDW's inability to absorb the added charge, and corroborates the
local GRPA stability of the paired saddle.

\begin{table}[h]
\centering
\caption{Self-consistent $T=0$ grand potential per orbital
[Eq.~\eqref{sm:cross:omega}] for the uniform SF and staggered CDW ans\"atze at
matched chemical potential ($N_k=36$; identical at $N_k=48$ to the quoted digits).
$\Omega_{\rm CDW}-\Omega_{\rm SF}>0$
at every tested point, so the SF is selected; the coexistence ansatz relaxes above
the pure SF (no energetically favoured coexistence). SF and CDW are exactly
degenerate only at $n=1$.}
\label{sm:cross:tab:energetics}
\begin{tabular}{ccccc}
\hline\hline
$|U|/t$ & $n$ & $\Omega_{\rm SF}$ & $\Omega_{\rm CDW}$ & $\Omega_{\rm CDW}-\Omega_{\rm SF}$\\
\hline
$8$  & $1.05$ & $-0.39027$ & $-0.38523$ & $+5.0\times10^{-3}$\\
$8$  & $1.10$ & $-0.40538$ & $-0.38523$ & $+2.0\times10^{-2}$\\
$10$ & $1.05$ & $-0.32359$ & $-0.31732$ & $+6.3\times10^{-3}$\\
$10$ & $1.10$ & $-0.34241$ & $-0.31732$ & $+2.5\times10^{-2}$\\
\hline\hline
\end{tabular}
\end{table}

\section{LSWT-normalised lattice comparison}
\label{sm:coord}
The strong-coupling mapping of Sec.~\ref{sm:sw} sends the layer-odd mode onto the
optical branch of an $\eta$-pseudospin antiferromagnet, whose frequency is fixed by the
intralayer and rung superexchange couplings $J=4t^2/|U|$ and $J_\perp=4t_h^2/|U|$,
\begin{equation}
  \Omega_{\rm SC}=\sqrt{z\,J\,J_\perp}\,F(n)=\frac{4\sqrt{z}\,t\,t_h}{|U|}\,F(n),
  \qquad
  F^2(n)=1-d^2\!\left(1-\frac{t_h^2}{z\,t^2}\right),\quad d\equiv n-1 .
  \label{sm:coord:law}
\end{equation}
Within this LSWT result, lattice dependence enters through the coordination number
$z$ (three for the honeycomb, four for the square lattice), both in the leading
$\sqrt z$ factor and weakly through $F(n)$; everything else is set by the rung and
by the filling. The law is therefore not specific to the honeycomb lattice within
this class of nearest-neighbour bipartite bilayers, and it invites a clean test:
strip off the known $z$- and $n$-dependence and check that what remains is unity.
The quantities compared below come from two approximate descriptions: GRPA applied
to the original fermionic Hubbard model, and LSWT applied to its second-order
strong-coupling $\eta$-pseudospin Hamiltonian. Their agreement is therefore an
asymptotic consistency check between the two descriptions, not a nonperturbative
validation of the exact quantum strong-coupling scale.

\begin{figure}[t]
  \centering
  \includegraphics[width=0.72\linewidth]{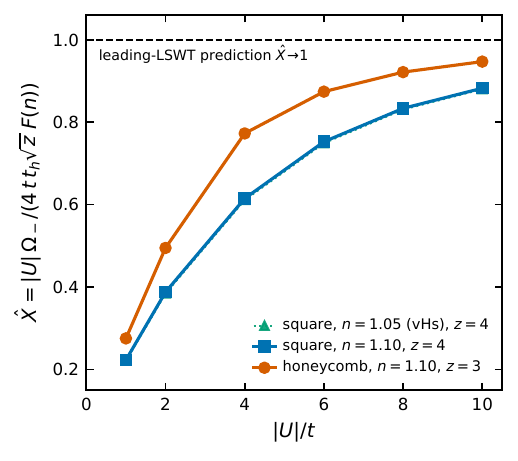}
  \caption{\textbf{LSWT-normalised lattice comparison.}
  $\hat X\equiv|U|\,\Omega_-/(4t\,t_h\sqrt{z}\,F(n))$ [Eq.~\eqref{sm:coord:Xhat}]
  versus $|U|/t$ for the honeycomb ($z=3$; $\hat X=0.947$ at $|U|=10$) and square
  ($z=4$; $\hat X=0.882$) lattices. Both curves move towards the leading-LSWT
  reference $\hat X=1$, with slower convergence on the square lattice. The
  $n=1.05$ square-lattice curve, including the van Hove filling, nearly overlaps
  the $n=1.10$ result, showing weak filling dependence over this comparison. This
  behaviour differs from the bare-hybridisation scaling,
  $\hat X=|U|/[2t\sqrt{z}\,F(n)]$, and from the isolated-rung scaling. GRPA
  treats the original fermionic Hubbard model, whereas LSWT treats its
  second-order strong-coupling $\eta$-pseudospin Hamiltonian; their agreement is
  therefore an asymptotic consistency check between the two approximate
  descriptions, not an independent test of the exact quantum strong-coupling
  scale.}
  \label{fig:coord}
\end{figure}

\paragraph{A coordination-independent discriminant.} Define the raw combination
\begin{equation}
  X\equiv\frac{|U|\,\Omega_-}{4\,t\,t_h}\;\xrightarrow{\ |U|\to\infty\ }\;\sqrt{z}\,F(n),
  \label{sm:coord:X}
\end{equation}
which isolates the collective scale: at weak coupling $\Omega_-\to2t_h$ and
$X\to|U|/(2t)$ carries no lattice information, whereas in the paired regime the leading
linear-spin-wave coordination law predicts $X\to\sqrt{z}\,F(n)$. Dividing out that asymptote gives the coordination-independent
quantity
\begin{equation}
  \hat X\equiv\frac{X}{\sqrt{z}\,F(n)}
  =\frac{|U|\,\Omega_-}{4\,t\,t_h\,\sqrt{z}\,F(n)}\;\longrightarrow\;1 ,
  \label{sm:coord:Xhat}
\end{equation}
so that, if the leading LSWT coordination law were exact, honeycomb and square data (which differ by the factor $\sqrt{4/3}$ in the raw scale) would approach a common value
of one. It is the collapse of $\hat X$ towards
unity, not any single frequency, that tests the coordination law (Fig.~\ref{fig:coord}).

\paragraph{Honeycomb versus square.} We evaluate $\Omega_-$ from the same GRPA analysis
code on both lattices at matched filling $n=1.10$, $t_h/t=0.6$, changing only the lattice
module (its $z$, orbital count, and interlayer bonds). Table~\ref{sm:coord:tab}
summarises the results. The bare scale $X$ is indeed larger on the square lattice
($\sqrt{4}$ versus $\sqrt{3}$), and dividing by $\sqrt{z}\,F(n)$ brings both towards the
same curve.

\begin{table}[t]
\centering
\caption{Coordination-law check at $n=1.10$, $t=1$, $t_h=0.6$.
Rows at $|U|=2,8$ are from the $N_k=24$ mesh; the $|U|=10$ entries marked $^{\ast}$ are
the finer-mesh ($N_k=72$) endpoints shown in Fig.~\ref{fig:coord}. $X$ and $\hat X$ are
defined in Eqs.~\eqref{sm:coord:X}--\eqref{sm:coord:Xhat}.}
\label{sm:coord:tab}
\begin{tabular}{cccccc}
\hline\hline
Lattice & $z$ & $|U|$ & $\Omega_-$ & $X$ & $\hat X$ \\
\hline
Honeycomb & 3 & 2            & 1.024 & 0.854 & 0.495 \\
Honeycomb & 3 & 8            & 0.477 & 1.589 & 0.921 \\
Honeycomb & 3 & $10^{\ast}$  & 0.392 & 1.633 & 0.947 \\
Square    & 4 & 2            & 0.928 & 0.773 & 0.388 \\
Square    & 4 & 8            & 0.498 & 1.659 & 0.833 \\
Square    & 4 & $10^{\ast}$  & 0.422 & 1.757 & 0.882 \\
\hline\hline
\end{tabular}
\end{table}

Two internal consistency checks accompany each row. The common-phase Goldstone stays
exact, $K_{\theta_+\theta_+}(0)\sim10^{-11}$--$10^{-14}$ with
$\Pi_{\theta_+\theta_+}(0)=-\mathrm{CN}/|U|$ on
both lattices, and the collective pole nearly exhausts the $T_z$ first-moment sum
rule: its fraction $f_{T_z}\equiv\Omega_-R_{T_zT_z}/M_1[T_z]$ is $0.97$--$1.00$, with
the numerical and analytic first moments agreeing to $0.2$--$0.3\%$. The closed form $F(n)=\sqrt{1-d^2(1-t_h^2/z t^2)}$ used in
Eq.~\eqref{sm:coord:Xhat} is validated against an independent Colpa
diagonalisation of the spin-wave problem, which reproduces all four magnon frequencies
and the $T_z$ residue to machine precision and gives $f_{T_z}=1$ exactly for $n\neq1$.

\paragraph{Convergence.} As $|U|$ grows the mode moves off the hybridisation scale
$2t_h$ and towards the superexchange scale, and $\hat X$ rises monotonically on both
lattices: $0.50\to0.92$ (honeycomb) and $0.39\to0.83$ (square) between $|U|=2$ and $8$;
at weak coupling $\hat X$ instead follows the bare-hybridisation scaling and lies far
below the LSWT value.
At the strongest coupling computed, $|U|=10$, the honeycomb reaches
$\hat X=0.947$, i.e.\ within $5.3\%$ of the leading-LSWT value of one, while the square
lattice lags at $\hat X=0.882$, about $12\%$ below that value and converging more
slowly. The separation between the two lattices decreases over the computed range --- the
gap $\hat X^{\rm hc}-\hat X^{\rm sq}$ falls from $0.107$ at $|U|=2$ to $0.088$ at $|U|=8$
--- consistent with a common approach towards the leading-LSWT value $\hat X=1$. We do not
claim a completed collapse, nor that the exact quantum asymptote equals one: the value
$\hat X=1$ is the prediction of the leading (linear-spin-wave) coordination law, and the
residual $\sim\!5\%$ (honeycomb) and $\sim\!12\%$ (square) deficits have two physically
distinct origins, which we do not disentangle here: finite-$|U|$ Hubbard corrections to
the effective model [relative $O(t^2/U^2,\,t_h^2/U^2)$, which \emph{vanish} as
$|U|\to\infty$], and
quantum $1/S$ corrections to the spin-wave dispersion [which do \emph{not} vanish as
$|U|\to\infty$, the LSWT form being only the leading semiclassical approximation to the
$S=\tfrac12$ magnet]. What the comparison establishes is the qualitative signature the law
predicts: dividing by $\sqrt{z}\,F(n)$ substantially reduces the lattice-dependent
separation and moves both curves towards the leading-LSWT reference value of one,
\emph{consistent with the leading-LSWT geometric-mean scaling
$\sqrt{zJJ_\perp}$ and inconsistent with the bare-$2t_h$ and isolated-rung ($J_\perp$)
scalings}. The comparison is between GRPA applied to the fermionic Hubbard model and
LSWT applied to its strong-coupling $\eta$-pseudospin Hamiltonian; it is an
asymptotic consistency check between the two descriptions, not a nonperturbative
validation of the exact quantum strong-coupling scale.

\section{Damping and parity}
\label{sm:damp}
The layer-odd mode sits low in the spectrum, and the natural worry is that it simply
melts into a fermionic or collective continuum. It does not, at least at Gaussian
order. Two facts do the work: a kinematic threshold that is set by the
\emph{layer-imbalance-active} pair continuum rather than by the absolute gap edge,
and a layer-parity selection rule that the entire coupled odd block obeys. We stress
at the outset that all statements below---that the mode is undamped below the
inter-parity threshold, dark to linear $T_z$ decay, or lying below the absolute
pair continuum---are properties of the $T=0$ Gaussian kernel. Finite-temperature (Landau) damping, in
which thermally populated quasiparticles broaden the mode, applies both to the
$\beta=12$ correlator measured by DQMC and to the experimental mode; we compute it
at Gaussian order at the end of this section.

\paragraph{Fermionic (pair-breaking) channel: two thresholds, not one.}
The linear $T_z$ response couples to the two-quasiparticle continuum only through
layer-off-diagonal (bonding--antibonding) processes. Its lowest edge is the
\emph{$T_z$-active threshold}
\begin{equation}
  \omega^{T_z}_{\rm 2QP}=\min_{\mathbf k}\big[E_a(\mathbf k)+E_b(\mathbf k)\big]
  =2\sqrt{t_h^2+\Delta^2},
  \label{sm:damp:tzedge}
\end{equation}
with $E_{a,b}$ the bonding/antibonding Bogoliubov energies of
Sec.~\ref{sm:kernelzero}; the equality holds at the working filling, where the
symmetric point $\varepsilon_{\mathbf k}=\mu$ is attained within the band
(otherwise the right-hand side is a lower bound on the edge). This is \emph{not}
the absolute two-quasiparticle edge.
The lowest place to break a pair anywhere in the spectrum is the ordinary gap edge
$2\Delta$, which is reached by intra-parity (layer-even) processes; the
layer-imbalance operator is blind to it. For $|U|<|U|^\star$, the branch overlaps
the absolute pair continuum but remains below the $T_z$-active inter-parity
continuum. Consequently, the $T_z$ bubble has no pair-breaking imaginary part at
the pole. Whether the full coupled odd-sector mode is undamped requires the
stronger result established next. Beyond Gaussian order a fermionic pair-breaking
vertex can in principle open a finite width while the mode overlaps the absolute
two-quasiparticle continuum ($|U|<|U|^\star$); evaluating its magnitude requires a
controlled interacting self-energy and is not attempted here, so we make no
quantitative linewidth claim.

\paragraph{The whole coupled odd block is real below the inter-parity threshold.}
The physical excitation is not the bare $T_z$ response but the zero-eigenvalue
vector of the \emph{full} coupled layer-odd Gaussian kernel spanned by
$\{\sigma_-,\theta_-,T_z\}$, so darkness of the $T_z$ bubble alone does not settle
the question. What settles it is a parity selection rule that all three vertices
share. Under the layer-exchange parity $P$ (interchange $1\leftrightarrow2$ at equal
fillings) every one of these operators is odd,
$P\,\theta_-=-\theta_-$, $P\,\sigma_-=-\sigma_-$, $P\,T_z=-T_z$, while the
Bogoliubov quasiparticles organise into a $P$-even (bonding, $a$) and a $P$-odd
(antibonding, $b$) branch. A $P$-odd vertex has nonzero matrix elements only between
states of opposite parity; it can create one bonding and one antibonding
quasiparticle but never an intra-parity $(a,a)$ or $(b,b)$ pair. Because \emph{every}
entry of the coupled kernel---diagonal and off-diagonal (cross) alike---is built from
such $P$-odd vertices on both legs, the two-quasiparticle spectrum of the entire
$3\times3$ odd bubble matrix begins at the same inter-parity edge,
\begin{equation}
  \omega^{\rm odd}_{\rm 2QP}=\min_{\mathbf k}\big[E_a(\mathbf k)+E_b(\mathbf k)\big]
  =2\sqrt{t_h^2+\Delta^2}.
  \label{sm:damp:oddedge}
\end{equation}
As a function of $\varepsilon_{\mathbf k}$ the summand $E_a+E_b$ is a sum of two
convex functions and has a unique minimum at the symmetric point $\xi_a=-\xi_b$; for
the working near-half-filled density ($n=1.10$), where $\mu$ lies within the band,
that point ($\varepsilon_{\mathbf k}=\mu$) is attained, giving
$E_a=E_b=\sqrt{t_h^2+\Delta^2}$ and an edge independent of $\mu$. (Were $\mu$ to
leave the band, the right-hand side of Eq.~\eqref{sm:damp:oddedge} would only be a
lower bound on the edge, strengthening the conclusion.) The consequence is stronger
than the single-bubble statement of the previous paragraph: not merely
$\mathrm{Im}\,\Pi_{T_zT_z}=0$, but $\mathrm{Im}\,K_{\rm odd}(\omega)=0$ for
\emph{every} entry of the coupled kernel at $\omega<2\sqrt{t_h^2+\Delta^2}$---the
whole Gaussian odd kernel is real below the inter-parity threshold. Since the
collective pole satisfies $\Omega_-<2\sqrt{t_h^2+\Delta^2}$ throughout
(Fig.~\ref{fig:sm:continua}), $\mathrm{Im}\,K_{\rm odd}(\Omega_-)=0$: the mode is undamped at
Gaussian order as a property of the full coupled block, not just of its $T_z$
projection. The lower absolute edge $2\Delta$ arises from intra-parity processes
and therefore does not enter the linear layer-odd Gaussian bubble matrix.

\paragraph{Two-Goldstone (Beliaev) channel and its selection rule.}
The leading collective decay one would write down is $\theta_-\to\theta_+\theta_+$,
one layer-odd mode into two Goldstones. This channel is forbidden by the same
parity. With $P\,\theta_-=-\theta_-$ and $P\,\theta_+=+\theta_+$, the cubic vertex
carries parity $(-)(+)(+)=-$ and vanishes at the symmetric point,
\begin{equation}
  V_{\theta_-\theta_+\theta_+}\big|_{\rm equal\ fillings}=0 .
  \label{sm:damp:selrule}
\end{equation}
This closes \emph{only} the two-Goldstone (even$+$even) channel. Parity
\emph{permits} decay into one odd plus one even daughter (odd$\to$odd$+$even), whose
kinematic threshold at zero external momentum is
\begin{equation}
  \omega^{(-,+)}_{\rm th}(\mathbf q{=}0)
  =\min_{\mathbf k}\big[\Omega_-(\mathbf k)+\Omega_+(-\mathbf k)\big].
  \label{sm:damp:collthr}
\end{equation}
We do \emph{not} evaluate these parity-allowed odd$\to$odd$+$even collective
channels. Equation~\eqref{sm:damp:selrule} therefore establishes only that the
two-Goldstone route is symmetry-forbidden at exact parity; it does \emph{not} by
itself demonstrate a vanishing intrinsic collective width, and we make no such claim.

\paragraph{Away from exact parity.}
A layer mismatch $\delta_{\rm layer}$ (unequal fillings or fields between the two
layers) breaks $P$, so the forbidden cubic vertex becomes analytic in the
parity-breaking perturbation, $V_{\rm off}=O(\delta_{\rm layer})$ generically. The
corresponding golden-rule rate therefore begins generically as
\begin{equation}
  \Gamma_{\rm Beliaev}=O(\delta_{\rm layer}^{2}).
  \label{sm:damp:gamma}
\end{equation}
Symmetry fixes the absence of a constant term; the prefactor, and whether
additional cancellations raise the leading power, are not determined here---a
fully momentum-resolved cubic (sunset) self-energy would be required---so
Eq.~\eqref{sm:damp:gamma} is quoted purely as a scaling statement, and any Beliaev
rate we report elsewhere is a normalised proxy rather than an absolute linewidth. The bosonic counterpart, Beliaev damping of
a gapped relative mode in a two-component condensate, has been analysed in
Ref.~\cite{Wu2024}; Eq.~\eqref{sm:damp:selrule} is its fermionic lattice analogue.

\paragraph{Robustness to further-neighbour hopping.}
A layer-symmetric next-nearest-neighbour hopping does not upset any of this: at
$t'=0.05$ it shifts $\Omega_-$ by well under a percent and leaves the parity
selection rule Eq.~\eqref{sm:damp:selrule} intact, since $t'$ preserves the
layer-exchange symmetry that the rule rests on. More generally, because the
weak-coupling identity holds for arbitrary layer-identical single-particle
dispersion $h_0(\mathbf k)$ (Sec.~\ref{sm:kernelzero}), layer-symmetric
further-neighbour or anisotropic intralayer hoppings leave the $2t_h$ Gaussian
anchor unchanged, while in the paired regime they generate additional
superexchange couplings of order $t_r^2/|U|$. The resulting strong-coupling
coefficient is model-dependent, but the contrast underlying the Hamiltonian
fingerprint is not.

\paragraph{Finite-temperature (Landau) linewidth at Gaussian order.}
The $T=0$ statements above leave one quantitative question: how wide is the mode at
the temperature of the calculation? We answer it at the same Gaussian order by
evaluating the layer-odd bubble with full Fermi occupations at $\beta=12$, keeping
both the pair-creation terms $(1-f-f')$ and the thermal scattering terms $(f-f')$
that vanish at $T=0$ [as a consistency check on the occupation factors, the
$T\to0$ limit of this bubble reproduces the zero-temperature kernel identically, and
$\mathrm{Im}\,\Pi(\omega)=-\mathrm{Im}\,\Pi(-\omega)$
holds to $10^{-18}$; the $T=0$ gap $\Delta(0)$ is used as input, justified by
$\Delta/T\ge3$ at all quoted couplings]. Writing the complex pole as $\omega_p=\Omega_--i\Gamma/2$, linearising the
relevant kernel eigenvalue gives
$\Gamma=2|\mathrm{Im}\,\lambda(\Omega_-)|/|\partial_\omega
\mathrm{Re}\,\lambda(\Omega_-)|$. The result is exponentially activated,
$\Gamma/\Omega_-\simeq A\,e^{-E_{\min}/T}$, where $E_{\min}$ is the minimal
quasiparticle energy \emph{on the thermal-scattering resonance surface}
$E_a-E_b=\Omega_-$, which sits up to $\sim10\%$ above the gap ($E_{\min}/\Delta=1.00$--$1.10$ across
$|U|=2$--$6$);
the empirical log-slope against $E_{\min}/T$ is $-0.90$ (against $\Delta/T$,
$-0.94$) because the prefactor drifts across couplings. At $\beta=12$ and $n=1.10$ we find
$\Gamma/\Omega_-=3\times10^{-2},\ 9\times10^{-5},\ 1\times10^{-7},\ 3\times10^{-13}$
at $|U|=2,3,4,6$
(for $|U|\ge3$ these are mesh-stable at $N_k=48\to72$ to $\lesssim2\%$ and
converged in the numerical spectral broadening $\eta$ used to evaluate the
finite-$T$ bubble; the $|U|=2$ value is \emph{not} $\eta$-converged---the
golden-rule and direct spectral-peak estimators there span
$\Gamma/\Omega_-\approx0.007$--$0.03$, so we quote it only as an order of
magnitude). As quality factors we
quote $Q_{\rm Gauss}\sim30$--$150$ at $|U|=2$ (estimator- and
$\Delta(T)$-limited), $\simeq10^{4}$ at $|U|=3$, and simply
$\gtrsim10^{6}$ for $|U|\ge4$: these are optimistic Gaussian estimates of the
quality factor (the deep-activation values are Boltzmann-factor evaluations);
additional beyond-Gaussian and experimental damping channels can only reduce the
observable $Q$. Two consequences follow. First, Gaussian quasiparticle Landau
damping is negligible over the DQMC window $|U|=3$--$6$ at $\beta=12$.
Second, because the suppression is controlled by $\Delta/T$, the Landau channel is
negligible once $T\lesssim\Delta/3$ (giving
$Q_{\rm Gauss}$ of order $10$--$10^{2}$ at the boundary; the constant is anchored
by the $|U|=2$ point, whose estimator spread is quoted above, and the
extrapolation treats the prefactor $A$ as $T$-independent);
the criterion uses $\Delta(0)$, and self-consistent
$\Delta(T)$ suppression near the boundary reduces $Q$ further, so it is optimistic
by construction. This bounds only the quasiparticle channel. At strong coupling,
observing a coherent relative-phase collective oscillation additionally requires
pair coherence: in two dimensions this is controlled by the
Berezinskii--Kosterlitz--Thouless coherence scale, parametrically
$T_{\rm BKT}\sim t^2/|U|$ in the strong-coupling
regime~\cite{Paiva2010,Fontenele2024}, which lies \emph{below} $\Delta/3$ there.
The practical condition is therefore $T\lesssim\min(\Delta/3,\,T_{\rm BKT})$,
while the weak-coupling hybridisation resonance requires no superfluid order.
Beyond-Gaussian collective channels (e.g.\ Beliaev-type processes,
Eq.~\eqref{sm:damp:gamma}) are controlled by the symmetry arguments above but their
absolute prefactors are not computed here.

\section{Experimental feasibility}
\label{sm:expt}
The layer-odd branch is directly visible in the layer-imbalance response, and the
two ingredients needed to excite and detect it---an interlayer bias and
layer-resolved readout---are available in cold-atom bilayer platforms. We give
only order-of-magnitude estimates, keeping separate the quantities calculated
within GRPA, $\Omega_-(|U|)$ and $R_{T_zT_z}(|U|)$, from experiment-specific
parameters. The Gaussian quasiparticle linewidth is evaluated in
Sec.~\ref{sm:damp} (exponentially activated; operating window
$T\lesssim\min(\Delta/3,\,T_{\rm BKT})$); observing a coherent relative-phase
oscillation additionally requires phase coherence, whose two-dimensional BKT
scale is not calculated here. Trap profile and technical decoherence remain
experiment-specific. The loading metric in a real
experiment is entropy per particle rather than temperature, and
interaction-driven entropy redistribution can play an important role in
fermionic optical lattices~\cite{Werner2005}; the entropy capacity of the
present gapped bilayer near half filling is untested, and we do not model
entropy redistribution here.

\paragraph{Frequency scale.} Fixing the intralayer hopping to a standard optical-lattice
value $t/h=500$~Hz gives $t/k_B\simeq24$~nK and $\hbar/t\simeq0.318$~ms, so a mode at
$\Omega_-/t\approx0.6$--$0.8$ in the paired regime sits at
$\Omega_-/h\approx300$--$400$~Hz. This hundreds-of-hertz scale is far below interband
frequencies but can overlap low-lying radial trap or sloshing modes in a harmonic
trap; a box (flat-bottom) potential, or choosing $t$ to detune the mode from the
measured trap frequencies, avoids parametric cross-driving.

\paragraph{Drive and readout.} A pulsed interlayer bias
$H'(t)=V_0\cos(\Omega_- t)\,T_z$ drives the layer-imbalance mode on resonance; switching
the drive off and recording the free oscillation of $\langle T_z\rangle(t)=(N_1-N_2)/2$
by layer-resolved imaging measures $\Omega_-$ directly. The linear-response
amplitude scales with the calculated residue $R_{T_zT_z}(|U|)$ (quoted per bilayer
cell, Sec.~\ref{sm:char}), which grows from $\approx0.15$ at $|U|=2$ to
$\approx0.30$ at $|U|=6$; the measured amplitude additionally depends on drive
strength, duration, damping, and detection noise, while atom-resolved imaging
makes the imbalance readout itself standard. The Gaussian quality factor
$Q_{\rm Gauss}\equiv\Omega_-/\Gamma_{\rm Gauss}$ of Sec.~\ref{sm:damp}
($\sim30$--$150$ at $|U|=2$, $\ge10^4$ for $|U|\ge3$ at $T=t/12$) is a Gaussian
quasiparticle-channel estimate, not a total width; experimental visibility
depends additionally on detection and technical decoherence, so within the
paired regime, and below the coherence temperature, the practical limit is
technical decoherence and uncomputed collective channels, not quasiparticle
damping.
Beyond-Gaussian collective decay channels are symmetry-suppressed
(Sec.~\ref{sm:damp}) but their absolute prefactors are not computed here.

\paragraph{Hamiltonian-sensitivity fingerprints.} The two Hamiltonian
sensitivities of the main text translate into in-situ protocols of unequal
difficulty, and we order them accordingly. The near-term protocol is $\Lambda_U$:
varying $|U|$ at fixed hoppings (a Feshbach-field ramp at constant lattice depth,
the standard BCS--BEC knob) and tracking
$\Lambda_U=-\partial\ln\Omega/\partial\ln|U|$ tests the evolution towards the $1/|U|$
second-order pair scaling; the $U=3\to6$ softening produces an approximately
$29\%$ change in the moment ratio $\Omega_{1,-1}$. We stress that this is a change
in the \emph{moment ratio}, i.e.\ in the low-frequency-weighted spectral
distribution, and not a demonstrated displacement of a spectral peak: the forward
$\chi^2$ test of Sec.~\ref{sm:disc:peak} finds a shared peak position
statistically acceptable ($\Delta\chi^2\simeq1.2$ for one degree of freedom).
The quantity to be measured is therefore $\Omega_{1,-1}$ itself, through the
static two-observable route of Sec.~\ref{sm:moments}
($\Omega_{1,-1}^2=2t_h\langle T_x\rangle/\chi_{zz}(0)$), which needs no peak
identification, no line-shape model and no analytic continuation. The harder follow-up is $\Lambda_t$: in an
optical lattice the intralayer hopping is set by lattice depth, but depth also
changes the on-site Wannier overlap and hence $U$, so holding $|U|$ fixed while $t$
varies is a coordinated two-knob operation---ramp the depth to move $t$, and
compensate the induced change of the on-site interaction with a Feshbach retune of the
scattering length at every point. At strong coupling the Wannier-integral estimate
of $U$ carries known higher-band corrections, so the per-point calibration should
be spectroscopic (rf or lattice-modulation measurement of $U$) rather than
computed; in a three-beam honeycomb lattice the depth ramp must also keep the three
intralayer tunnellings balanced and leave the interlayer coupling $t_h$ unchanged.
This requires a usable (preferably broad) Feshbach resonance
and per-point calibration, and is standard but not turnkey. The signal itself is
modest: at $\Lambda_t\simeq0.45$, the $t=0.9\to1.1$ scan changes
$\Omega_{1,-1}$ end-to-end by approximately $9\%$ ($\sim30$~Hz at
$t/h=500$~Hz). As above this is a shift of the moment ratio, not of a resolved
peak, so the requirement is percent-level resolution on the two static
observables entering $\Omega_{1,-1}$ rather than on a line centre; the actual requirement is set by technical stability and by
uncomputed collective broadening. Density inhomogeneity is the other practical broadener:
the mode frequency is only weakly filling-sensitive---the GRPA density
sensitivity is $|\partial\ln\Omega_-/\partial\ln n|\approx0.13$ at strong coupling
(cf.\ Sec.~\ref{sm:dqmc})---so a $5\%$ rms density spread blurs $\Omega_-$ by
$\lesssim0.7\%$ in a box potential, whereas a harmonic trap would require
explicit treatment or local-density selection to control inhomogeneous
broadening. At the largest interactions, higher-band effects must also remain
controlled: the interaction, drive and relevant many-body energy scales should
remain sufficiently below the interband separation for the single-band Hubbard
description to apply. This generally favours deeper lattices, although the
resulting reduction of $t$ must be included in the experimental calibration; the
intermediate window $|U|\simeq3$--$4$ is where the protocol is cleanest. As for platforms, the cited experiments establish the required ingredients
separately---honeycomb Fermi gases, bilayer (two-plane) Hubbard systems with
layer-resolved microscopy, and attractive Hubbard gases under a quantum-gas
microscope \cite{Gall2021,Hartke2023,Rydow2025}; combining attractive
interactions, honeycomb bilayer geometry and layer-resolved detection in one
apparatus remains an experimental integration challenge. A concrete route would use
$^6$Li, whose broad Feshbach resonance near $832$~G~\cite{Zurn2013} provides the attractive-side
scattering-length control assumed throughout, in the three-beam honeycomb
geometry of Ref.~\cite{Tarruell2012} combined with a $z$-axis double-well
superlattice defining the two layers (the bilayer analogue of the two-plane
control of Ref.~\cite{Gall2021}); the experiment would prepare an average
filling near $n\simeq1.1$, calibrated from the measured density profile, and
two-component fermions on the attractive side benefit from Pauli suppression of
three-body loss~\cite{Petrov2004}.


\bibliographystyle{apsrev4-2}
\bibliography{Ref}

\begin{thebibliography}{59}%
\makeatletter
\providecommand \@ifxundefined [1]{%
 \@ifx{#1\undefined}
}%
\providecommand \@ifnum [1]{%
 \ifnum #1\expandafter \@firstoftwo
 \else \expandafter \@secondoftwo
 \fi
}%
\providecommand \@ifx [1]{%
 \ifx #1\expandafter \@firstoftwo
 \else \expandafter \@secondoftwo
 \fi
}%
\providecommand \natexlab [1]{#1}%
\providecommand \enquote  [1]{``#1''}%
\providecommand \bibnamefont  [1]{#1}%
\providecommand \bibfnamefont [1]{#1}%
\providecommand \citenamefont [1]{#1}%
\providecommand \href@noop [0]{\@secondoftwo}%
\providecommand \href [0]{\begingroup \@sanitize@url \@href}%
\providecommand \@href[1]{\@@startlink{#1}\@@href}%
\providecommand \@@href[1]{\endgroup#1\@@endlink}%
\providecommand \@sanitize@url [0]{\catcode `\\12\catcode `\$12\catcode
  `\&12\catcode `\#12\catcode `\^12\catcode `\_12\catcode `\%12\relax}%
\providecommand \@@startlink[1]{}%
\providecommand \@@endlink[0]{}%
\providecommand \url  [0]{\begingroup\@sanitize@url \@url }%
\providecommand \@url [1]{\endgroup\@href {#1}{\urlprefix }}%
\providecommand \urlprefix  [0]{URL }%
\providecommand \Eprint [0]{\href }%
\providecommand \doibase [0]{https://doi.org/}%
\providecommand \selectlanguage [0]{\@gobble}%
\providecommand \bibinfo  [0]{\@secondoftwo}%
\providecommand \bibfield  [0]{\@secondoftwo}%
\providecommand \translation [1]{[#1]}%
\providecommand \BibitemOpen [0]{}%
\providecommand \bibitemStop [0]{}%
\providecommand \bibitemNoStop [0]{.\EOS\space}%
\providecommand \EOS [0]{\spacefactor3000\relax}%
\providecommand \BibitemShut  [1]{\csname bibitem#1\endcsname}%
\let\auto@bib@innerbib\@empty
\bibitem [{\citenamefont {Eagles}(1969)}]{Eagles1969}%
  \BibitemOpen
  \bibfield  {author} {\bibinfo {author} {\bibfnamefont {D.~M.}\ \bibnamefont
  {Eagles}},\ }\href {https://doi.org/10.1103/PhysRev.186.456} {\bibfield
  {journal} {\bibinfo  {journal} {Phys. Rev.}\ }\textbf {\bibinfo {volume}
  {186}},\ \bibinfo {pages} {456} (\bibinfo {year} {1969})}\BibitemShut
  {NoStop}%
\bibitem [{\citenamefont {Leggett}(1980)}]{Leggett1980}%
  \BibitemOpen
  \bibfield  {author} {\bibinfo {author} {\bibfnamefont {A.~J.}\ \bibnamefont
  {Leggett}},\ }in\ \href {https://doi.org/10.1007/BFb0120125} {\emph {\bibinfo
  {booktitle} {Modern Trends in the Theory of Condensed Matter}}},\ \bibinfo
  {series} {Lecture Notes in Physics}, Vol.\ \bibinfo {volume} {115}\ (\bibinfo
   {publisher} {Springer},\ \bibinfo {year} {1980})\ pp.\ \bibinfo {pages}
  {13--27}\BibitemShut {NoStop}%
\bibitem [{\citenamefont {Nozi\`eres}\ and\ \citenamefont
  {Schmitt-Rink}(1985)}]{NozieresSchmittRink1985}%
  \BibitemOpen
  \bibfield  {author} {\bibinfo {author} {\bibfnamefont {P.}~\bibnamefont
  {Nozi\`eres}}\ and\ \bibinfo {author} {\bibfnamefont {S.}~\bibnamefont
  {Schmitt-Rink}},\ }\href {https://doi.org/10.1007/BF00683774} {\bibfield
  {journal} {\bibinfo  {journal} {J. Low Temp. Phys.}\ }\textbf {\bibinfo
  {volume} {59}},\ \bibinfo {pages} {195} (\bibinfo {year} {1985})}\BibitemShut
  {NoStop}%
\bibitem [{\citenamefont {Randeria}\ and\ \citenamefont
  {Taylor}(2014)}]{Randeria2014}%
  \BibitemOpen
  \bibfield  {author} {\bibinfo {author} {\bibfnamefont {M.}~\bibnamefont
  {Randeria}}\ and\ \bibinfo {author} {\bibfnamefont {E.}~\bibnamefont
  {Taylor}},\ }\href {https://doi.org/10.1146/annurev-conmatphys-031113-133829}
  {\bibfield  {journal} {\bibinfo  {journal} {Annu. Rev. Condens. Matter
  Phys.}\ }\textbf {\bibinfo {volume} {5}},\ \bibinfo {pages} {209} (\bibinfo
  {year} {2014})}\BibitemShut {NoStop}%
\bibitem [{\citenamefont {Bloch}\ \emph {et~al.}(2008)\citenamefont {Bloch},
  \citenamefont {Dalibard},\ and\ \citenamefont {Zwerger}}]{Bloch2008}%
  \BibitemOpen
  \bibfield  {author} {\bibinfo {author} {\bibfnamefont {I.}~\bibnamefont
  {Bloch}}, \bibinfo {author} {\bibfnamefont {J.}~\bibnamefont {Dalibard}},\
  and\ \bibinfo {author} {\bibfnamefont {W.}~\bibnamefont {Zwerger}},\ }\href
  {https://doi.org/10.1103/RevModPhys.80.885} {\bibfield  {journal} {\bibinfo
  {journal} {Rev. Mod. Phys.}\ }\textbf {\bibinfo {volume} {80}},\ \bibinfo
  {pages} {885} (\bibinfo {year} {2008})}\BibitemShut {NoStop}%
\bibitem [{\citenamefont {Toschi}\ \emph {et~al.}(2005)\citenamefont {Toschi},
  \citenamefont {Capone},\ and\ \citenamefont {Castellani}}]{Toschi2005}%
  \BibitemOpen
  \bibfield  {author} {\bibinfo {author} {\bibfnamefont {A.}~\bibnamefont
  {Toschi}}, \bibinfo {author} {\bibfnamefont {M.}~\bibnamefont {Capone}},\
  and\ \bibinfo {author} {\bibfnamefont {C.}~\bibnamefont {Castellani}},\
  }\href {https://doi.org/10.1103/PhysRevB.72.235118} {\bibfield  {journal}
  {\bibinfo  {journal} {Phys. Rev. B}\ }\textbf {\bibinfo {volume} {72}},\
  \bibinfo {pages} {235118} (\bibinfo {year} {2005})}\BibitemShut {NoStop}%
\bibitem [{\citenamefont {Astrakharchik}\ \emph {et~al.}(2005)\citenamefont
  {Astrakharchik}, \citenamefont {Combescot}, \citenamefont {Leyronas},\ and\
  \citenamefont {Stringari}}]{Astrakharchik2005}%
  \BibitemOpen
  \bibfield  {author} {\bibinfo {author} {\bibfnamefont {G.~E.}\ \bibnamefont
  {Astrakharchik}}, \bibinfo {author} {\bibfnamefont {R.}~\bibnamefont
  {Combescot}}, \bibinfo {author} {\bibfnamefont {X.}~\bibnamefont
  {Leyronas}},\ and\ \bibinfo {author} {\bibfnamefont {S.}~\bibnamefont
  {Stringari}},\ }\href {https://doi.org/10.1103/PhysRevLett.95.030404}
  {\bibfield  {journal} {\bibinfo  {journal} {Phys. Rev. Lett.}\ }\textbf
  {\bibinfo {volume} {95}},\ \bibinfo {pages} {030404} (\bibinfo {year}
  {2005})}\BibitemShut {NoStop}%
\bibitem [{\citenamefont {Park}\ and\ \citenamefont
  {Choi}(2024)}]{ParkChoi2024}%
  \BibitemOpen
  \bibfield  {author} {\bibinfo {author} {\bibfnamefont {T.-H.}\ \bibnamefont
  {Park}}\ and\ \bibinfo {author} {\bibfnamefont {H.-Y.}\ \bibnamefont
  {Choi}},\ }\href@noop {} {\bibfield  {journal} {\bibinfo  {journal}
  {arXiv:2411.14782}\ } (\bibinfo {year} {2024})}\BibitemShut {NoStop}%
\bibitem [{\citenamefont {Trotzky}\ \emph {et~al.}(2008)\citenamefont
  {Trotzky}, \citenamefont {Cheinet}, \citenamefont {F{\"o}lling},
  \citenamefont {Feld}, \citenamefont {Schnorrberger}, \citenamefont {Rey},
  \citenamefont {Polkovnikov}, \citenamefont {Demler}, \citenamefont {Lukin},\
  and\ \citenamefont {Bloch}}]{Trotzky2008}%
  \BibitemOpen
  \bibfield  {author} {\bibinfo {author} {\bibfnamefont {S.}~\bibnamefont
  {Trotzky}}, \bibinfo {author} {\bibfnamefont {P.}~\bibnamefont {Cheinet}},
  \bibinfo {author} {\bibfnamefont {S.}~\bibnamefont {F{\"o}lling}}, \bibinfo
  {author} {\bibfnamefont {M.}~\bibnamefont {Feld}}, \bibinfo {author}
  {\bibfnamefont {U.}~\bibnamefont {Schnorrberger}}, \bibinfo {author}
  {\bibfnamefont {A.~M.}\ \bibnamefont {Rey}}, \bibinfo {author} {\bibfnamefont
  {A.}~\bibnamefont {Polkovnikov}}, \bibinfo {author} {\bibfnamefont {E.~A.}\
  \bibnamefont {Demler}}, \bibinfo {author} {\bibfnamefont {M.~D.}\
  \bibnamefont {Lukin}},\ and\ \bibinfo {author} {\bibfnamefont
  {I.}~\bibnamefont {Bloch}},\ }\href {https://doi.org/10.1126/science.1150841}
  {\bibfield  {journal} {\bibinfo  {journal} {Science}\ }\textbf {\bibinfo
  {volume} {319}},\ \bibinfo {pages} {295} (\bibinfo {year}
  {2008})}\BibitemShut {NoStop}%
\bibitem [{\citenamefont {Micnas}\ \emph {et~al.}(1990)\citenamefont {Micnas},
  \citenamefont {Ranninger},\ and\ \citenamefont {Robaszkiewicz}}]{Micnas1990}%
  \BibitemOpen
  \bibfield  {author} {\bibinfo {author} {\bibfnamefont {R.}~\bibnamefont
  {Micnas}}, \bibinfo {author} {\bibfnamefont {J.}~\bibnamefont {Ranninger}},\
  and\ \bibinfo {author} {\bibfnamefont {S.}~\bibnamefont {Robaszkiewicz}},\
  }\href@noop {} {\bibfield  {journal} {\bibinfo  {journal} {Rev. Mod. Phys.}\
  }\textbf {\bibinfo {volume} {62}},\ \bibinfo {pages} {113} (\bibinfo {year}
  {1990})}\BibitemShut {NoStop}%
\bibitem [{\citenamefont {Robaszkiewicz}\ \emph {et~al.}(1981)\citenamefont
  {Robaszkiewicz}, \citenamefont {Micnas},\ and\ \citenamefont
  {Chao}}]{Robaszkiewicz1981}%
  \BibitemOpen
  \bibfield  {author} {\bibinfo {author} {\bibfnamefont {S.}~\bibnamefont
  {Robaszkiewicz}}, \bibinfo {author} {\bibfnamefont {R.}~\bibnamefont
  {Micnas}},\ and\ \bibinfo {author} {\bibfnamefont {K.~A.}\ \bibnamefont
  {Chao}},\ }\href {https://doi.org/10.1103/PhysRevB.23.1447} {\bibfield
  {journal} {\bibinfo  {journal} {Phys. Rev. B}\ }\textbf {\bibinfo {volume}
  {23}},\ \bibinfo {pages} {1447} (\bibinfo {year} {1981})}\BibitemShut
  {NoStop}%
\bibitem [{\citenamefont {Leggett}(1966)}]{Leggett1966}%
  \BibitemOpen
  \bibfield  {author} {\bibinfo {author} {\bibfnamefont {A.~J.}\ \bibnamefont
  {Leggett}},\ }\href {https://doi.org/10.1143/PTP.36.901} {\bibfield
  {journal} {\bibinfo  {journal} {Prog. Theor. Phys.}\ }\textbf {\bibinfo
  {volume} {36}},\ \bibinfo {pages} {901} (\bibinfo {year} {1966})}\BibitemShut
  {NoStop}%
\bibitem [{\citenamefont {Sharapov}\ \emph {et~al.}(2002)\citenamefont
  {Sharapov}, \citenamefont {Gusynin},\ and\ \citenamefont
  {Beck}}]{SharapovGusyninBeck2002}%
  \BibitemOpen
  \bibfield  {author} {\bibinfo {author} {\bibfnamefont {S.~G.}\ \bibnamefont
  {Sharapov}}, \bibinfo {author} {\bibfnamefont {V.~P.}\ \bibnamefont
  {Gusynin}},\ and\ \bibinfo {author} {\bibfnamefont {H.}~\bibnamefont
  {Beck}},\ }\href {https://doi.org/10.1140/epjb/e2002-00356-9} {\bibfield
  {journal} {\bibinfo  {journal} {Eur. Phys. J. B}\ }\textbf {\bibinfo {volume}
  {30}},\ \bibinfo {pages} {45} (\bibinfo {year} {2002})}\BibitemShut {NoStop}%
\bibitem [{\citenamefont {Anishchanka}\ \emph {et~al.}(2007)\citenamefont
  {Anishchanka}, \citenamefont {Volkov},\ and\ \citenamefont
  {Efetov}}]{Anishchanka2007}%
  \BibitemOpen
  \bibfield  {author} {\bibinfo {author} {\bibfnamefont {A.}~\bibnamefont
  {Anishchanka}}, \bibinfo {author} {\bibfnamefont {A.~F.}\ \bibnamefont
  {Volkov}},\ and\ \bibinfo {author} {\bibfnamefont {K.~B.}\ \bibnamefont
  {Efetov}},\ }\href {https://doi.org/10.1103/PhysRevB.76.104504} {\bibfield
  {journal} {\bibinfo  {journal} {Phys. Rev. B}\ }\textbf {\bibinfo {volume}
  {76}},\ \bibinfo {pages} {104504} (\bibinfo {year} {2007})}\BibitemShut
  {NoStop}%
\bibitem [{\citenamefont {Burnell}\ \emph {et~al.}(2010)\citenamefont
  {Burnell}, \citenamefont {Hu}, \citenamefont {Parish},\ and\ \citenamefont
  {Bernevig}}]{BurnellHuLin2010}%
  \BibitemOpen
  \bibfield  {author} {\bibinfo {author} {\bibfnamefont {F.~J.}\ \bibnamefont
  {Burnell}}, \bibinfo {author} {\bibfnamefont {J.}~\bibnamefont {Hu}},
  \bibinfo {author} {\bibfnamefont {M.~M.}\ \bibnamefont {Parish}},\ and\
  \bibinfo {author} {\bibfnamefont {B.~A.}\ \bibnamefont {Bernevig}},\ }\href
  {https://doi.org/10.1103/PhysRevB.82.144506} {\bibfield  {journal} {\bibinfo
  {journal} {Phys. Rev. B}\ }\textbf {\bibinfo {volume} {82}},\ \bibinfo
  {pages} {144506} (\bibinfo {year} {2010})}\BibitemShut {NoStop}%
\bibitem [{\citenamefont {Blumberg}\ \emph {et~al.}(2007)\citenamefont
  {Blumberg} \emph {et~al.}}]{Blumberg2007}%
  \BibitemOpen
  \bibfield  {author} {\bibinfo {author} {\bibfnamefont {G.}~\bibnamefont
  {Blumberg}} \emph {et~al.},\ }\href
  {https://doi.org/10.1103/PhysRevLett.99.227002} {\bibfield  {journal}
  {\bibinfo  {journal} {Phys. Rev. Lett.}\ }\textbf {\bibinfo {volume} {99}},\
  \bibinfo {pages} {227002} (\bibinfo {year} {2007})}\BibitemShut {NoStop}%
\bibitem [{\citenamefont {Cuozzo}\ \emph {et~al.}(2024)\citenamefont {Cuozzo},
  \citenamefont {Yu}, \citenamefont {Davids} \emph {et~al.}}]{Cuozzo2024}%
  \BibitemOpen
  \bibfield  {author} {\bibinfo {author} {\bibfnamefont {J.~J.}\ \bibnamefont
  {Cuozzo}}, \bibinfo {author} {\bibfnamefont {W.}~\bibnamefont {Yu}}, \bibinfo
  {author} {\bibfnamefont {P.}~\bibnamefont {Davids}}, \emph {et~al.},\ }\href
  {https://doi.org/10.1038/s41567-024-02412-4} {\bibfield  {journal} {\bibinfo
  {journal} {Nat. Phys.}\ }\textbf {\bibinfo {volume} {20}},\ \bibinfo {pages}
  {1118} (\bibinfo {year} {2024})}\BibitemShut {NoStop}%
\bibitem [{\citenamefont {Iskin}\ and\ \citenamefont {S\'a~de
  Melo}(2005)}]{IskinSadeMelo2005}%
  \BibitemOpen
  \bibfield  {author} {\bibinfo {author} {\bibfnamefont {M.}~\bibnamefont
  {Iskin}}\ and\ \bibinfo {author} {\bibfnamefont {C.~A.~R.}\ \bibnamefont
  {S\'a~de Melo}},\ }\href {https://doi.org/10.1103/PhysRevB.72.024512}
  {\bibfield  {journal} {\bibinfo  {journal} {Phys. Rev. B}\ }\textbf {\bibinfo
  {volume} {72}},\ \bibinfo {pages} {024512} (\bibinfo {year}
  {2005})}\BibitemShut {NoStop}%
\bibitem [{\citenamefont {Hackner}\ and\ \citenamefont
  {Brydon}(2023)}]{Hackner2023}%
  \BibitemOpen
  \bibfield  {author} {\bibinfo {author} {\bibfnamefont {N.~A.}\ \bibnamefont
  {Hackner}}\ and\ \bibinfo {author} {\bibfnamefont {P.~M.~R.}\ \bibnamefont
  {Brydon}},\ }\href {https://doi.org/10.1103/PhysRevB.108.L220505} {\bibfield
  {journal} {\bibinfo  {journal} {Phys. Rev. B}\ }\textbf {\bibinfo {volume}
  {108}},\ \bibinfo {pages} {L220505} (\bibinfo {year} {2023})}\BibitemShut
  {NoStop}%
\bibitem [{\citenamefont {Gall}\ \emph {et~al.}(2021)\citenamefont {Gall},
  \citenamefont {Wurz}, \citenamefont {Samland}, \citenamefont {Chan},\ and\
  \citenamefont {K\"ohl}}]{Gall2021}%
  \BibitemOpen
  \bibfield  {author} {\bibinfo {author} {\bibfnamefont {M.}~\bibnamefont
  {Gall}}, \bibinfo {author} {\bibfnamefont {N.}~\bibnamefont {Wurz}}, \bibinfo
  {author} {\bibfnamefont {J.}~\bibnamefont {Samland}}, \bibinfo {author}
  {\bibfnamefont {C.~F.}\ \bibnamefont {Chan}},\ and\ \bibinfo {author}
  {\bibfnamefont {M.}~\bibnamefont {K\"ohl}},\ }\href
  {https://doi.org/10.1038/s41586-020-03058-x} {\bibfield  {journal} {\bibinfo
  {journal} {Nature}\ }\textbf {\bibinfo {volume} {589}},\ \bibinfo {pages}
  {40} (\bibinfo {year} {2021})}\BibitemShut {NoStop}%
\bibitem [{\citenamefont {Hartke}\ \emph {et~al.}(2023)\citenamefont {Hartke},
  \citenamefont {Oreg}, \citenamefont {Turnbaugh}, \citenamefont {Jia},\ and\
  \citenamefont {Zwierlein}}]{Hartke2023}%
  \BibitemOpen
  \bibfield  {author} {\bibinfo {author} {\bibfnamefont {T.}~\bibnamefont
  {Hartke}}, \bibinfo {author} {\bibfnamefont {B.}~\bibnamefont {Oreg}},
  \bibinfo {author} {\bibfnamefont {C.}~\bibnamefont {Turnbaugh}}, \bibinfo
  {author} {\bibfnamefont {N.}~\bibnamefont {Jia}},\ and\ \bibinfo {author}
  {\bibfnamefont {M.}~\bibnamefont {Zwierlein}},\ }\href
  {https://doi.org/10.1126/science.ade4245} {\bibfield  {journal} {\bibinfo
  {journal} {Science}\ }\textbf {\bibinfo {volume} {381}},\ \bibinfo {pages}
  {82} (\bibinfo {year} {2023})}\BibitemShut {NoStop}%
\bibitem [{\citenamefont {Rydow}\ \emph {et~al.}(2025)\citenamefont {Rydow}
  \emph {et~al.}}]{Rydow2025}%
  \BibitemOpen
  \bibfield  {author} {\bibinfo {author} {\bibfnamefont {E.}~\bibnamefont
  {Rydow}} \emph {et~al.},\ }\href {https://doi.org/10.1038/s41467-025-62277-w}
  {\bibfield  {journal} {\bibinfo  {journal} {Nat. Commun.}\ }\textbf {\bibinfo
  {volume} {16}},\ \bibinfo {pages} {7201} (\bibinfo {year}
  {2025})}\BibitemShut {NoStop}%
\bibitem [{\citenamefont {Abad}\ and\ \citenamefont
  {Recati}(2013)}]{AbadRecati2013}%
  \BibitemOpen
  \bibfield  {author} {\bibinfo {author} {\bibfnamefont {M.}~\bibnamefont
  {Abad}}\ and\ \bibinfo {author} {\bibfnamefont {A.}~\bibnamefont {Recati}},\
  }\href {https://doi.org/10.1140/epjd/e2013-40053-2} {\bibfield  {journal}
  {\bibinfo  {journal} {Eur. Phys. J. D}\ }\textbf {\bibinfo {volume} {67}},\
  \bibinfo {pages} {148} (\bibinfo {year} {2013})}\BibitemShut {NoStop}%
\bibitem [{\citenamefont {Forsthofer}\ \emph {et~al.}(1996)\citenamefont
  {Forsthofer}, \citenamefont {Kind},\ and\ \citenamefont {Keller}}]{FKK1996}%
  \BibitemOpen
  \bibfield  {author} {\bibinfo {author} {\bibfnamefont {F.}~\bibnamefont
  {Forsthofer}}, \bibinfo {author} {\bibfnamefont {S.}~\bibnamefont {Kind}},\
  and\ \bibinfo {author} {\bibfnamefont {J.}~\bibnamefont {Keller}},\ }\href
  {https://doi.org/10.1103/PhysRevB.53.14481} {\bibfield  {journal} {\bibinfo
  {journal} {Phys. Rev. B}\ }\textbf {\bibinfo {volume} {53}},\ \bibinfo
  {pages} {14481} (\bibinfo {year} {1996})}\BibitemShut {NoStop}%
\bibitem [{\citenamefont {Yang}(1989)}]{Yang1989}%
  \BibitemOpen
  \bibfield  {author} {\bibinfo {author} {\bibfnamefont {C.~N.}\ \bibnamefont
  {Yang}},\ }\href {https://doi.org/10.1103/PhysRevLett.63.2144} {\bibfield
  {journal} {\bibinfo  {journal} {Phys. Rev. Lett.}\ }\textbf {\bibinfo
  {volume} {63}},\ \bibinfo {pages} {2144} (\bibinfo {year}
  {1989})}\BibitemShut {NoStop}%
\bibitem [{\citenamefont {Zhang}(1990)}]{Zhang1990}%
  \BibitemOpen
  \bibfield  {author} {\bibinfo {author} {\bibfnamefont {S.-C.}\ \bibnamefont
  {Zhang}},\ }\href {https://doi.org/10.1103/PhysRevLett.65.120} {\bibfield
  {journal} {\bibinfo  {journal} {Phys. Rev. Lett.}\ }\textbf {\bibinfo
  {volume} {65}},\ \bibinfo {pages} {120} (\bibinfo {year} {1990})}\BibitemShut
  {NoStop}%
\bibitem [{\citenamefont {Anderson}(1958)}]{Anderson1958}%
  \BibitemOpen
  \bibfield  {author} {\bibinfo {author} {\bibfnamefont {P.~W.}\ \bibnamefont
  {Anderson}},\ }\href {https://doi.org/10.1103/PhysRev.112.1900} {\bibfield
  {journal} {\bibinfo  {journal} {Phys. Rev.}\ }\textbf {\bibinfo {volume}
  {112}},\ \bibinfo {pages} {1900} (\bibinfo {year} {1958})}\BibitemShut
  {NoStop}%
\bibitem [{\citenamefont {Dornheim}\ \emph {et~al.}(2023)\citenamefont
  {Dornheim}, \citenamefont {Wicaksono}, \citenamefont {Suarez-Cardona},
  \citenamefont {Tolias}, \citenamefont {B{\"o}hme}, \citenamefont
  {Moldabekov}, \citenamefont {Hamann},\ and\ \citenamefont
  {Vorberger}}]{Dornheim2023moments}%
  \BibitemOpen
  \bibfield  {author} {\bibinfo {author} {\bibfnamefont {T.}~\bibnamefont
  {Dornheim}}, \bibinfo {author} {\bibfnamefont {D.~C.}\ \bibnamefont
  {Wicaksono}}, \bibinfo {author} {\bibfnamefont {J.~E.}\ \bibnamefont
  {Suarez-Cardona}}, \bibinfo {author} {\bibfnamefont {P.}~\bibnamefont
  {Tolias}}, \bibinfo {author} {\bibfnamefont {M.~P.}\ \bibnamefont
  {B{\"o}hme}}, \bibinfo {author} {\bibfnamefont {Z.~A.}\ \bibnamefont
  {Moldabekov}}, \bibinfo {author} {\bibfnamefont {J.}~\bibnamefont {Hamann}},\
  and\ \bibinfo {author} {\bibfnamefont {J.}~\bibnamefont {Vorberger}},\ }\href
  {https://doi.org/10.1103/PhysRevB.107.155148} {\bibfield  {journal} {\bibinfo
   {journal} {Phys. Rev. B}\ }\textbf {\bibinfo {volume} {107}},\ \bibinfo
  {pages} {155148} (\bibinfo {year} {2023})}\BibitemShut {NoStop}%
\bibitem [{\citenamefont {Prasad}(2022)}]{Prasad2022}%
  \BibitemOpen
  \bibfield  {author} {\bibinfo {author} {\bibfnamefont {Y.}~\bibnamefont
  {Prasad}},\ }\href {https://doi.org/10.1103/PhysRevB.106.184506} {\bibfield
  {journal} {\bibinfo  {journal} {Phys. Rev. B}\ }\textbf {\bibinfo {volume}
  {106}},\ \bibinfo {pages} {184506} (\bibinfo {year} {2022})}\BibitemShut
  {NoStop}%
\bibitem [{\citenamefont {Zhao}\ and\ \citenamefont
  {Paramekanti}(2006)}]{ZhaoParamekanti2006}%
  \BibitemOpen
  \bibfield  {author} {\bibinfo {author} {\bibfnamefont {E.}~\bibnamefont
  {Zhao}}\ and\ \bibinfo {author} {\bibfnamefont {A.}~\bibnamefont
  {Paramekanti}},\ }\href {https://doi.org/10.1103/PhysRevLett.97.230404}
  {\bibfield  {journal} {\bibinfo  {journal} {Phys. Rev. Lett.}\ }\textbf
  {\bibinfo {volume} {97}},\ \bibinfo {pages} {230404} (\bibinfo {year}
  {2006})}\BibitemShut {NoStop}%
\bibitem [{\citenamefont {Iskin}(2019)}]{Iskin2019HoneycombStiffness}%
  \BibitemOpen
  \bibfield  {author} {\bibinfo {author} {\bibfnamefont {M.}~\bibnamefont
  {Iskin}},\ }\href {https://doi.org/10.1103/PhysRevA.99.023608} {\bibfield
  {journal} {\bibinfo  {journal} {Phys. Rev. A}\ }\textbf {\bibinfo {volume}
  {99}},\ \bibinfo {pages} {023608} (\bibinfo {year} {2019})}\BibitemShut
  {NoStop}%
\bibitem [{\citenamefont {Kohmoto}\ and\ \citenamefont
  {Takada}(1990)}]{KohmotoTakada1990}%
  \BibitemOpen
  \bibfield  {author} {\bibinfo {author} {\bibfnamefont {M.}~\bibnamefont
  {Kohmoto}}\ and\ \bibinfo {author} {\bibfnamefont {Y.}~\bibnamefont
  {Takada}},\ }\href {https://doi.org/10.1143/JPSJ.59.1541} {\bibfield
  {journal} {\bibinfo  {journal} {J. Phys. Soc. Jpn.}\ }\textbf {\bibinfo
  {volume} {59}},\ \bibinfo {pages} {1541} (\bibinfo {year}
  {1990})}\BibitemShut {NoStop}%
\bibitem [{\citenamefont {Nozi\`eres}\ and\ \citenamefont
  {Pistolesi}(1999)}]{NozieresPistolesi1999}%
  \BibitemOpen
  \bibfield  {author} {\bibinfo {author} {\bibfnamefont {P.}~\bibnamefont
  {Nozi\`eres}}\ and\ \bibinfo {author} {\bibfnamefont {F.}~\bibnamefont
  {Pistolesi}},\ }\href {https://doi.org/10.1007/s100510050897} {\bibfield
  {journal} {\bibinfo  {journal} {Eur. Phys. J. B}\ }\textbf {\bibinfo {volume}
  {10}},\ \bibinfo {pages} {649} (\bibinfo {year} {1999})}\BibitemShut
  {NoStop}%
\bibitem [{\citenamefont {Prasad}\ \emph {et~al.}(2014)\citenamefont {Prasad},
  \citenamefont {Medhi},\ and\ \citenamefont {Shenoy}}]{PrasadShenoyPRA89}%
  \BibitemOpen
  \bibfield  {author} {\bibinfo {author} {\bibfnamefont {Y.}~\bibnamefont
  {Prasad}}, \bibinfo {author} {\bibfnamefont {A.}~\bibnamefont {Medhi}},\ and\
  \bibinfo {author} {\bibfnamefont {V.~B.}\ \bibnamefont {Shenoy}},\ }\href
  {https://doi.org/10.1103/PhysRevA.89.043605} {\bibfield  {journal} {\bibinfo
  {journal} {Phys. Rev. A}\ }\textbf {\bibinfo {volume} {89}},\ \bibinfo
  {pages} {043605} (\bibinfo {year} {2014})}\BibitemShut {NoStop}%
\bibitem [{\citenamefont {Ghadimi}\ \emph {et~al.}(2024)\citenamefont
  {Ghadimi}, \citenamefont {Mondal}, \citenamefont {Kim},\ and\ \citenamefont
  {Yang}}]{Ghadimi2024}%
  \BibitemOpen
  \bibfield  {author} {\bibinfo {author} {\bibfnamefont {R.}~\bibnamefont
  {Ghadimi}}, \bibinfo {author} {\bibfnamefont {C.}~\bibnamefont {Mondal}},
  \bibinfo {author} {\bibfnamefont {S.}~\bibnamefont {Kim}},\ and\ \bibinfo
  {author} {\bibfnamefont {B.-J.}\ \bibnamefont {Yang}},\ }\href
  {https://doi.org/10.1103/PhysRevLett.133.196603} {\bibfield  {journal}
  {\bibinfo  {journal} {Phys. Rev. Lett.}\ }\textbf {\bibinfo {volume} {133}},\
  \bibinfo {pages} {196603} (\bibinfo {year} {2024})}\BibitemShut {NoStop}%
\bibitem [{\citenamefont {Mondal}\ \emph {et~al.}(2026)\citenamefont {Mondal},
  \citenamefont {Ghadimi},\ and\ \citenamefont {Yang}}]{Mondal2026}%
  \BibitemOpen
  \bibfield  {author} {\bibinfo {author} {\bibfnamefont {C.}~\bibnamefont
  {Mondal}}, \bibinfo {author} {\bibfnamefont {R.}~\bibnamefont {Ghadimi}},\
  and\ \bibinfo {author} {\bibfnamefont {B.-J.}\ \bibnamefont {Yang}},\ }\href
  {https://doi.org/10.1103/3pnm-76hh} {\bibfield  {journal} {\bibinfo
  {journal} {Phys. Rev. B}\ }\textbf {\bibinfo {volume} {113}},\ \bibinfo
  {pages} {L081101} (\bibinfo {year} {2026})}\BibitemShut {NoStop}%
\bibitem [{\citenamefont {Feynman}(1954)}]{Feynman1954}%
  \BibitemOpen
  \bibfield  {author} {\bibinfo {author} {\bibfnamefont {R.~P.}\ \bibnamefont
  {Feynman}},\ }\href {https://doi.org/10.1103/PhysRev.94.262} {\bibfield
  {journal} {\bibinfo  {journal} {Phys. Rev.}\ }\textbf {\bibinfo {volume}
  {94}},\ \bibinfo {pages} {262} (\bibinfo {year} {1954})}\BibitemShut
  {NoStop}%
\bibitem [{\citenamefont {Girvin}\ \emph {et~al.}(1986)\citenamefont {Girvin},
  \citenamefont {MacDonald},\ and\ \citenamefont {Platzman}}]{Girvin1986}%
  \BibitemOpen
  \bibfield  {author} {\bibinfo {author} {\bibfnamefont {S.~M.}\ \bibnamefont
  {Girvin}}, \bibinfo {author} {\bibfnamefont {A.~H.}\ \bibnamefont
  {MacDonald}},\ and\ \bibinfo {author} {\bibfnamefont {P.~M.}\ \bibnamefont
  {Platzman}},\ }\href {https://doi.org/10.1103/PhysRevB.33.2481} {\bibfield
  {journal} {\bibinfo  {journal} {Phys. Rev. B}\ }\textbf {\bibinfo {volume}
  {33}},\ \bibinfo {pages} {2481} (\bibinfo {year} {1986})}\BibitemShut
  {NoStop}%
\bibitem [{\citenamefont {Tarruell}\ \emph {et~al.}(2012)\citenamefont
  {Tarruell}, \citenamefont {Greif}, \citenamefont {Uehlinger}, \citenamefont
  {Jotzu},\ and\ \citenamefont {Esslinger}}]{Tarruell2012}%
  \BibitemOpen
  \bibfield  {author} {\bibinfo {author} {\bibfnamefont {L.}~\bibnamefont
  {Tarruell}}, \bibinfo {author} {\bibfnamefont {D.}~\bibnamefont {Greif}},
  \bibinfo {author} {\bibfnamefont {T.}~\bibnamefont {Uehlinger}}, \bibinfo
  {author} {\bibfnamefont {G.}~\bibnamefont {Jotzu}},\ and\ \bibinfo {author}
  {\bibfnamefont {T.}~\bibnamefont {Esslinger}},\ }\href
  {https://doi.org/10.1038/nature10871} {\bibfield  {journal} {\bibinfo
  {journal} {Nature}\ }\textbf {\bibinfo {volume} {483}},\ \bibinfo {pages}
  {302} (\bibinfo {year} {2012})}\BibitemShut {NoStop}%
\bibitem [{\citenamefont {Esslinger}(2010)}]{Esslinger2010}%
  \BibitemOpen
  \bibfield  {author} {\bibinfo {author} {\bibfnamefont {T.}~\bibnamefont
  {Esslinger}},\ }\href
  {https://doi.org/10.1146/annurev-conmatphys-070909-104059} {\bibfield
  {journal} {\bibinfo  {journal} {Annu. Rev. Condens. Matter Phys.}\ }\textbf
  {\bibinfo {volume} {1}},\ \bibinfo {pages} {129} (\bibinfo {year}
  {2010})}\BibitemShut {NoStop}%
\bibitem [{\citenamefont {Gall}\ \emph {et~al.}(2020)\citenamefont {Gall},
  \citenamefont {Chan}, \citenamefont {Wurz},\ and\ \citenamefont
  {K{\"o}hl}}]{Gall2020}%
  \BibitemOpen
  \bibfield  {author} {\bibinfo {author} {\bibfnamefont {M.}~\bibnamefont
  {Gall}}, \bibinfo {author} {\bibfnamefont {C.~F.}\ \bibnamefont {Chan}},
  \bibinfo {author} {\bibfnamefont {N.}~\bibnamefont {Wurz}},\ and\ \bibinfo
  {author} {\bibfnamefont {M.}~\bibnamefont {K{\"o}hl}},\ }\href
  {https://doi.org/10.1103/PhysRevLett.124.010403} {\bibfield  {journal}
  {\bibinfo  {journal} {Phys. Rev. Lett.}\ }\textbf {\bibinfo {volume} {124}},\
  \bibinfo {pages} {010403} (\bibinfo {year} {2020})}\BibitemShut {NoStop}%
\bibitem [{\citenamefont {Fontenele}\ \emph {et~al.}(2024)\citenamefont
  {Fontenele}, \citenamefont {Costa}, \citenamefont {Paiva},\ and\
  \citenamefont {dos Santos}}]{Fontenele2024}%
  \BibitemOpen
  \bibfield  {author} {\bibinfo {author} {\bibfnamefont {R.~A.}\ \bibnamefont
  {Fontenele}}, \bibinfo {author} {\bibfnamefont {N.~C.}\ \bibnamefont
  {Costa}}, \bibinfo {author} {\bibfnamefont {T.}~\bibnamefont {Paiva}},\ and\
  \bibinfo {author} {\bibfnamefont {R.~R.}\ \bibnamefont {dos Santos}},\ }\href
  {https://doi.org/10.1103/PhysRevA.110.053315} {\bibfield  {journal} {\bibinfo
   {journal} {Phys. Rev. A}\ }\textbf {\bibinfo {volume} {110}},\ \bibinfo
  {pages} {053315} (\bibinfo {year} {2024})},\ \Eprint
  {https://arxiv.org/abs/2408.17405} {arXiv:2408.17405 [cond-mat.quant-gas]}
  \BibitemShut {NoStop}%
\bibitem [{\citenamefont {Levy}\ \emph {et~al.}(2007)\citenamefont {Levy},
  \citenamefont {Lahoud}, \citenamefont {Shomroni},\ and\ \citenamefont
  {Steinhauer}}]{Levy2007}%
  \BibitemOpen
  \bibfield  {author} {\bibinfo {author} {\bibfnamefont {S.}~\bibnamefont
  {Levy}}, \bibinfo {author} {\bibfnamefont {E.}~\bibnamefont {Lahoud}},
  \bibinfo {author} {\bibfnamefont {I.}~\bibnamefont {Shomroni}},\ and\
  \bibinfo {author} {\bibfnamefont {J.}~\bibnamefont {Steinhauer}},\ }\href
  {https://doi.org/10.1038/nature06186} {\bibfield  {journal} {\bibinfo
  {journal} {Nature}\ }\textbf {\bibinfo {volume} {449}},\ \bibinfo {pages}
  {579} (\bibinfo {year} {2007})}\BibitemShut {NoStop}%
\bibitem [{\citenamefont {Zibold}\ \emph {et~al.}(2010)\citenamefont {Zibold},
  \citenamefont {Nicklas}, \citenamefont {Gross},\ and\ \citenamefont
  {Oberthaler}}]{Zibold2010}%
  \BibitemOpen
  \bibfield  {author} {\bibinfo {author} {\bibfnamefont {T.}~\bibnamefont
  {Zibold}}, \bibinfo {author} {\bibfnamefont {E.}~\bibnamefont {Nicklas}},
  \bibinfo {author} {\bibfnamefont {C.}~\bibnamefont {Gross}},\ and\ \bibinfo
  {author} {\bibfnamefont {M.~K.}\ \bibnamefont {Oberthaler}},\ }\href
  {https://doi.org/10.1103/PhysRevLett.105.204101} {\bibfield  {journal}
  {\bibinfo  {journal} {Phys. Rev. Lett.}\ }\textbf {\bibinfo {volume} {105}},\
  \bibinfo {pages} {204101} (\bibinfo {year} {2010})}\BibitemShut {NoStop}%
\bibitem [{\citenamefont {Veeravalli}\ \emph {et~al.}(2008)\citenamefont
  {Veeravalli}, \citenamefont {Kuhnle}, \citenamefont {Dyke},\ and\
  \citenamefont {Vale}}]{Veeravalli2008}%
  \BibitemOpen
  \bibfield  {author} {\bibinfo {author} {\bibfnamefont {G.}~\bibnamefont
  {Veeravalli}}, \bibinfo {author} {\bibfnamefont {E.}~\bibnamefont {Kuhnle}},
  \bibinfo {author} {\bibfnamefont {P.}~\bibnamefont {Dyke}},\ and\ \bibinfo
  {author} {\bibfnamefont {C.~J.}\ \bibnamefont {Vale}},\ }\href
  {https://doi.org/10.1103/PhysRevLett.101.250403} {\bibfield  {journal}
  {\bibinfo  {journal} {Phys. Rev. Lett.}\ }\textbf {\bibinfo {volume} {101}},\
  \bibinfo {pages} {250403} (\bibinfo {year} {2008})}\BibitemShut {NoStop}%
\bibitem [{\citenamefont {Chin}\ \emph {et~al.}(2004)\citenamefont {Chin} \emph
  {et~al.}}]{Chin2004}%
  \BibitemOpen
  \bibfield  {author} {\bibinfo {author} {\bibfnamefont {C.}~\bibnamefont
  {Chin}} \emph {et~al.},\ }\href {https://doi.org/10.1126/science.1100818}
  {\bibfield  {journal} {\bibinfo  {journal} {Science}\ }\textbf {\bibinfo
  {volume} {305}},\ \bibinfo {pages} {1128} (\bibinfo {year}
  {2004})}\BibitemShut {NoStop}%
\bibitem [{\citenamefont {Cea}\ and\ \citenamefont
  {Benfatto}(2016)}]{CeaBenfatto2016}%
  \BibitemOpen
  \bibfield  {author} {\bibinfo {author} {\bibfnamefont {T.}~\bibnamefont
  {Cea}}\ and\ \bibinfo {author} {\bibfnamefont {L.}~\bibnamefont {Benfatto}},\
  }\href {https://doi.org/10.1103/PhysRevB.94.064512} {\bibfield  {journal}
  {\bibinfo  {journal} {Phys. Rev. B}\ }\textbf {\bibinfo {volume} {94}},\
  \bibinfo {pages} {064512} (\bibinfo {year} {2016})}\BibitemShut {NoStop}%
\bibitem [{\citenamefont {Blankenbecler}\ \emph {et~al.}(1981)\citenamefont
  {Blankenbecler}, \citenamefont {Scalapino},\ and\ \citenamefont
  {Sugar}}]{BSS1981}%
  \BibitemOpen
  \bibfield  {author} {\bibinfo {author} {\bibfnamefont {R.}~\bibnamefont
  {Blankenbecler}}, \bibinfo {author} {\bibfnamefont {D.~J.}\ \bibnamefont
  {Scalapino}},\ and\ \bibinfo {author} {\bibfnamefont {R.~L.}\ \bibnamefont
  {Sugar}},\ }\href {https://doi.org/10.1103/PhysRevD.24.2278} {\bibfield
  {journal} {\bibinfo  {journal} {Phys. Rev. D}\ }\textbf {\bibinfo {volume}
  {24}},\ \bibinfo {pages} {2278} (\bibinfo {year} {1981})}\BibitemShut
  {NoStop}%
\bibitem [{\citenamefont {Hirsch}(1985)}]{Hirsch1985}%
  \BibitemOpen
  \bibfield  {author} {\bibinfo {author} {\bibfnamefont {J.~E.}\ \bibnamefont
  {Hirsch}},\ }\href {https://doi.org/10.1103/PhysRevB.31.4403} {\bibfield
  {journal} {\bibinfo  {journal} {Phys. Rev. B}\ }\textbf {\bibinfo {volume}
  {31}},\ \bibinfo {pages} {4403} (\bibinfo {year} {1985})}\BibitemShut
  {NoStop}%
\bibitem [{\citenamefont {Bercx}\ \emph {et~al.}(2017)\citenamefont {Bercx},
  \citenamefont {Goth}, \citenamefont {Hofmann},\ and\ \citenamefont
  {Assaad}}]{ALF2017}%
  \BibitemOpen
  \bibfield  {author} {\bibinfo {author} {\bibfnamefont {M.}~\bibnamefont
  {Bercx}}, \bibinfo {author} {\bibfnamefont {F.}~\bibnamefont {Goth}},
  \bibinfo {author} {\bibfnamefont {J.~S.}\ \bibnamefont {Hofmann}},\ and\
  \bibinfo {author} {\bibfnamefont {F.~F.}\ \bibnamefont {Assaad}},\ }\href
  {https://doi.org/10.21468/SciPostPhys.3.2.013} {\bibfield  {journal}
  {\bibinfo  {journal} {SciPost Phys.}\ }\textbf {\bibinfo {volume} {3}},\
  \bibinfo {pages} {013} (\bibinfo {year} {2017})}\BibitemShut {NoStop}%
\bibitem [{\citenamefont {Assaad}\ \emph {et~al.}(2022)\citenamefont {Assaad},
  \citenamefont {Bercx}, \citenamefont {Goth}, \citenamefont {G{\"o}tz},
  \citenamefont {Hofmann}, \citenamefont {Huffman}, \citenamefont {Liu},
  \citenamefont {Parisen~Toldin}, \citenamefont {Portela},\ and\ \citenamefont
  {Schwab}}]{ALF2022}%
  \BibitemOpen
  \bibfield  {author} {\bibinfo {author} {\bibfnamefont {F.~F.}\ \bibnamefont
  {Assaad}}, \bibinfo {author} {\bibfnamefont {M.}~\bibnamefont {Bercx}},
  \bibinfo {author} {\bibfnamefont {F.}~\bibnamefont {Goth}}, \bibinfo {author}
  {\bibfnamefont {A.}~\bibnamefont {G{\"o}tz}}, \bibinfo {author}
  {\bibfnamefont {J.~S.}\ \bibnamefont {Hofmann}}, \bibinfo {author}
  {\bibfnamefont {E.}~\bibnamefont {Huffman}}, \bibinfo {author} {\bibfnamefont
  {Z.}~\bibnamefont {Liu}}, \bibinfo {author} {\bibfnamefont {F.}~\bibnamefont
  {Parisen~Toldin}}, \bibinfo {author} {\bibfnamefont {J.~S.~E.}\ \bibnamefont
  {Portela}},\ and\ \bibinfo {author} {\bibfnamefont {J.}~\bibnamefont
  {Schwab}},\ }\bibfield  {journal} {\bibinfo  {journal} {SciPost Phys.
  Codebases}\ }\textbf {\bibinfo {volume} {1}},\ \href
  {https://doi.org/10.21468/SciPostPhysCodeb.1} {10.21468/SciPostPhysCodeb.1}
  (\bibinfo {year} {2022})\BibitemShut {NoStop}%
\bibitem [{\citenamefont {Prasad}\ and\ \citenamefont
  {Lee}(2024)}]{Prasad2024}%
  \BibitemOpen
  \bibfield  {author} {\bibinfo {author} {\bibfnamefont {Y.}~\bibnamefont
  {Prasad}}\ and\ \bibinfo {author} {\bibfnamefont {H.}~\bibnamefont {Lee}},\
  }\href {https://doi.org/10.1103/PhysRevB.109.064506} {\bibfield  {journal}
  {\bibinfo  {journal} {Phys. Rev. B}\ }\textbf {\bibinfo {volume} {109}},\
  \bibinfo {pages} {064506} (\bibinfo {year} {2024})}\BibitemShut {NoStop}%
\bibitem [{\citenamefont {Scalettar}\ \emph {et~al.}(1989)\citenamefont
  {Scalettar}, \citenamefont {Loh}, \citenamefont {Gubernatis}, \citenamefont
  {Moreo}, \citenamefont {White}, \citenamefont {Scalapino}, \citenamefont
  {Sugar},\ and\ \citenamefont {Dagotto}}]{Scalettar1989}%
  \BibitemOpen
  \bibfield  {author} {\bibinfo {author} {\bibfnamefont {R.~T.}\ \bibnamefont
  {Scalettar}}, \bibinfo {author} {\bibfnamefont {E.~Y.}\ \bibnamefont {Loh}},
  \bibinfo {author} {\bibfnamefont {J.~E.}\ \bibnamefont {Gubernatis}},
  \bibinfo {author} {\bibfnamefont {A.}~\bibnamefont {Moreo}}, \bibinfo
  {author} {\bibfnamefont {S.~R.}\ \bibnamefont {White}}, \bibinfo {author}
  {\bibfnamefont {D.~J.}\ \bibnamefont {Scalapino}}, \bibinfo {author}
  {\bibfnamefont {R.~L.}\ \bibnamefont {Sugar}},\ and\ \bibinfo {author}
  {\bibfnamefont {E.}~\bibnamefont {Dagotto}},\ }\href
  {https://doi.org/10.1103/PhysRevLett.62.1407} {\bibfield  {journal} {\bibinfo
   {journal} {Phys. Rev. Lett.}\ }\textbf {\bibinfo {volume} {62}},\ \bibinfo
  {pages} {1407} (\bibinfo {year} {1989})}\BibitemShut {NoStop}%
\bibitem [{\citenamefont {Jarrell}\ and\ \citenamefont
  {Gubernatis}(1996)}]{JarrellGubernatis1996}%
  \BibitemOpen
  \bibfield  {author} {\bibinfo {author} {\bibfnamefont {M.}~\bibnamefont
  {Jarrell}}\ and\ \bibinfo {author} {\bibfnamefont {J.~E.}\ \bibnamefont
  {Gubernatis}},\ }\href {https://doi.org/10.1016/0370-1573(95)00074-7}
  {\bibfield  {journal} {\bibinfo  {journal} {Phys. Rep.}\ }\textbf {\bibinfo
  {volume} {269}},\ \bibinfo {pages} {133} (\bibinfo {year}
  {1996})}\BibitemShut {NoStop}%
\bibitem [{\citenamefont {Wu}(2024)}]{Wu2024}%
  \BibitemOpen
  \bibfield  {author} {\bibinfo {author} {\bibfnamefont {R.}~\bibnamefont
  {Wu}},\ }\href {https://doi.org/10.1103/PhysRevA.109.063314} {\bibfield
  {journal} {\bibinfo  {journal} {Phys. Rev. A}\ }\textbf {\bibinfo {volume}
  {109}},\ \bibinfo {pages} {063314} (\bibinfo {year} {2024})}\BibitemShut
  {NoStop}%
\bibitem [{\citenamefont {Paiva}\ \emph {et~al.}(2010)\citenamefont {Paiva},
  \citenamefont {Scalettar}, \citenamefont {Randeria},\ and\ \citenamefont
  {Trivedi}}]{Paiva2010}%
  \BibitemOpen
  \bibfield  {author} {\bibinfo {author} {\bibfnamefont {T.}~\bibnamefont
  {Paiva}}, \bibinfo {author} {\bibfnamefont {R.~T.}\ \bibnamefont
  {Scalettar}}, \bibinfo {author} {\bibfnamefont {M.}~\bibnamefont
  {Randeria}},\ and\ \bibinfo {author} {\bibfnamefont {N.}~\bibnamefont
  {Trivedi}},\ }\href {https://doi.org/10.1103/PhysRevLett.104.066406}
  {\bibfield  {journal} {\bibinfo  {journal} {Phys. Rev. Lett.}\ }\textbf
  {\bibinfo {volume} {104}},\ \bibinfo {pages} {066406} (\bibinfo {year}
  {2010})}\BibitemShut {NoStop}%
\bibitem [{\citenamefont {Werner}\ \emph {et~al.}(2005)\citenamefont {Werner},
  \citenamefont {Parcollet}, \citenamefont {Georges},\ and\ \citenamefont
  {Hassan}}]{Werner2005}%
  \BibitemOpen
  \bibfield  {author} {\bibinfo {author} {\bibfnamefont {F.}~\bibnamefont
  {Werner}}, \bibinfo {author} {\bibfnamefont {O.}~\bibnamefont {Parcollet}},
  \bibinfo {author} {\bibfnamefont {A.}~\bibnamefont {Georges}},\ and\ \bibinfo
  {author} {\bibfnamefont {S.~R.}\ \bibnamefont {Hassan}},\ }\href
  {https://doi.org/10.1103/PhysRevLett.95.056401} {\bibfield  {journal}
  {\bibinfo  {journal} {Phys. Rev. Lett.}\ }\textbf {\bibinfo {volume} {95}},\
  \bibinfo {pages} {056401} (\bibinfo {year} {2005})}\BibitemShut {NoStop}%
\bibitem [{\citenamefont {Z{\"u}rn}\ \emph {et~al.}(2013)\citenamefont
  {Z{\"u}rn}, \citenamefont {Lompe}, \citenamefont {Wenz}, \citenamefont
  {Jochim}, \citenamefont {Julienne},\ and\ \citenamefont {Hutson}}]{Zurn2013}%
  \BibitemOpen
  \bibfield  {author} {\bibinfo {author} {\bibfnamefont {G.}~\bibnamefont
  {Z{\"u}rn}}, \bibinfo {author} {\bibfnamefont {T.}~\bibnamefont {Lompe}},
  \bibinfo {author} {\bibfnamefont {A.~N.}\ \bibnamefont {Wenz}}, \bibinfo
  {author} {\bibfnamefont {S.}~\bibnamefont {Jochim}}, \bibinfo {author}
  {\bibfnamefont {P.~S.}\ \bibnamefont {Julienne}},\ and\ \bibinfo {author}
  {\bibfnamefont {J.~M.}\ \bibnamefont {Hutson}},\ }\href
  {https://doi.org/10.1103/PhysRevLett.110.135301} {\bibfield  {journal}
  {\bibinfo  {journal} {Phys. Rev. Lett.}\ }\textbf {\bibinfo {volume} {110}},\
  \bibinfo {pages} {135301} (\bibinfo {year} {2013})}\BibitemShut {NoStop}%
\bibitem [{\citenamefont {Petrov}\ \emph {et~al.}(2004)\citenamefont {Petrov},
  \citenamefont {Salomon},\ and\ \citenamefont {Shlyapnikov}}]{Petrov2004}%
  \BibitemOpen
  \bibfield  {author} {\bibinfo {author} {\bibfnamefont {D.~S.}\ \bibnamefont
  {Petrov}}, \bibinfo {author} {\bibfnamefont {C.}~\bibnamefont {Salomon}},\
  and\ \bibinfo {author} {\bibfnamefont {G.~V.}\ \bibnamefont {Shlyapnikov}},\
  }\href {https://doi.org/10.1103/PhysRevLett.93.090404} {\bibfield  {journal}
  {\bibinfo  {journal} {Phys. Rev. Lett.}\ }\textbf {\bibinfo {volume} {93}},\
  \bibinfo {pages} {090404} (\bibinfo {year} {2004})}\BibitemShut {NoStop}%
\end{thebibliography}%

\end{document}